\documentclass[final,5p,times,twocolumn]{elsarticle}

\usepackage[utf8]{inputenc}
\usepackage{graphicx}
\usepackage{amssymb}
\usepackage{pifont}
\usepackage{subcaption}
\usepackage[all]{nowidow}
\usepackage{booktabs}
\usepackage{tabularx}
\usepackage{threeparttable}
\usepackage[svgnames]{xcolor}
\usepackage[most]{tcolorbox}
\usepackage[hidelinks]{hyperref}
\usepackage[capitalise]{cleveref}

\usepackage{lineno}

\biboptions{comma,square,sort}
\makeatletter
\def\ps@pprintTitle{%
	\let\@oddhead\@empty
	\let\@evenhead\@empty
	\def\@oddfoot{\footnotesize\itshape{Accepted for publication in Journal of Systems and Software on 12 October 2023. \hfill}}
	\let\@evenfoot\@oddfoot
}
\makeatother

\crefrangelabelformat{subfigure}{#3#1#4--#5\crefstripprefix{#1}{#2}#6}
\journal{Journal of Systems and Software}

\begin{document}

\begin{frontmatter}

\title{Benchmarking scalability of stream processing frameworks\\deployed as microservices in the cloud}

\author[jku]{Sören Henning\corref{cor1}}
\ead{soeren.henning@jku.at}

\author[cau]{Wilhelm Hasselbring}
\ead{hasselbring@email.uni-kiel.de}

\address[jku]{JKU/Dynatrace Co-Innovation Lab, LIT CPS Lab, Johannes Kepler University Linz, Linz, Austria}
\address[cau]{Software Engineering Group, Kiel University, 24098 Kiel, Germany}

\cortext[cor1]{Corresponding author}

\begin{abstract}\noindent%
\textbf{Context:} The combination of distributed stream processing with microservice architectures is an emerging pattern for building data-intensive software systems. In such systems, stream processing frameworks such as Apache Flink, Apache Kafka Streams, Apache Samza, Hazelcast Jet, or the Apache Beam SDK are used inside microservices to continuously process massive amounts of data in a distributed fashion. While all of these frameworks promote scalability as a core feature, there is only little empirical research evaluating and comparing their scalability.\\
\textbf{Objective:}
The goal of this study to obtain evidence about the scalability of state-of-the-art stream processing framework in different execution environments and regarding different scalability dimensions.\\
\textbf{Method:} We benchmark five modern stream processing frameworks regarding their scalability using a systematic method. We conduct over 740~hours of experiments on Kubernetes clusters in the Google cloud and in a private cloud, where we deploy up to 110 simultaneously running microservice instances, which process up to one million messages per second.\\
\textbf{Results:} All benchmarked frameworks exhibit approximately linear scalability as long as sufficient cloud resources are provisioned. However, the frameworks show considerable differences in the rate at which resources have to be added to cope with increasing load. There is no clear superior framework, but the ranking of the frameworks depends on the use case. Using Apache Beam as an abstraction layer still comes at the cost of significantly higher resource requirements regardless of the use case.
We observe our results regardless of scaling load on a microservice, scaling the computational work performed inside the microservice, and the selected cloud environment.
Moreover, vertical scaling can be a complementary measure to achieve scalability of stream processing frameworks.\\
\textbf{Conclusion:} While scalable microservices can be designed with all evaluated frameworks, the choice of a framework and its deployment has a considerable impact on the cost of operating it.

\end{abstract}


\end{frontmatter}

\newcommand{\rqbase}{RQ\,1}
\newcommand{\rqbeam}{RQ\,2}
\newcommand{\rqwindows}{RQ\,3}
\newcommand{\rqvertical}{RQ\,4}
\newcommand{\rqpublic}{RQ\,5}
\newcommand{\rqshift}{RQ\,6}

\section{Introduction}

Over the last decade, microservices became a frequently adopted software architecture pattern for building scalable, cloud-native software systems~\cite{Hasselbring2017, Soldani2018, Fritzsch2019, Knoche2019}.
More recently, a shift toward adopting distributed stream processing techniques in microservice architectures can be observed~\cite{Laigner2021}. In such systems, different microservices communicate with each other through asynchronous messages, which are sent via scalable messaging systems such as Apache Kafka~\cite{Bellemare2020}. Especially data-intensive applications and big data analytics systems are increasingly designed as microservices building upon frameworks that process continuous data streams in a scalable manner~\cite{Davoudian2020}.

Combining microservice architectures and large-scale stream processing leads to a new way of operating stream processing frameworks. %
Whereas traditional big data systems rely on a single, heavyweight stream processing platform, %
which runs several data analysis jobs, microservice architectures allow running stream processing frameworks inside individual microservices, embedded as a library.
This allows for choosing a suitable, usually more lightweight stream processing framework for each microservice.
From an operational perspective, such microservices are deployed and scaled using established microservice orchestration tooling such as Kubernetes.
Internally, the stream processing frameworks perform the necessary coordination among microservice instances to (re)partition data streams for state locality and, thus, enable scalability.

While there are several open-source stream processing frameworks promoting scalability as a core feature, there is only little empirical research evaluating and comparing their scalability.
However, the need for systematic scalability evaluations has been recognized \cite{vanDongen2020, Hesse2021}.
Benchmarking is a well-established method in software engineering to assess and compare the quality of software systems and services~\cite{PROPSER2021}. As such, it is used both in research and engineering~\cite{Kounev2020} to choose among competing software solutions, to evaluate the quality of new ones, or to assure quality levels over time.

In previous work, we presented and empirically evaluated a scalability benchmarking method for cloud-native applications in general \cite{EMSE2022} and stream processing frameworks in particular \cite{LTB2021}. Further, we presented specific benchmarks %
for stream processing frameworks \cite{BDR2021}.
In this work, we use the presented benchmarking method and the benchmarks to experimentally evaluate the scalability of stream processing frameworks, particularly suited to be used within microservices.
Specifically, we address the following research questions:

\begin{description}
	\item[\rqbase:] \textbf{How do different stream processing frameworks deployed as microservices compete regarding their scalability?}
	Scalability is a main driver for adopting microservices and stream processing-based architectures.
	The choice of a stream processing framework is thus likely to have a crucial impact on the scalability of the overall system.
	\item[\rqbeam:] \textbf{Can the observed performance limitations of Apache Beam's abstraction layer be overcome with recently proposed performance optimization configurations?}
	Our results show that Apache Beam scales with significantly higher resource demands, which has also been reported in related work for an older version of the framework~\cite{Hesse2019}. Since a couple of configuration settings have been proposed for improved performance, we evaluate whether such configurations can improve scalability.
	\item[\rqwindows:] \textbf{How do stream processing frameworks scale with increasing computational work performed inside the microservice?}
	The most common way to assess scalability is by evaluating how a system can handle increasing external load as addressed by \rqbase.
	Likewise, however, it can be important to assess how a framework scales with increasing the complexity of the performed computation, for example, to improve the quality of its results.
	\item[\rqvertical:] \textbf{Can vertical scaling be a complementary measure to achieve scalability of stream processing frameworks?}
	A common way to scale microservices is to increase the number of instances deployed, which may require additional underlying cluster nodes (horizontal scaling). Within the bounds of the individual cluster nodes, microservices can also be scaled by increasing their provided resources, such as CPU cores or memory (vertical scaling). We benchmark both alternatives regarding their scalability on a single large node.
	\item[\rqpublic:] \textbf{Do scalability results differ between public and private clouds?}
	Public clouds and private clouds are both representative environments for operating microservice-based systems. We evaluate whether our scalability benchmark results differ between both environments.
	\item[\rqshift:] \textbf{Can scalability limits be raised by using larger clusters?}
	To investigate whether observed scalability limits are due to the frameworks or just due to high utilization of the underlying cluster resources, we repeat selected experiments with clusters of different sizes.
\end{description}

We address these research questions by conducting over 740~hours of scalability experiments on Kubernetes clusters in Google Cloud and in a private cloud environment.
We benchmark the frameworks Apache Flink, Apache Kafka Streams, Hazelcast Jet, and Apache Beam with the Flink and the Samza runners, for which we deploy up to 110 simultaneously running instances, which process up to one million messages per second.
We provide a replication package and the collected data of all experiments as supplemental material \cite{ReplicationPackage}, allowing other researchers to repeat and extend our work.

\paragraph{Outline}
The remainder of this work starts by discussing the relation of microservices and stream processing as well as the foundations of scalability benchmarking in \cref{sec:background}.
\Cref{sec:frameworks} introduces the stream processing frameworks benchmarked in this study and \cref{sec:clouds} gives an overview of the cloud environments for our experiments.
\Cref{sec:experimental-setup} describes our experimental setup and \cref{sec:experimental-results} presents and discusses the results of our experiments.
Afterward, \cref{sec:threats-to-validity} discusses threats to validity, followed by a discussion of related work in \cref{sec:related-work}.
Finally, \cref{sec:conclusions} concludes this works.

\section{Background}\label{sec:background}

In the following, we first discuss the concepts of stream processing within microservice architectures before outlining our Theodolite scalability benchmarking method used for this research.

\subsection{Stream Processing within Microservice Architectures}

\begin{figure}
	\centering
	\includegraphics[width=\linewidth]{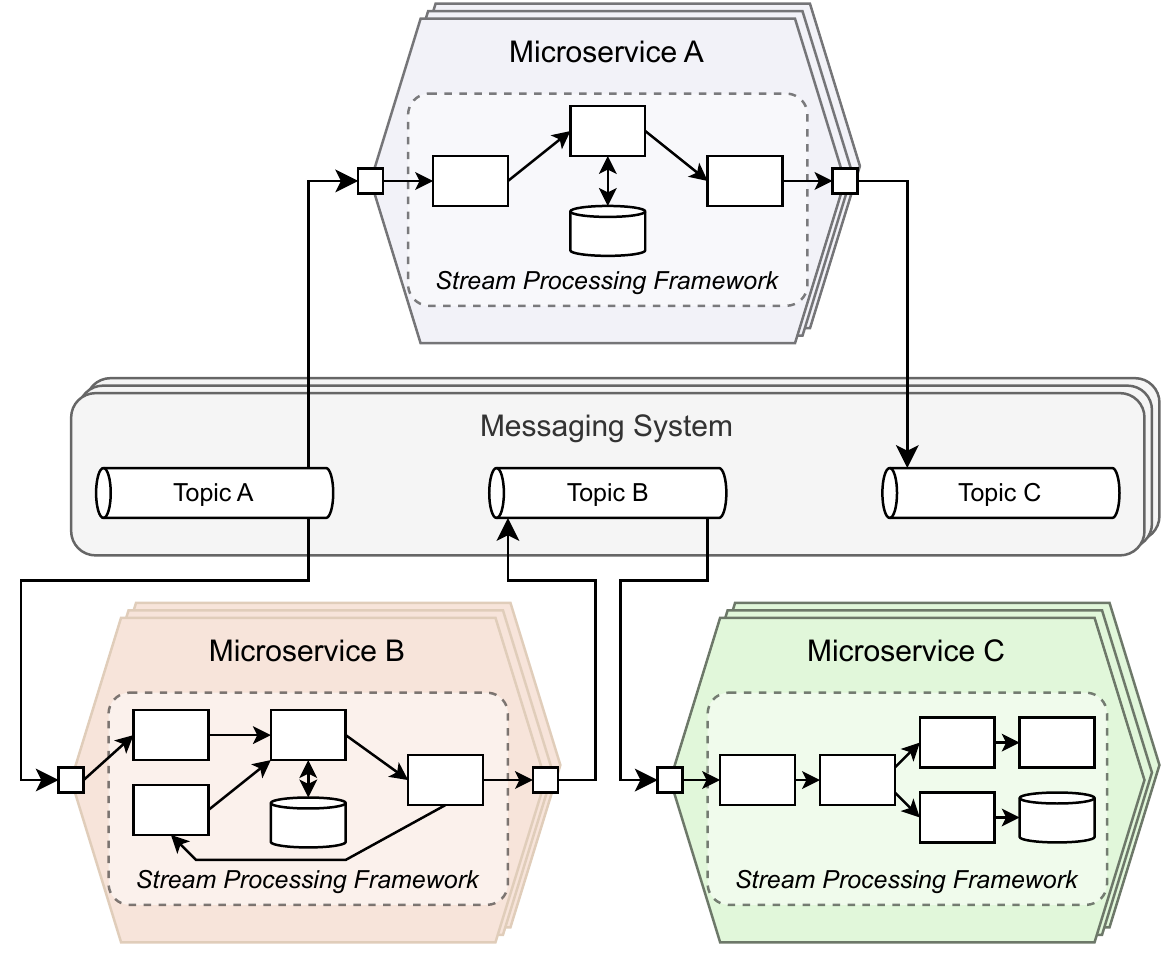}
	\caption{Example microservice architecture of a software system, in which microservices communicate via asynchronous messages and process these messages using stream processing frameworks. With such frameworks, the dataflow inside microservices is modeled as directed graphs of (potentially stateful) \textit{operators}. In this study, we benchmark different state-of-the-art stream processing frameworks with task samples, representing microservices such as those depicted in this example architecture.}
	\label{fig:background:example-architecture}
\end{figure}

At the time of this research, combining microservice architectures and distributed stream processing is covered only superficially in scientific literature.
While there are a couple of case studies reporting on stream processing-based microservices, research is still lacking a systematic evaluation of this new architectural style.
On the other hand, some textbooks for practitioners~\cite{Bellemare2020, Stopford2018} have recently been published, which also serve as references for this work.
Despite the lack of systematic studies, stream processing within microservices is named an emerging trend~\cite{Davoudian2020,Fragkoulis2023,KarabeyAksakalli2021} and the need for further research on this topic is recognized~\cite{Katsifodimos2019,Laigner2021}.
\Cref{fig:background:example-architecture} shows an exemplary architecture with microservices applying stream processing frameworks.\footnote{In practice, one can typically observe a combination of different communication methods, such as additional HTTP-based communication)~\cite{KarabeyAksakalli2021}. Depending on the microservice-specific use case, such architectures may also include microservices relying on other mechanisms to process data streams, such as simple producers/consumers or Function-as-a-Service deployments~\cite{Bellemare2020}.} %
The following attributes can often be observed for such microservices and are particularly relevant to this study.

\paragraph{Self-contained and loosely coupled}
In microservice architectures, a software system is composed of multiple small services that are built around business capabilities~\cite{Hasselbring2018}. Individual microservices run in their own processes, may use different technology stacks, and communicate via lightweight, fault-tolerant mechanisms over the network.
For microservices employing distributed stream processing techniques, this means each service can use its own stream processing framework.
In contrast to traditional big data stream processing systems running on top of resource management systems such as Apache Hadoop YARN or Apache Mesos, %
this also leads to smaller jobs and a single job per stream processing cluster.

\paragraph{Independently scalable}

In contrast to monolithic systems, individual microservices can be scaled independently due to their loose coupling. %
In fact, scalability has been reported as one of the most important drivers for and benefits of adopting microservice architectures in several systematic literature reviews \cite{Pahl2016, Li2021, Kratzke2017, Soldani2018, Laigner2021}, interview studies \cite{Taibi2017, Fritzsch2019, Knoche2019, Laigner2021, Zhou2023}, and experience reports \cite{Balalaie2016b, Hasselbring2017, Bucchiarone2018}.
Typically, horizontal duplication, data partitioning, and function decomposition are distinguished as methods for scaling microservice architectures. 
From an operation's perspective, also microservices apoting stream processing frameworks can be scaled by horizontal duplication. Internally, however, duplicating services leads to data partitioning, meaning that each instance of a service handles only messages with certain \textit{keys}.
The actual data partitioning as well as the necessary state management and fault tolerance is managed by modern stream processing frameworks.

\paragraph{Cloud-native deployment}
Microservice architectures are a pattern, particularly suited for building cloud-native applications~\cite{Balalaie2016,Gannon2017,Pahl2018}.
They are mostly deployed as containers in public or private cloud environments, with Kubernetes~\cite{Burns2016} being the de-facto standard orchestration tool for cloud-native applications~\cite{CNCF2022}.
With Kubernetes, microservice deployments including, for example, resource restrictions or numbers of replicas, are defined purely declaratively.

\paragraph{Asynchronous communication via messaging system}

Microservices that process and consume streams of messages often employ log-based messaging systems for their communication~\cite{KarabeyAksakalli2021}.
To eventually reach consistency among individual microservices, the log must be durable, append-only, fault-tolerant, partitioned, and it must support sequential reads~\cite{Kleppmann2019}.
Probably the most prominent messaging system fulfilling these properties is Apache Kafka~\cite{Kreps2011,Wang2015}, which is intensively used in industry.\footnote{\url{https://kafka.apache.org/powered-by}}
Log-based messaging systems such as Kafka are required by modern stream processing frameworks to provide strict processing guarantees and fault tolerance while scaling out~\cite{Fragkoulis2023}.

\paragraph{Adoption of distributed stream processing techniques}

Modern stream processing frameworks are designed to run in a distributed fashion on commodity hardware in order to scale with massive amounts of data~\cite{Fragkoulis2023}. Besides high throughput, these systems focus on low latency, fault tolerance, and coping with out-of-order streams.
Modern stream processing frameworks process data in jobs, where a job is defined as a dataflow graph of processing operators. They can be started with multiple instances (e.g., on different compute nodes, containers, or with multiple threads). For each job, each instance processes only a portion of the data. Whereas isolated processing of data records is not affected by the assignment of data portions to instances, processing that relies on previous data records (e.g., aggregations over time windows) requires the management of state. Similar to the MapReduce programming model, keys are assigned to records and the stream processing frameworks guarantee that all records with the same key are processed by the same instance. Hence, no state synchronization among instances is required. When a processing operator changes the record key and a subsequent operator performs a stateful operation, the stream processing framework splits the dataflow graph into subgraphs that can be processed independently by different instances.
We refer to the recent surveys of \citet{Fragkoulis2023} and \citet{Margara2022} for detailed information on state-of-the-art stream processing models and patterns.

\subsection{Scalability Benchmarking with Theodolite}

Our study builds upon our Theodolite scalability benchmarking method for cloud-native applications~\cite{EMSE2022}.
According to the ACM SIGSOFT Empirical Standard for benchmarking as a software engineering research method~\cite{Ralph2021,PROPSER2021}, we briefly summarize the quality, metric, measurement method, and task samples used in this study.\footnote{\url{https://acmsigsoft.github.io/EmpiricalStandards/docs/?standard=Benchmarking}}

\paragraph{Quality}
The quality evaluated in this study is scalability. Scalability is defined as ``the ability of [a] system to sustain increasing workloads by making use of additional resources''~\cite{Herbst2013}.
It should not be confused with elasticity, which describes how fast or how precise a system (automatically) adapts to varying workloads~\cite{Herbst2013,Lehrig2015}.

\paragraph{Metric}
The Theodolite scalability benchmarking method provides two alternative scalability metrics, the \textit{resource demand} metric and the \textit{load capacity} metric. In this study, we focus on the \textit{resource demand} metric.
It is a function, mapping the load intensity on a system under test (SUT) to the minimal amount of resources that must be provisioned for the SUT such that the SUT fulfills all specified service-level objectives (SLOs).
In \cref{sec:experimental-setup:method}, we describe how load, resources, and SLOs are defined in this study.

\paragraph{Measurement method}
To measure scalability according to the metric, we chose discrete subsets of the load and resource domains and run isolated performance experiments for different load and resource combinations to assess whether the specified SLOs can be fulfilled.
Experiment duration, warm-up periods, and the number of repetitions are configurable and have to be adjusted to the context. By using appropriate search strategies, not all combinations of load and resources have to be executed.

\paragraph{Task samples}
Our Theodolite benchmarking method is not restricted to specific task samples, but allows also using existing benchmarks for cloud-native applications.
Based on real-world use cases for Industrial Internet of Things analytics~\cite{DIMA2021}, we proposed four task samples of different complexity for stream processing frameworks in previous work~\cite{BDR2021}.
As we discuss in \cref{sec:related-work}, these are the only benchmarks focusing on modern stream processing frameworks deployed as cloud-native microservices.
In \cref{sec:experimental-setup:task-samples},
we describe how we configure these task samples for our evaluation.

\section{Evaluated Stream Processing Frameworks}\label{sec:frameworks}

In their textbook, \citet{Bellemare2020} distinguishes between lightweight and heavyweight stream processing frameworks.
Lightweight frameworks are embedded as a programming library into the source code of independently deployable components such as microservices.
This way, the stream processing framework does not require any specific way to build or deploy the microservice. This allows the service to also perform other tasks beyond stream processing such as providing a REST API. Individual instances of a service discover each other (e.g., via features of the messaging system or Kubernetes) and perform the necessary coordination internally.
Heavyweight frameworks on the contrary are provided as deployable software systems, which can be configured by one or more stream processing jobs to be executed. They are typically designed as a master--worker architecture. In this study, we mainly focus on lightweight frameworks.
All modern stream processing frameworks can be deployed containerized on commodity hardware with Kubernetes.
In the following, we give a brief overview of frameworks, particularly suited to be deployed as microservices and which we decided to benchmark in this study.
For a detailed comparison of the framework's features, see the works of, for example, \citet{Hesse2015}, %
\citet{Fragkoulis2023}, and \citet{vanDongen2021c}.

\paragraph{Apache Flink}
Originating from a scientific research project~\cite{Alexandrov2014}, Apache Flink~\cite{Carbone2015} has been extensively used, evaluated, and extended in research and became increasingly popular in industry.
It offers one of the most elaborated dataflow models, providing precise control of time and state~\cite{Carbone2015,Carbone2017,Akidau2021}.
Moreover, Flink provides different abstraction layers and a rich feature set regarding the integration with external systems.
Flink clearly falls into the category of heavyweight frameworks~\cite{Bellemare2020}. Its deployment consists of one or---for fault-tolerance---more coordinating \textit{JobManagers} and a scalable amount of \textit{TaskManagers}.
Although heavyweight, we consider Flink in this study due to its widespread adoption and since we observe recent trends to more lightweight deployments of Flink.

\paragraph{Apache Kafka Streams}
Kafka Streams~\cite{Sax2018,Wang2021} is a stream processing framework built on top of Apache Kafka.
It is available as a Java library and, thus, aligns with the idea of incorporating stream processing in standalone microservices.
Compared to most other stream processing frameworks, it has a restricted set of features, in particular, concerning the integration with external systems.
Kafka Streams only supports Kafka topics as data sources and sinks.
Integrating external systems always requires transferring data via Kafka topics.

\paragraph{Apache Samza}
Similar to Kafka Streams, Apache Samza~\cite{Noghabi2017} can be embedded as a library in standalone applications. Individual instances of the same application use Apache Zookeeper and Apache Kafka for coordination, data transfer, and fault tolerance.
Although still maintained, Samza is sometimes considered a predecessor of Kafka Streams~\cite{Kleppmann2015}.
In this study, we use Samza as a runner for Apache Beam pipelines (see below), which allows implementing more complex use cases~\cite{Zhang2020}. 

\paragraph{Hazelcast Jet}
Hazelcast Jet~\cite{Gencer2021} is a stream processing framework built on top of the Hazelcast IMDG distributed, in-memory object store.
It can be embedded into Java applications and does not have any dependencies on an external system.
Instead, individual instances discover each other, form a cluster, and handle coordination and data replication internally.
Hazelcast Jet differs from other frameworks in its execution model, which is based on a concept similar to coroutines and cooperative threads~\cite{Gencer2021}.
With the release of Hazelcast~5.0 in 2021, Hazelcast Jet has been merged with Hazelcast IMDG into one unified product.

\paragraph{Apache Beam}

Apache Beam is not a stream processing system by itself, but instead, an SDK to implement stream processing jobs in a uniform model, which can be executed by several modern stream processing systems.
Apache Beam implements Google's Dataflow model \cite{Akidau2015}, which is also internally used by Google's cloud service \textit{Google Cloud Dataflow}.
A stream processing job implemented with Apache Beam is executed by a so-called runner. Runners can be seen as adapters for the actual stream processing systems. Besides a runner for Google Cloud Dataflow, Apache Beam provides also runners for several other systems, including the aforementioned Apache Flink, Apache Samza, and Hazelcast Jet.
Previous research found that using Apache Beam as an abstraction layer comes with a significant negative impact on performance~\cite{Hesse2019}.

\paragraph{Other Stream Processing Frameworks}
Apache Spark \cite{Zaharia2016}, Apache Storm \cite{Toshniwal2014}, and the successor of the latter, Apache Heron \cite{Kulkarni2015}, are considered heavyweight frameworks and, thus, fit less into the context of microservices~\cite{Bellemare2020}.
Spark differs from the frameworks discussed before in that it processes data streams in ``micro-batches''.
Storm and Heron provide less sophisticated programming models and weaker fault tolerance mechanisms~\cite{Fragkoulis2023}.
Moreover, there are several cloud services available for stream processing, which, however, are out of scope of our study.

\begin{table*}
	\caption{Configuration of Kubernetes clusters used for our evaluation, running at Google Cloud and a private cloud (SPEL).}
	\label{tab:clouds}
	\newcolumntype{L}{>{\raggedright\arraybackslash}X}
	\newcolumntype{R}{>{\raggedleft\arraybackslash}X}
	\newcolumntype{C}{>{\centering\arraybackslash}X}
	\newcolumntype{o}{p{0pt}}
	\begin{tabularx}{\linewidth}{LLLLL}
		\toprule
		& \cref{sec:experimental-results:base,sec:experimental-results:beam-config,sec:experimental-results:windows,sec:experimental-results:public-private} & \cref{sec:experimental-results:vertical} & \cref{sec:experimental-results:public-private}& \cref{sec:experimental-results:large-cluster} \\
		\midrule
		Cloud platform & Private cloud & Private cloud & Google Cloud & Google Cloud\\
		Nodes & 5 & 5 & 5 & 8--12 \\
		CPU cores & $2 \times 16$ & $2 \times 16$ & $32$ & $16$ \\
		RAM & $384$\,GB & $384$\,GB & $128$\,GB  & $64$\,GB \\
		Machine type &  Intel Xeon Gold 6130 & Intel Xeon Gold 6130  & e2-standard-32 & e2-standard-16 \\
		Kubernetes & 1.23.7 & 1.23.7 & 1.24.11-gke.1000 & 1.24.11-gke.1000 \\
		Kafka brokers & 5 & 3 & 5 & 4 \\
		\bottomrule
	\end{tabularx}
\end{table*}

\section{Evaluated Cloud Platforms}\label{sec:clouds}

We run our experiments in a private and a public cloud, representing two common execution environments for microservice-based software systems. On both platforms, we operate Kubernetes clusters.
For the public cloud infrastructure, we chose Google Cloud, representing one of the largest cloud vendors %
with a Kubernetes offering that provides multiple configuration options.
As private cloud infrastructure, we use the Software Performance Engineering Lab (SPEL) at Kiel University.
\Cref{tab:clouds} summarizes the configuration of our Kubernetes clusters for all experiments.

\paragraph{Google Cloud}
The Google Cloud clusters consist of \textit{e2-standard} virtual machines.
This machine type is promoted by Google as cost-optimized and suitable for general purposes such as microservices.
In the course of our experiments, we use clusters with different numbers and sizes of such VMs.
All clusters are set up and managed with the Google Kubernetes Engine (GKE) service and run in the \textit{us-central1} region.

\paragraph{Private cloud (SPEL)}
In the private cloud, a self-operated Kubernetes is installed via Kubespray on five bare metal nodes, connected via 10~Gigabit Ethernet. Each node is equipped with two 16-core CPUs and 384\,GB of memory.

\section{Experimental Setup}\label{sec:experimental-setup}

\begin{figure}
	\centering
	\includegraphics[width=\linewidth]{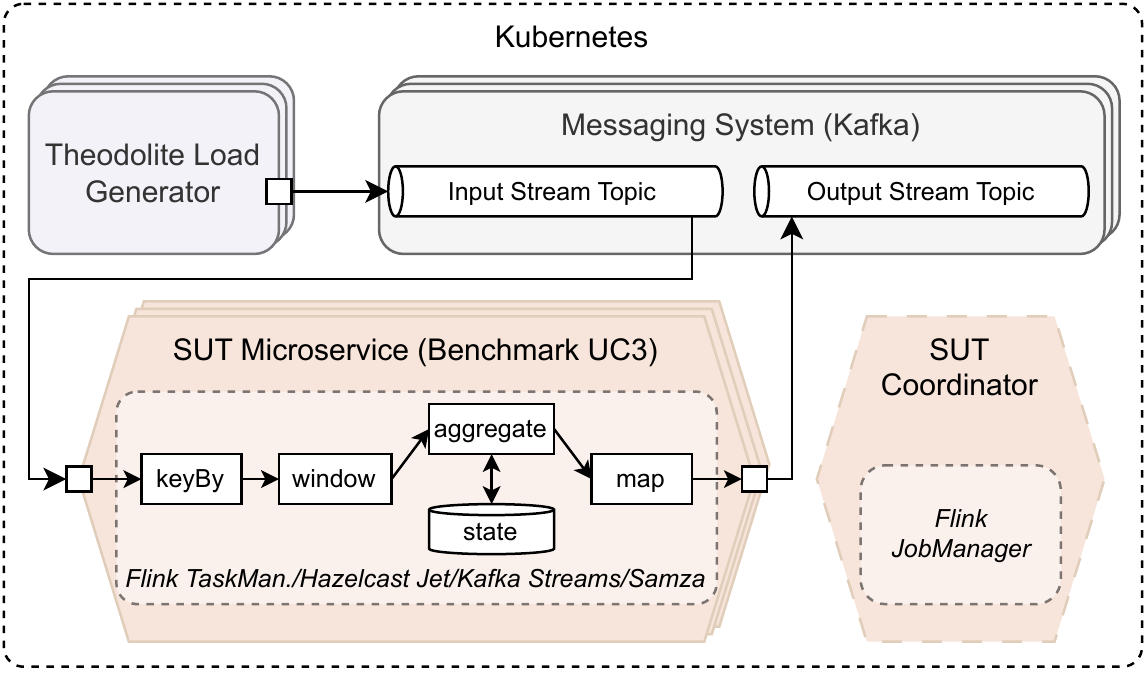}
	\caption{Benchmark deployment illustrated for the case of benchmark UC3. Our Theodolite load generator sends messages with a constant, but configurable frequency to the input stream topic in Kafka. From there, our SUT microservice consumes the messages, aggregates them and sends the results to the output stream topic in Kafka. The actual processing inside the SUT microservice is implemented with a stream processing framework. For each evaluated stream processing framework, we implemented dedicated microservice. The implementation of Apache Flink requires an additional coordinating instance running the Flink JobManager. The deployment for the other task samples look similar, however, with different dataflow architectures~\cite{BDR2021} and potentially different input and output streams.}
	\label{fig:experimental-setup:benchmark-architecture}
\end{figure}

We run our experiments with our Theodolite scalability benchmarking framework\footnote{\url{https://www.theodolite.rocks/}} (version 0.9).
Theodolite is installed as a Kubernetes Operator inside the Kubernetes clusters and controls the execution of benchmarks according to our scalability benchmarking method~\cite{IC2E2022Demo}.
Unless stated differently, we run 5~Kafka brokers, one on each node, with Kafka version 3.2 and 100~partitions per Kafka topic.
All implementations of the benchmarks and the benchmarking tool itself are available as open-source software.\footnote{\url{https://github.com/cau-se/theodolite}}
In our replication package \cite{ReplicationPackage}, we provide the declarative Theodolite files used for executing the benchmarks and the collected data of all experiments along with analysis scripts, allowing other researchers to repeat and extend our work.
In the following, we summarize the configuration of the benchmarked stream processing frameworks, the selected benchmark task samples, and the benchmarking method.
\Cref{fig:experimental-setup:benchmark-architecture} illustrates our benchmark deployment.

\subsection{Configuration of Stream Processing Frameworks}\label{sec:experimental-setup:frameworks}

We benchmark the stream processing frameworks Apache Beam (version 2.35) with the Flink (version 1.13) and the Samza runner (version 1.5), Apache Flink (version 1.13), Hazelcast Jet (version 4.5), and Apache Kafka Streams (version 3.1). For a fair comparison, we evaluate all frameworks with mostly their default configuration.

We enable committing read offsets to Kafka in all frameworks. This allows us to monitor the consumer lag via Kafka metrics, which is required to evaluate our lag trend SLO. %
Enabling offset committing is also often done in production deployments to increase observability.
We set the commit interval to 5~seconds for all frameworks, which is the default configuration of Kafka consumers once offset committing is enabled.
Kafka Streams has a default commit interval of 30~seconds as in Kafka Streams, the commit interval also controls fault tolerance (comparable to the checkpointing interval in other frameworks).
For our experiments with Apache Beam and the Flink runner, we enable the \textit{FasterCopy} option as we further discuss and evaluate in \cref{sec:experimental-results:beam-config}.
Each microservice instance runs as a container in its own Kubernetes Pod with an additional sidecar container, exposing monitoring data specific to the stream processing framework.
Per default, we configure 1~CPU and 4\,GB of memory for each Pod, which is a common ratio of CPU and memory of cloud VMs.
For Flink, we always have one additional Pod running the JobManager (see \cref{fig:experimental-setup:benchmark-architecture}).

\subsection{Configuration of Task Samples}\label{sec:experimental-setup:task-samples}

As benchmark task samples, we use dataflow architectures representing typical use cases for analyzing IIoT sensor data streams.
The task samples consume messages representing simulated power consumption measurements and produce messages with aggregation results.
Our previous publication~\cite{BDR2021} describes our four benchmark task samples, named UC1--UC4, in detail.
Unless otherwise stated, we use the following configuration of our benchmark dataflow architectures.
\begin{itemize}
	\item Benchmark UC1 is configured to write each incoming message as a log statement to the standard output stream to simulate a database write operation (i.e., simulating a side effect in the dataflow architecture).
	\item Benchmark UC2 aggregates incoming messages over tumbling windows of one minute. Any out-of-order records arriving after the window has been closed are discarded. %
	\item Benchmark UC3 aggregates records by their hour of day attribute over a time window of three days with a slide period of one day. That means each incoming record belongs to three time windows. Early results (i.e., before the end of the time window has passed) are emitted every 5~seconds.
	For Kafka Streams, such emission cannot explicitly be configured. However, Kafka Streams continuously forwards aggregation results based on the configured \textit{commit interval} (which is 5~seconds as well).
	\item Benchmark UC4 aggregates incoming messages in nested groups of sensors. In contrast to previous work, we benchmark a simplified version of benchmark UC4, which omits the feedback loop. This allows for better predictability of the message volume and, hence, more comparable results.
\end{itemize}

\subsection{Configuration of the Benchmarking Method}\label{sec:experimental-setup:method}

Our Theodolite benchmarking method assesses scalability regarding a configurable load type, a resource type, and SLOs.

\paragraph{Load type}
If not stated differently, we evaluate scalability in regards to increasing the number of simulated sensors as load type. In benchmark UC4, this is indirectly controlled by increasing the number of nested groups, with each group containing $4$ sub-groups or sensors. This means for $n$ nested groups, we simulate $4^n$ sensors. Each simulated sensor generates one measurement per second.
When addressing \rqwindows, we evaluate scalability in regards to increasing the size of time windows for which data is aggregated.

\paragraph{Resource type}
Unless otherwise stated, we use the number of instances as resource type.
When addressing \rqvertical, we additionally use the CPU and memory resources of a single instance as resource type.

\begin{figure*}
	\centering
	\begin{subfigure}[b]{0.33\linewidth}%
		\centering%
		\includegraphics[width=\textwidth]{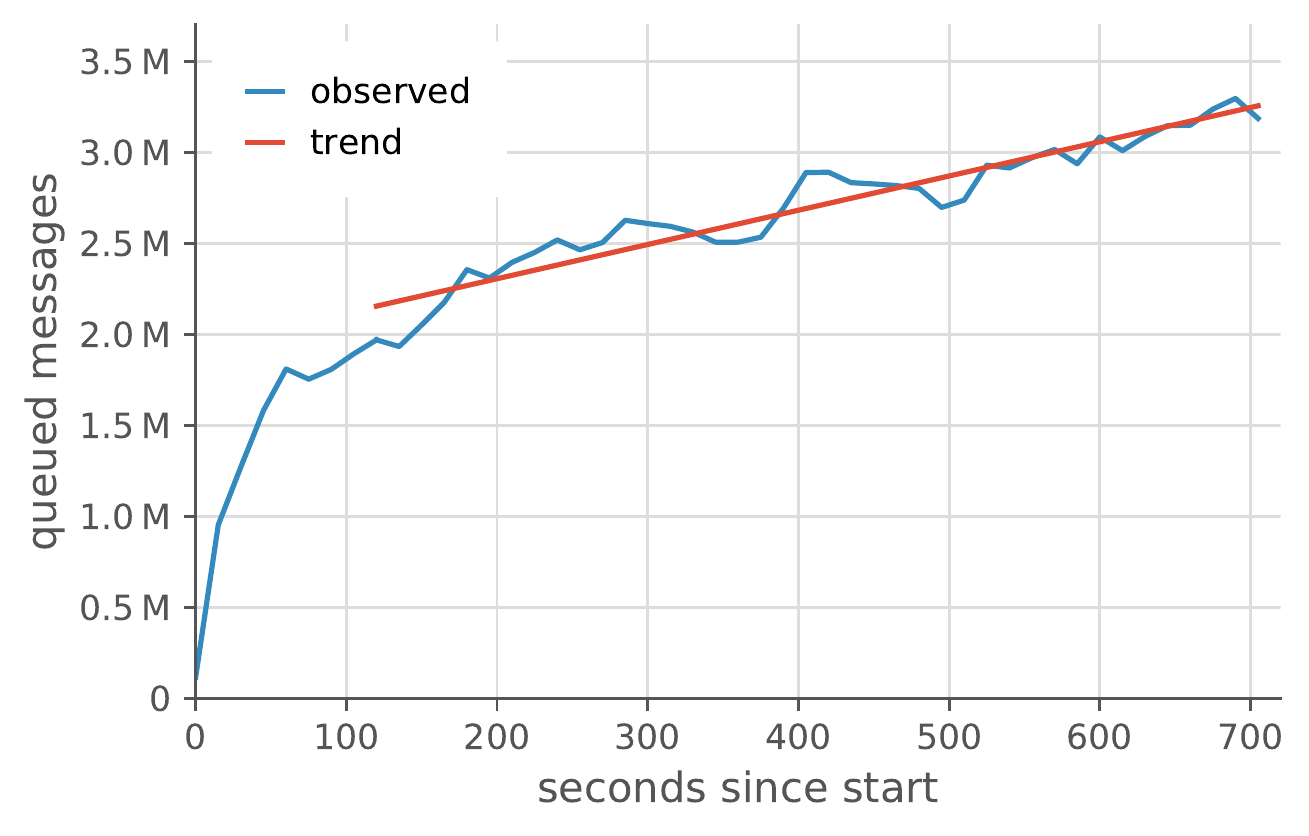}%
		\caption{6 SUT instances}%
	\end{subfigure}%
	\hfill%
	\begin{subfigure}[b]{0.33\linewidth}%
		\centering%
		\includegraphics[width=\textwidth]{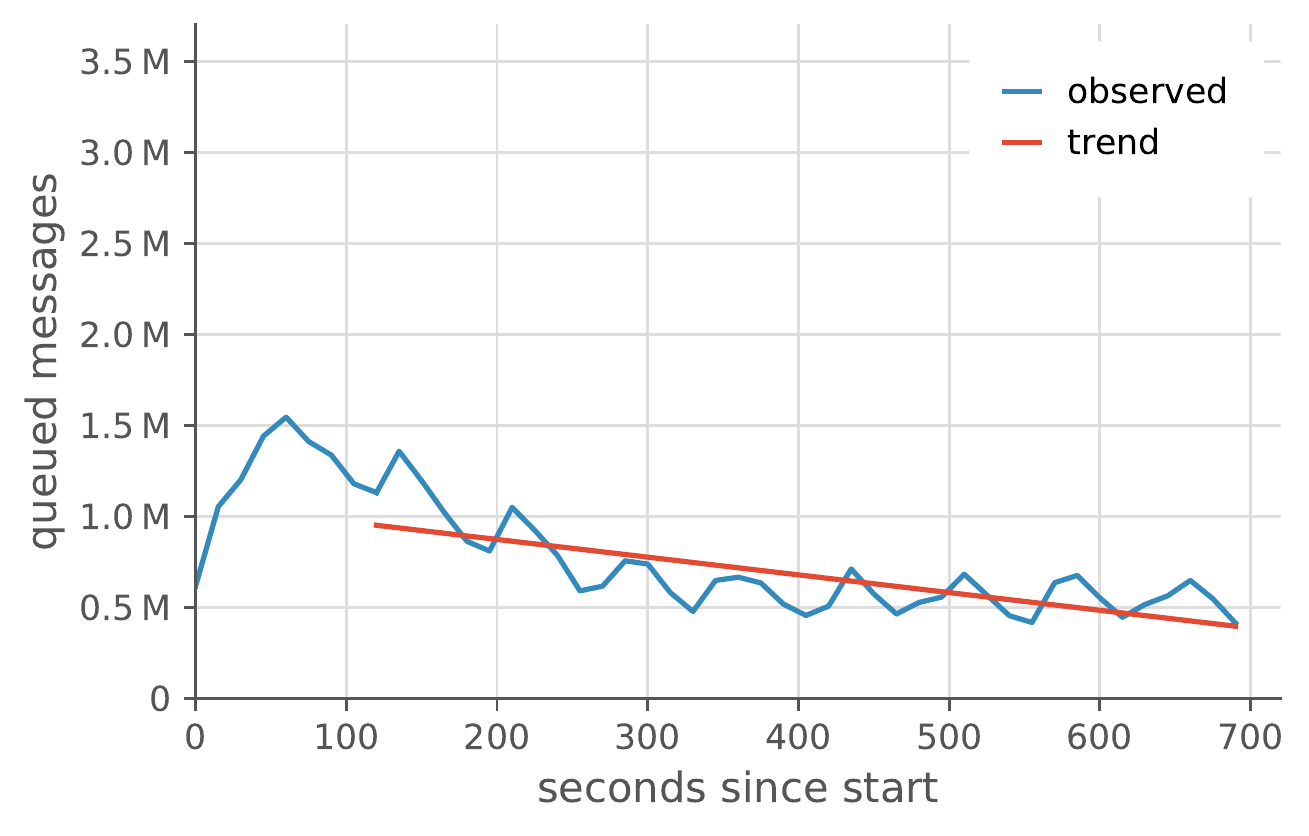}%
		\caption{7 SUT instances}%
	\end{subfigure}%
	\hfill%
	\begin{subfigure}[b]{0.33\linewidth}%
		\centering%
		\includegraphics[width=\textwidth]{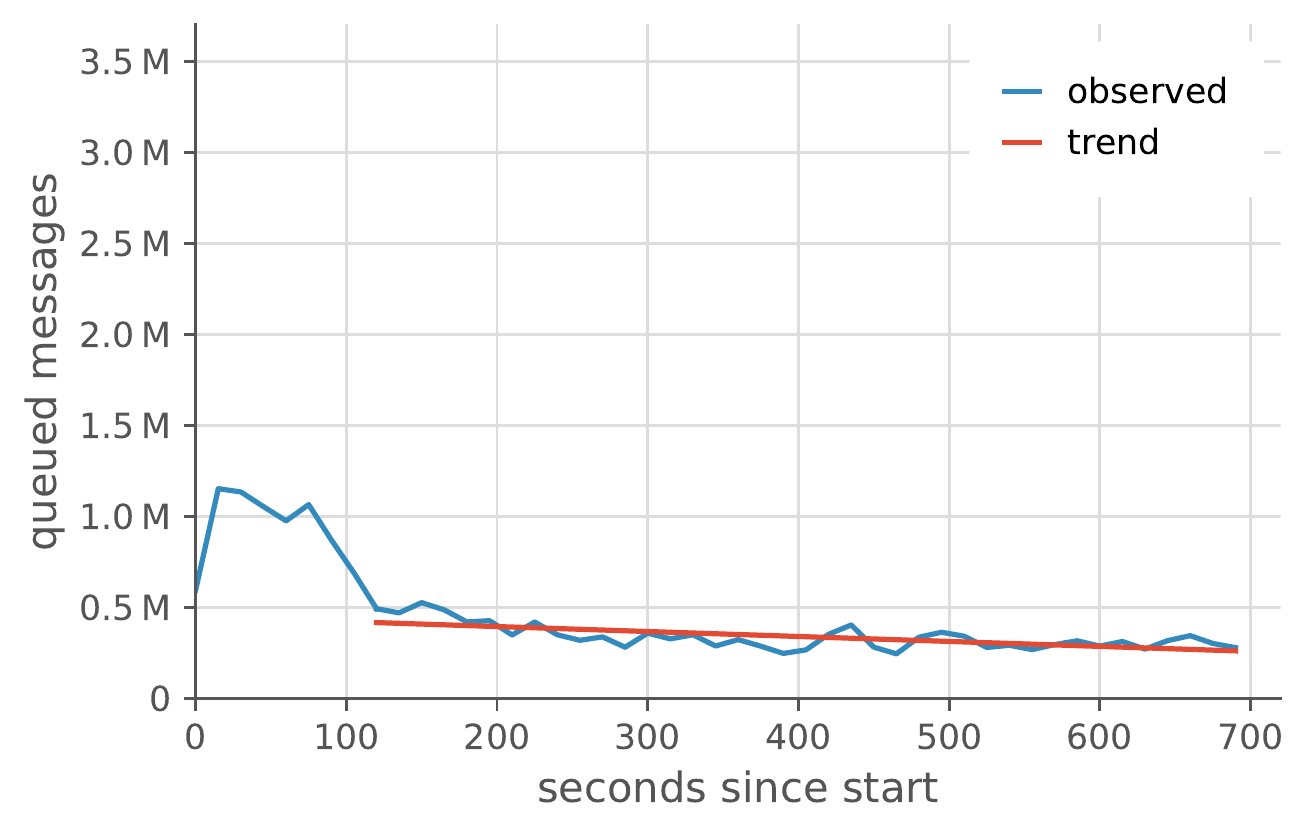}%
		\caption{8 SUT instances}%
	\end{subfigure}
	\caption{Illustration of the consumer lag trend metric for an exemplary benchmark execution (Theodolite's UC3 benchmark implemented with Kafka Streams and 50\,000 messages/second) with different numbers of SUT instances.
		Independent of the number of instances, we can observe a variable lag.
		However, computing a trend line (without considering the measurements from an initial warmup period), reveals that for 6 instances, the number of queued messages will steadily increase over time.
		Providing 7 instances leads to a decrease in queued messages after the warmup period, while 8 instances yields an almost constant trend line.}
	\label{fig:lag-trend-example}
\end{figure*}

\paragraph{SLOs}
In all benchmark executions, we use an SLO based on our consumer lag trend metric~\cite{BDR2021}. The consumer lag trend describes how many messages are queued in the messaging system, which have not been processed. Our consumer lag trend metric describes the average increase (or decrease)
of the lag per second. It can be measured by monitoring the lag and computing a trend line using linear regression. The slope of this line is the lag trend.
\Cref{fig:lag-trend-example} illustrates the concept of the lag trend.
For our experiments, we consider the SLO to be fulfilled if the lag trend does not increase by more than 1\% of the generated message volume.
For Apache Samza, we set this threshold to 5\% of the generated message volume since we observe a permanent slightly increase independent of the provided resources.

In certain cases, we observed that under high load the consumer lag does not substantially increase, but records were discarded due to lateness. In most stream processing frameworks, operations on time windows still accept out-of-order records for a configurable amount of time. If this time has elapsed, records are discarded and not further processed. Thus, records are still consumed from the messaging system, not causing a consumer lag increase, but results become incorrect.
To detect these cases, our benchmarks UC2 and UC4 (which aggregate data in short windows) contain a second SLO, which requires that no more than 1\% of the generated messages are discarded.
For the Apache Beam implementation with the Samza runner, no metrics concerning the number of dropped records are provided. With Beam's Flink runner, these metrics are only unreliably available.\footnote{We asked a corresponding question regarding the metrics of both runners at Beam's mailing list, but did not receive an answer.}
This means that for these two SUTs, we cannot definitely be sure whether the determined resource demand for UC2 and UC4 is sufficient to process all records successfully, but still indicates a lower bound.

\paragraph{Additional configuration}
According to the previous experimental evaluations of our benchmarking method~\cite{EMSE2022},
we run our experiments with benchmark UC1--UC3 for a duration of 5~minutes while considering the first 2~minutes as warm-up period. As benchmark UC4 shows a higher variability in the results, we run its experiments for 10~minutes including a 4-minute warm-up period.
We repeat all our experiments 3~times, which we experimentally observed to be a good trade-off between overall execution time (or costs in the public cloud) and statistical grounding~\cite{EMSE2022}.
We provide a replication package~\cite{ReplicationPackage} allowing to further repeat our experiments.
We quantify scalability with the resource demand metric and use the \textit{linear search} strategy in combination with the \textit{lower bound restriction}~\cite{EMSE2022}.

\section{Experimental Results}\label{sec:experimental-results}

In this section, we discuss and present the results of our benchmark executions.
\cref{sec:experimental-results:base} addresses \rqbase\ by running baseline experiments for each benchmark and framework.
\cref{sec:experimental-results:beam-config} addresses \rqbeam\ and investigates the impact of recently proposed performance optimization configurations for Apache Beam.
\cref{sec:experimental-results:windows} addresses \rqwindows\ and evaluates how different stream processing frameworks scale when increasing the duration of window aggregations.
\cref{sec:experimental-results:vertical} addresses \rqvertical\ by evaluating how stream processing frameworks scale on a single node.
\cref{sec:experimental-results:public-private} addresses \rqpublic\ and compares our baseline results in the private cloud with those of a public cloud.
Finally, \cref{sec:experimental-results:large-cluster} addresses \rqshift\ and repeats the same experiments for benchmark UC3 in Kubernetes clusters of different sizes.

\subsection{Baseline Comparison of Frameworks}\label{sec:experimental-results:base}

\begin{figure*}
	\centering
	\begin{subfigure}[b]{0.495\linewidth}%
		\centering
		\includegraphics[width=\textwidth]{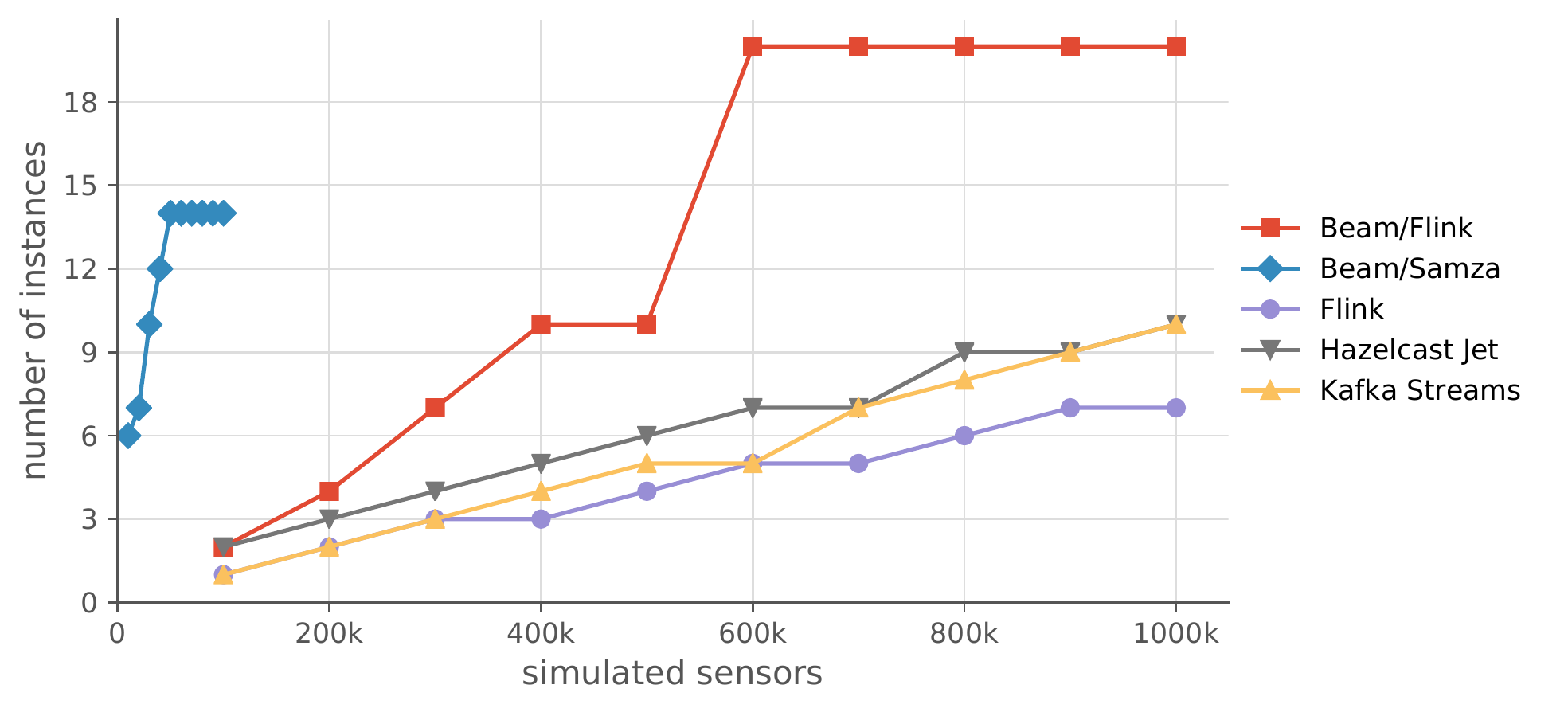}
		\caption{UC1}
		\label{fig:eval:frameworks:base:uc1}
	\end{subfigure}
	\hfill%
	\begin{subfigure}[b]{0.495\linewidth}%
		\centering
		\includegraphics[width=\textwidth]{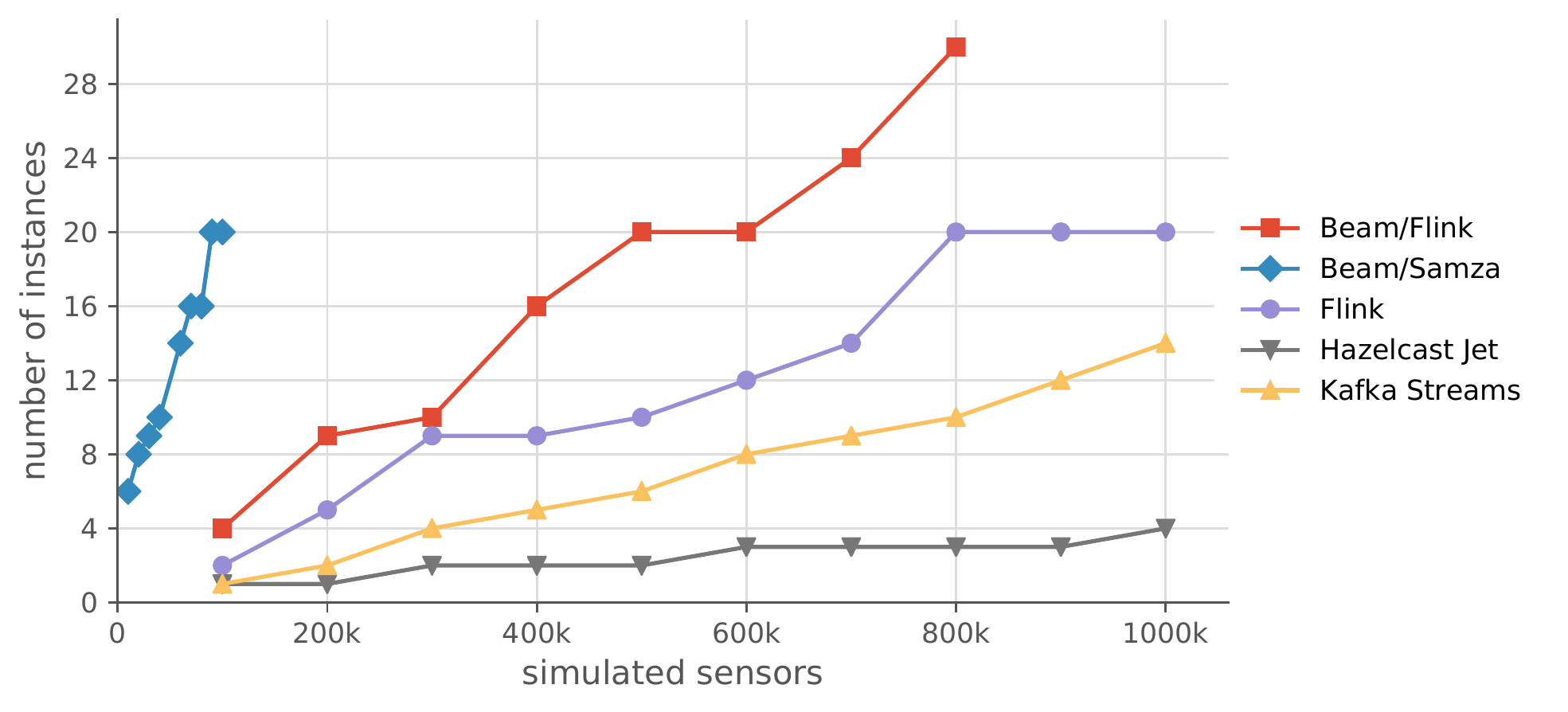}
		\caption{UC2}
		\label{fig:eval:frameworks:base:uc2}
	\end{subfigure}
	
	\vspace{1em}
	
	\begin{subfigure}[b]{0.495\linewidth}%
		\centering
		\includegraphics[width=\textwidth]{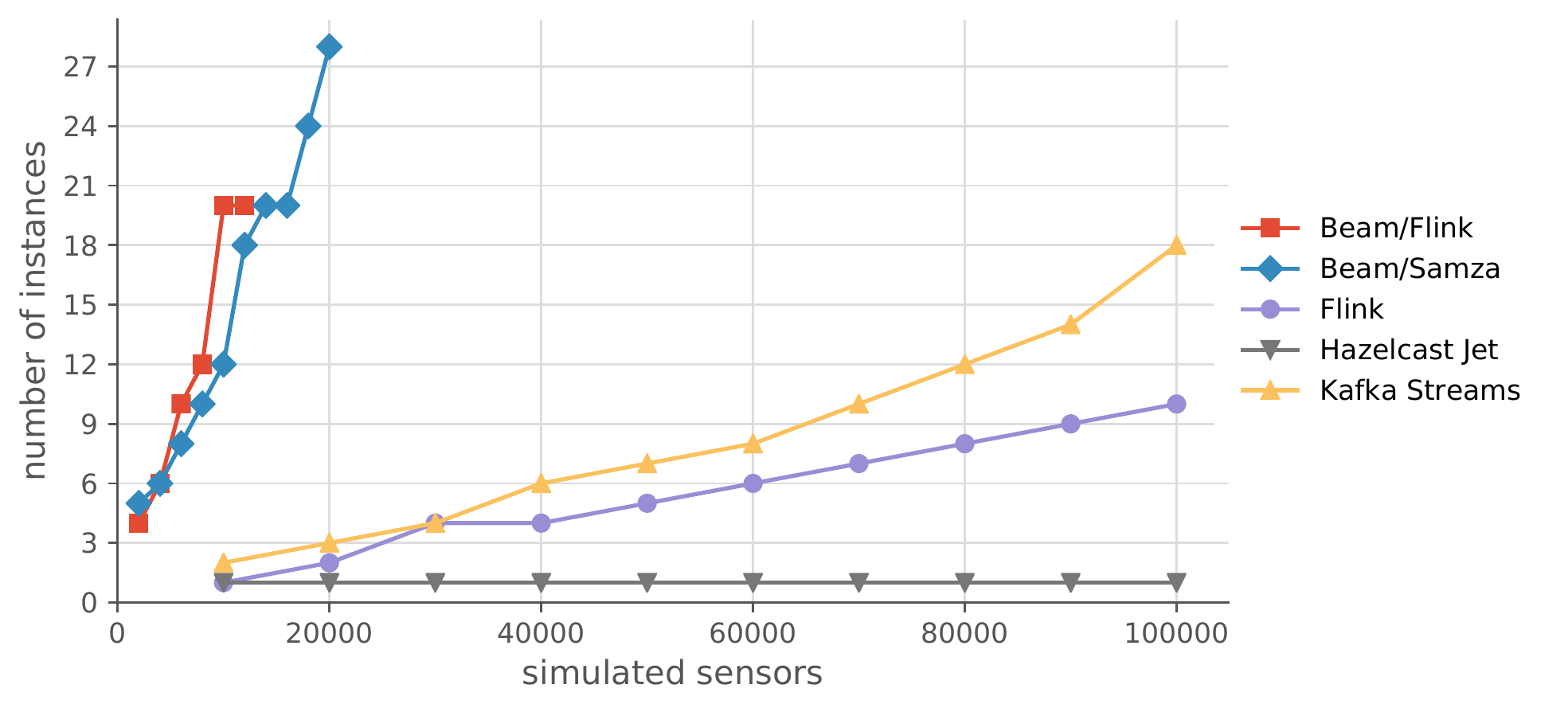}
		\caption{UC3}
		\label{fig:eval:frameworks:base:uc3}
	\end{subfigure}
	\hfill%
	\begin{subfigure}[b]{0.495\linewidth}%
		\centering
		\includegraphics[width=\textwidth]{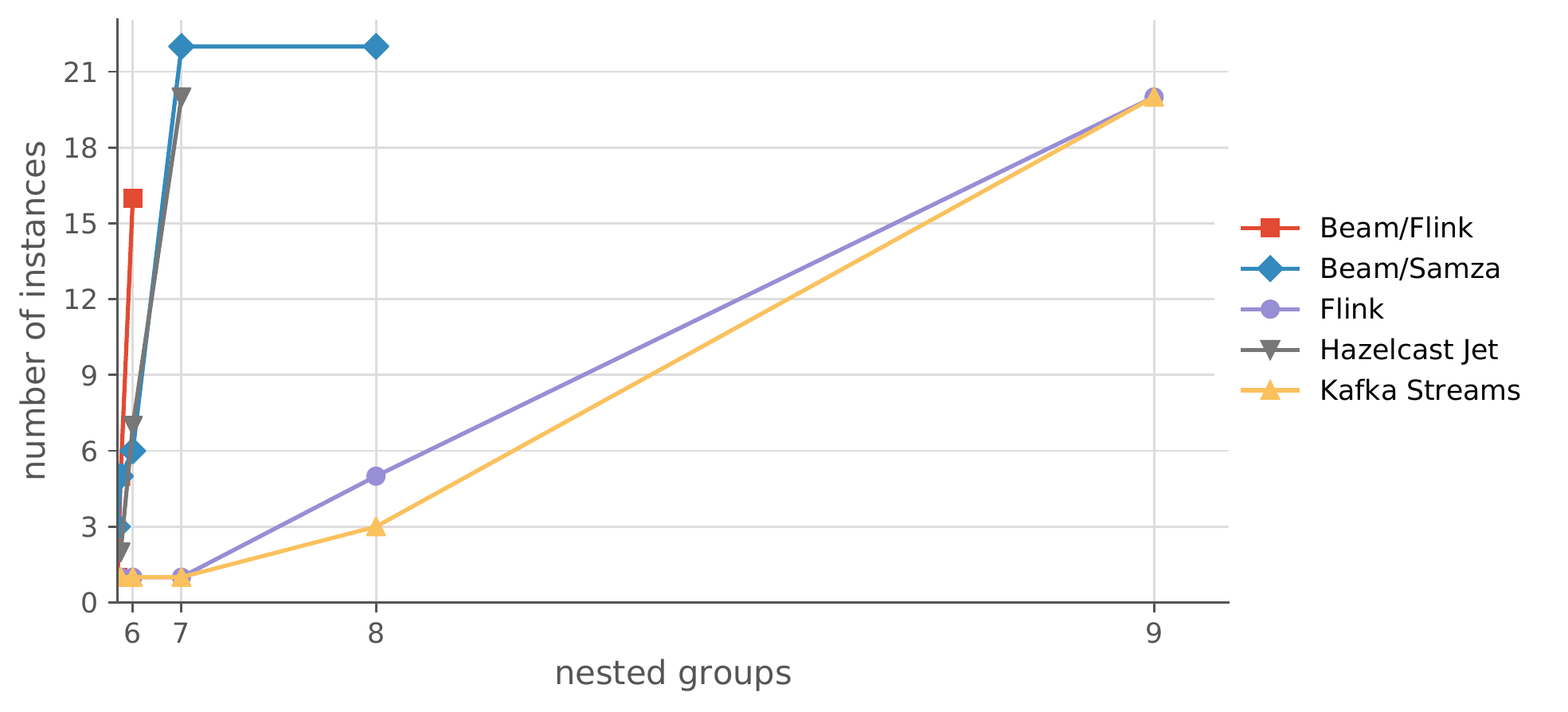}
		\caption{UC4}
		\label{fig:eval:frameworks:base:uc4}
	\end{subfigure}
	\caption{Scalability benchmark results according to our resource demand metric for the stream processing frameworks Apache Beam (with the Flink and Samza runners), Apache Flink, Hazelcast Jet, and Apache Kafka Streams.}
	\label{fig:eval:frameworks:base}
\end{figure*}

\begin{figure}
	\centering
	\begin{subfigure}[b]{0.495\linewidth}%
		\centering
		\includegraphics[width=\textwidth]{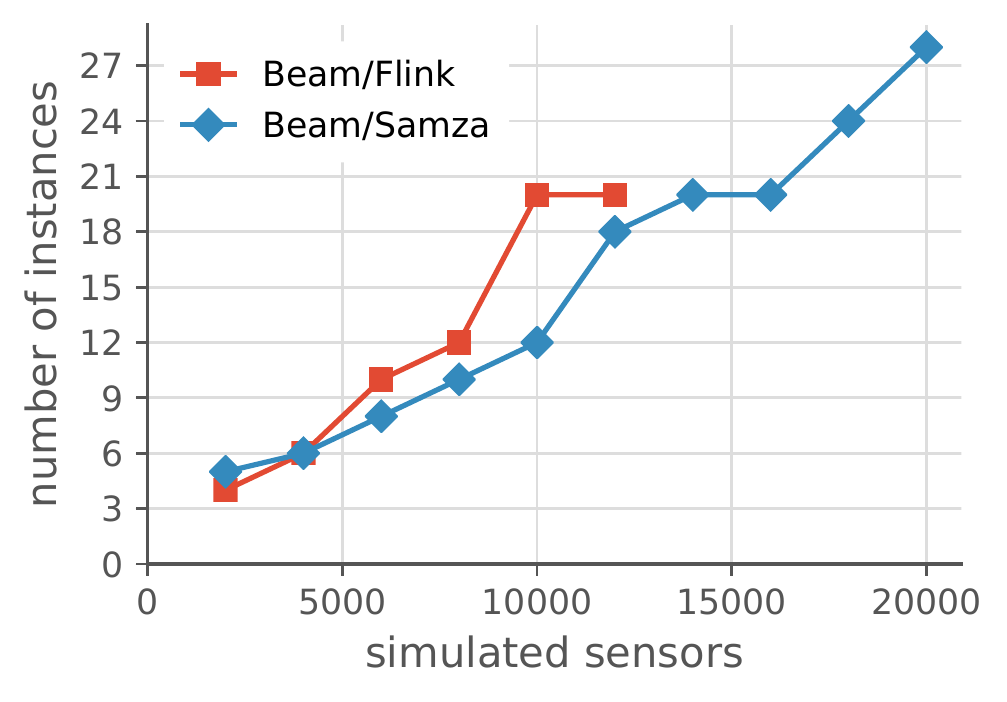}
		\caption{UC3}
		\label{fig:eval:frameworks:base-low:uc3}
	\end{subfigure}
	\hfill%
	\begin{subfigure}[b]{0.495\linewidth}%
		\centering
		\includegraphics[width=\textwidth]{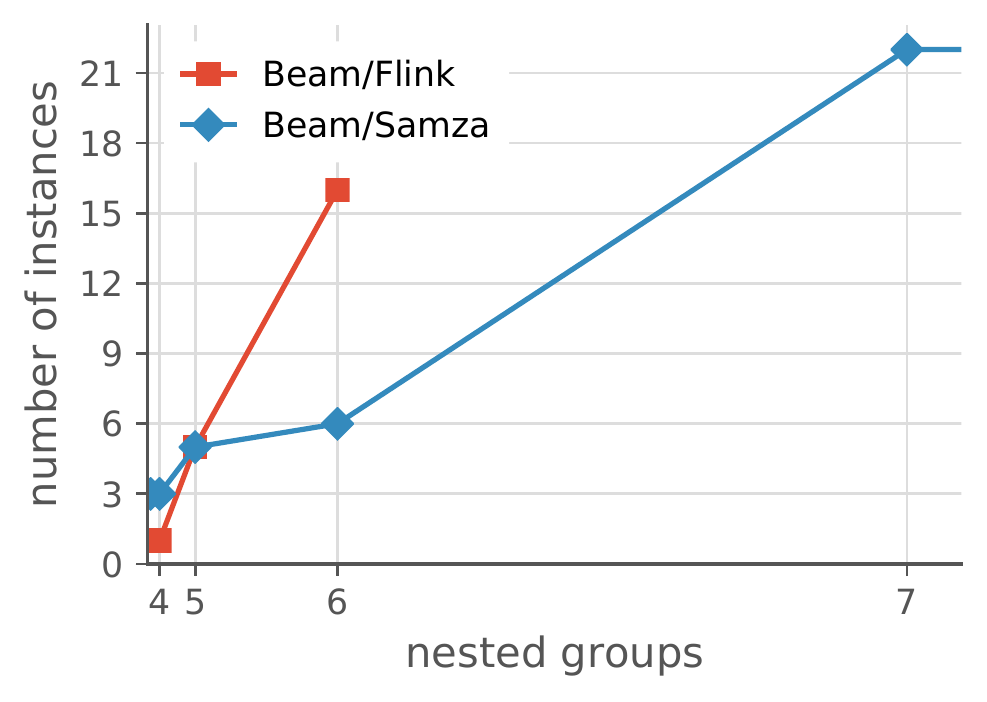}
		\caption{UC4}
		\label{fig:eval:frameworks:base-low:uc4}
	\end{subfigure}
	\caption{Repetition of scalability experiments shown in \crefrange{fig:eval:frameworks:base:uc3}{fig:eval:frameworks:base:uc4} with lower load intensities for Apache Beam with the Flink and the Samza runners.}
	\label{fig:eval:frameworks:base-low}
\end{figure}

For our baseline experiments (\rqbase), we use the private cloud environment (see \cref{tab:clouds}).
We benchmark load intensities between 100\,000 and 1\,000\,000 simulated sensors (and, thus, generated messages per second) for benchmark UC1 and UC2, 10\,000 to 100\,000 simulated sensors for UC3, and 5 to 9 nested groups (1\,024--262\,144 generated messages per second) for benchmark UC4.

\Cref{fig:eval:frameworks:base} shows the resource demand results for all evaluated frameworks and benchmarks. The results for benchmark UC4 (in the following figures as well) are visualized with an exponential scale with base 4 at the horizontal axis since the  number of generated messages grows exponentially with a linear increase in the number of nested groups.
We can observe that in almost all experiments, Flink, Hazelcast Jet, and Kafka Streams, have a considerably lower resource demand than the Beam deployments. Only Hazelcast Jet in UC4 and the Beam Flink runner for low loads in UC1 are exceptions to this.
As in some cases, the generated load intensities were too high for the Beam deployments, we repeat the corresponding experiments with lower load intensities (see \cref{fig:eval:frameworks:base-low} and \cref{fig:eval:frameworks:beam-config-samza}).

\begin{figure}
	\centering
	\begin{subfigure}[b]{0.495\linewidth}%
		\centering
		\includegraphics[width=\textwidth]{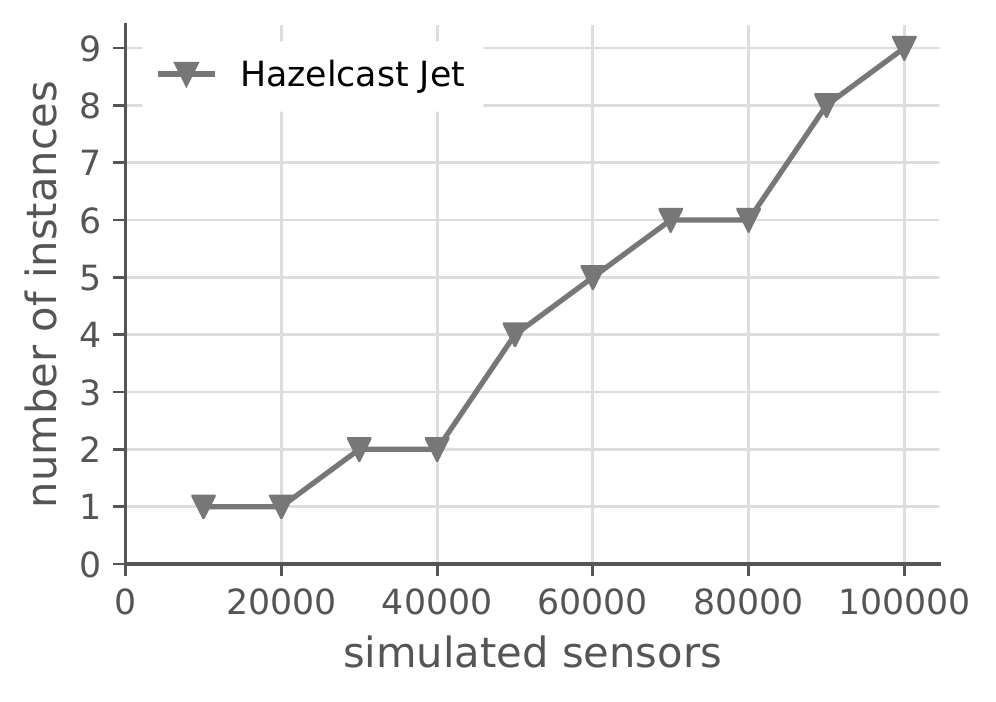}
		\caption{30 days aggregation period}
		\label{fig:eval:frameworks:hazelcastjet-uc3:base}
	\end{subfigure}
	\hfill%
	\begin{subfigure}[b]{0.495\linewidth}%
		\centering
		\includegraphics[width=\textwidth]{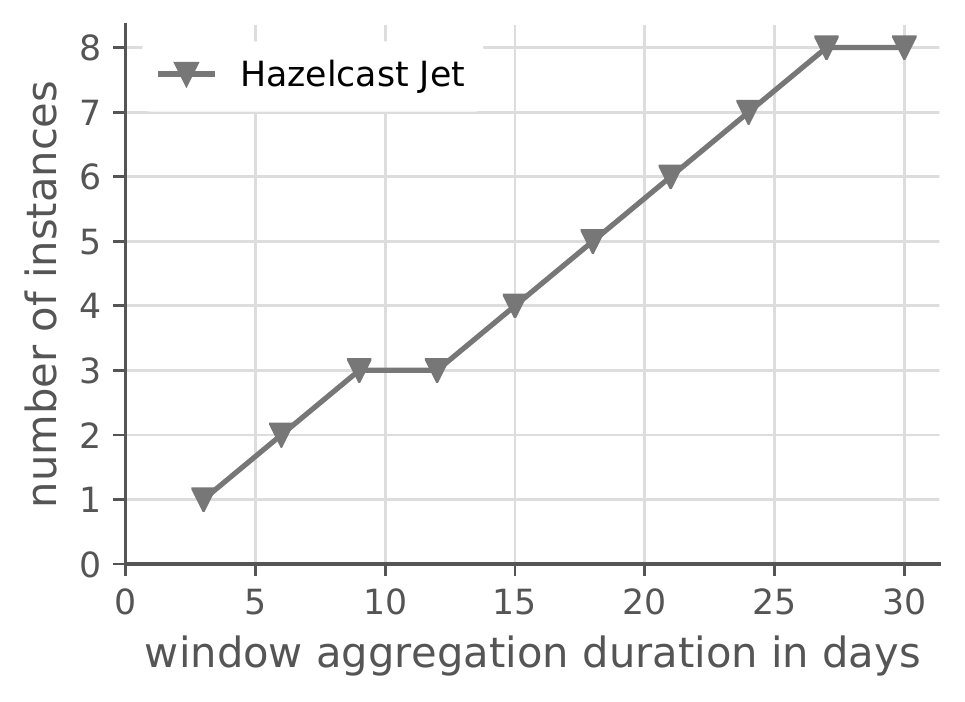}
		\caption{100\,000 simulated sensors}
		\label{fig:eval:frameworks:hazelcastjet-uc3:parallel-windows}
	\end{subfigure}
	\caption{Repetition of scalability experiments with Hazelcast Jet and benchmark UC3. \protect\subref{fig:eval:frameworks:hazelcastjet-uc3:base} evaluates scalability regarding the number of simulated sensors with a 30~days aggregation period in contrast to the 3~days period in \cref{fig:eval:frameworks:base:uc3}.
		\protect\subref{fig:eval:frameworks:hazelcastjet-uc3:parallel-windows} evaluates scalability regarding the aggregation period with a constant load of 100\,000 simulated sensors in contrast to 10\,000 simulated sensors in \cref{fig:eval:frameworks:windows:not-beam}.}
	\label{fig:eval:frameworks:hazelcastjet-uc3}
\end{figure}

Despite some outliers (see Beam/Flink in UC1 and Beam/Samza in UC4), all frameworks show linear scalability, yet with different rates. Whereas both SUTs based on Beam show the steepest increase in required resources, the results of Flink, Kafka Streams, and Hazelcast Jet vary depending on the benchmark.
In UC1 (see \cref{fig:eval:frameworks:base:uc1}), all frameworks behave similarly, with resource demands increasing slightly steeper for Hazelcast Jet compared to Kafka Streams and for Kafka Streams compared to Flink.
For UC2 (see \cref{fig:eval:frameworks:base:uc2}), we see a clear ranking with Hazelcast Jet showing the best results, followed by Kafka Streams and Flink.
For UC3 (see \cref{fig:eval:frameworks:base:uc3}), Hazelcast Jet appears to be even more superior.
A single Jet instance is sufficient for all evaluated load intensities. On the other hand, Flink requires up to 10~TaskManagers and Kafka Streams up to 18~instances. Overall, Kafka Streams' resource demand for UC3 increases at a steeper rate compared to Flink.
To further inspect the scalability of Hazelcast Jet for UC3, we repeat these experiments with an aggregation duration of 30~days in contrast to 3~days as used in the other experiments. \cref{fig:eval:frameworks:hazelcastjet-uc3:base} shows that Hazelcast Jet also scales linearly in this case. %
With UC4 (see \cref{fig:eval:frameworks:base:uc4}), we observe a comparable increase in resource demand for Kafka Streams and Flink. In contrast to the other benchmarks, Hazelcast Jet shows a significantly higher resource demand. Up to 30~instances are not able to handle load from more than 7 nested sensor groups.

For the frameworks used with Apache Beam, we observe a significantly steeper increase in resource demand of Samza compared to Flink in UC1 (cf. \cref{fig:eval:frameworks:base:uc1}) %
and UC2 (cf. \cref{fig:eval:frameworks:base:uc2}). %
For benchmark UC3 (see \cref{fig:eval:frameworks:base-low:uc3}), both frameworks scale at similar rates with Samza requiring slightly fewer instances.
For benchmark UC4 (see \cref{fig:eval:frameworks:base-low:uc4}), it appears that the resource demand of Flink increases at a steeper rate as well. %
For lower load intensities (less than 5~nested groups), Flink requires fewer instances than Samza. 

Worth mentioning is also the significant difference between native Apache Flink and the Flink runner of Apache Beam. In almost all experiments, the resource demand of Apache Beam with Flink is at least twice as high. For the more compute-intensive benchmarks UC3 and UC4, it is tremendously higher.
The performance overhead of using Apache Beam as an abstraction layer has also been observed in related research~\cite{Hesse2019}.

\begin{tcolorbox}
	\textbf{\rqbase:}
	All frameworks appear to be linearly scalable, however, with different resource usage. Depending on the task sample, Flink, Hazelcast Jet, or Kafka Streams have the smallest resource demand. In particular for the task samples of medium complexity, Hazelcast Jet's performance is outstanding.
	Apache Beam implementations, executed by Samza or Flink, are inferior, independent of the use case.
\end{tcolorbox}

\subsection{Impact of Apache Beam Configuration}\label{sec:experimental-results:beam-config}

As we have seen in \cref{sec:experimental-results:base},
Apache Flink and Apache Samza in combination with Apache Beam have a significantly higher resource demand compared to the other evaluated frameworks.
In this section, we address \rqbeam\ and take a closer look at the scalability of the Apache Beam SUTs and evaluate how scalability is affected by different configuration options.
Again, we use the private cloud environment (see \cref{tab:clouds}).
In the following, we first look at the Apache Flink runner and, afterward, at the Apache Samza runner.

\subsubsection{Apache Flink}

In their master's thesis, \citet{Spaeren2021} investigates possible reasons for the performance overhead of the Flink runner found by \citet{Hesse2019}.
They discovered unnecessary serialization and deserialization between operators and introduced the \textit{FasterCopy} option, which disables these copy operations. This option is integrated in Beam since version 2.26.
While the stream processing application must fulfill some requirements to run with the \textit{FasterCopy} option, \citet{Bensien2021} found that the Theodolite benchmarks fulfill these requirements.
As additionally this option might become the default in future releases, %
we decided to turn on this option in our benchmark implementations by default.
In this section, we evaluate how enabling and disabling \textit{FasterCopy} affects scalability.

Additionally, we observed that Beam's Kafka consumers generate a lot of log messages if not configured differently. This contrasts with the other frameworks.
As extensive logging can actually have an impact on performance (see the following \cref{sec:experimental-results:vertical}),
we evaluate whether disabling all logging results in lower resource demand.

\begin{figure}
	\centering
	\begin{subfigure}[b]{0.495\linewidth}%
		\centering
		\includegraphics[width=\textwidth]{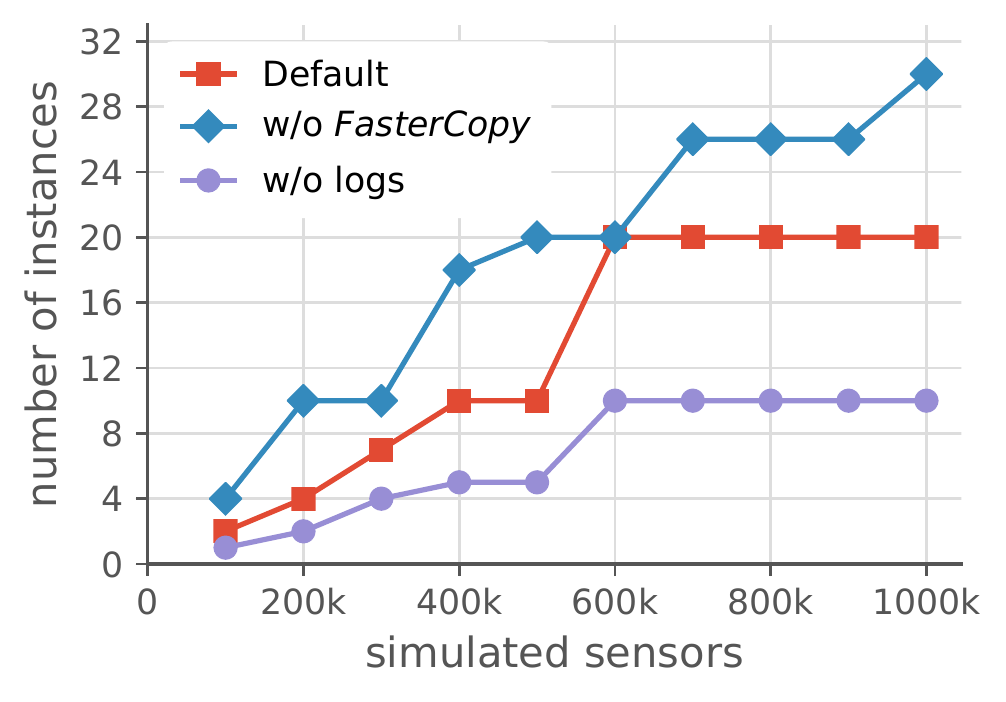}
		\caption{UC1}
		\label{fig:eval:frameworks:beam-config-flink:uc1}
	\end{subfigure}
	\hfill%
	\begin{subfigure}[b]{0.495\linewidth}%
		\centering
		\includegraphics[width=\textwidth]{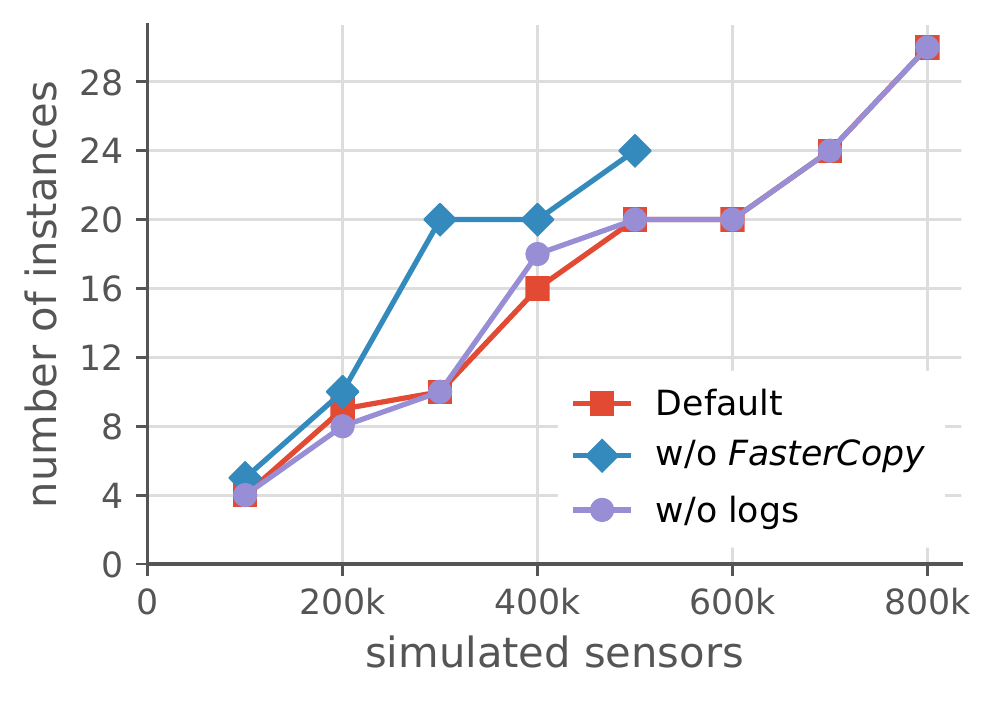}
		\caption{UC2}
		\label{fig:eval:frameworks:beam-config-flink:uc2}
	\end{subfigure}
	
	\vspace{1em}
	
	\begin{subfigure}[b]{0.495\linewidth}%
		\centering
		\includegraphics[width=\textwidth]{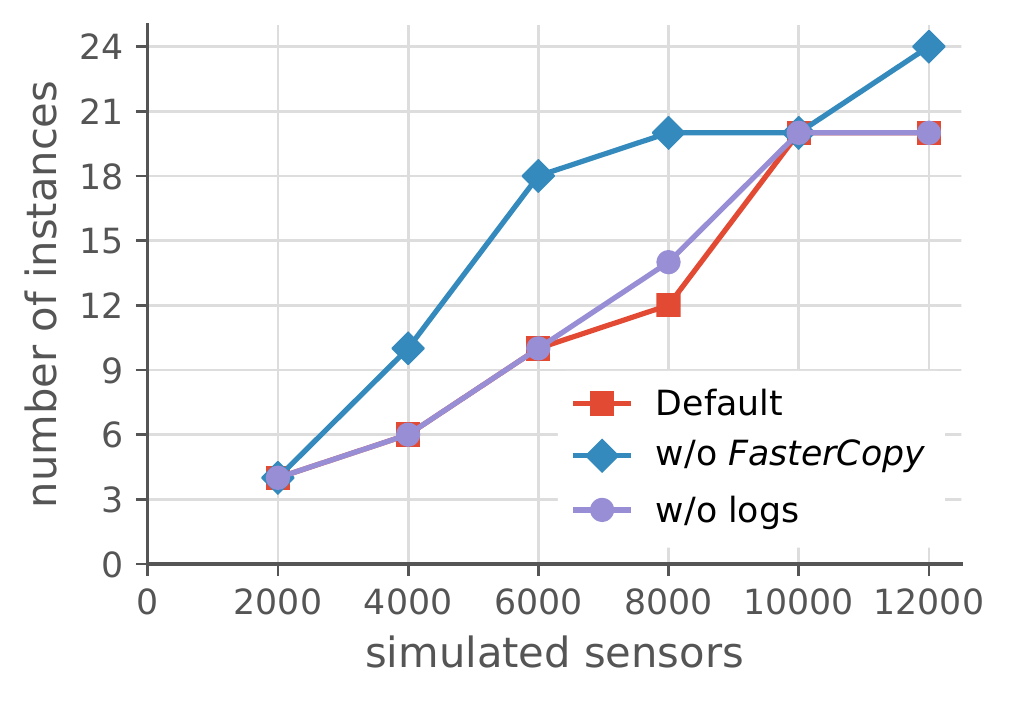}
		\caption{UC3}
		\label{fig:eval:frameworks:beam-config-flink:uc3}
	\end{subfigure}
	\hfill%
	\begin{subfigure}[b]{0.495\linewidth}%
		\centering
		\includegraphics[width=\textwidth]{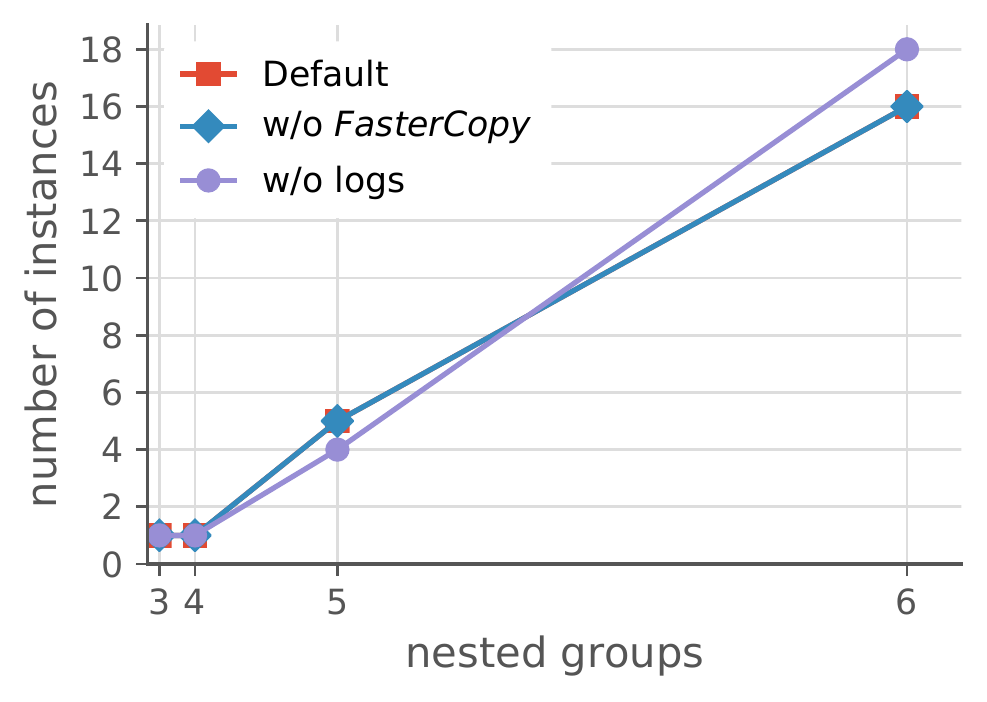}
		\caption{UC4}
		\label{fig:eval:frameworks:beam-config-flink:uc4}
	\end{subfigure}
	\caption{Scalability benchmark results for different configurations of Apache Beam with the Apache Flink runner.}
	\label{fig:eval:frameworks:beam-config-flink}
\end{figure}

\Cref{fig:eval:frameworks:beam-config-flink} shows our results of running the scalability benchmarks with the \textit{FasterCopy} option disabled and logging disabled compared to the experiments from \cref{sec:experimental-results:base}.
We can see that enabling \textit{FasterCopy} results in significantly lower resource demands for UC1--UC3 (see \crefrange{fig:eval:frameworks:beam-config-flink:uc1}{fig:eval:frameworks:beam-config-flink:uc3}). This is in line with the performance improvements reported by \citet{Spaeren2021}.
For benchmark UC4 (see \cref{fig:eval:frameworks:beam-config-flink:uc4}), enabling \textit{FasterCopy} seems not to have an effect on the resource demand.
A possible explanation is that the dataflow architecture of UC4 involves more data transfer among instances and, hence, actually requires serialization and deserialization between operators.

We can observe that disabling all logging only has a small impact on the resource demand of benchmark UC2--UC4 (see \crefrange{fig:eval:frameworks:beam-config-flink:uc2}{fig:eval:frameworks:beam-config-flink:uc4}), but significantly reduces the resource demand of benchmark UC1 (see \cref{fig:eval:frameworks:beam-config-flink:uc1}).
The latter is expected since benchmark UC1 logs each incoming message to simulate side effects such as writing records to a database.
We can conclude that the extensive logging of Beam's Kafka consumer contributes very little to the overhead introduced by Apache Beam.

\subsubsection{Apache Samza}

In a blog post, software engineers at LinkedIn \cite{Zhang2020} report how they tremendously improved the performance of Beam's Samza runner.
Primarily, this was achieved by exporting Beam metrics more efficiently.
Moreover, the authors observed that performance could further be improved when disabling the Beam metrics entirely.
Although this might not be an option in production \cite{Zhang2020}, we are still interested in how much performance could be further improved when disabling all Beam metrics.

\begin{figure}
	\centering
	\begin{subfigure}[b]{0.495\linewidth}%
		\centering
		\includegraphics[width=\textwidth]{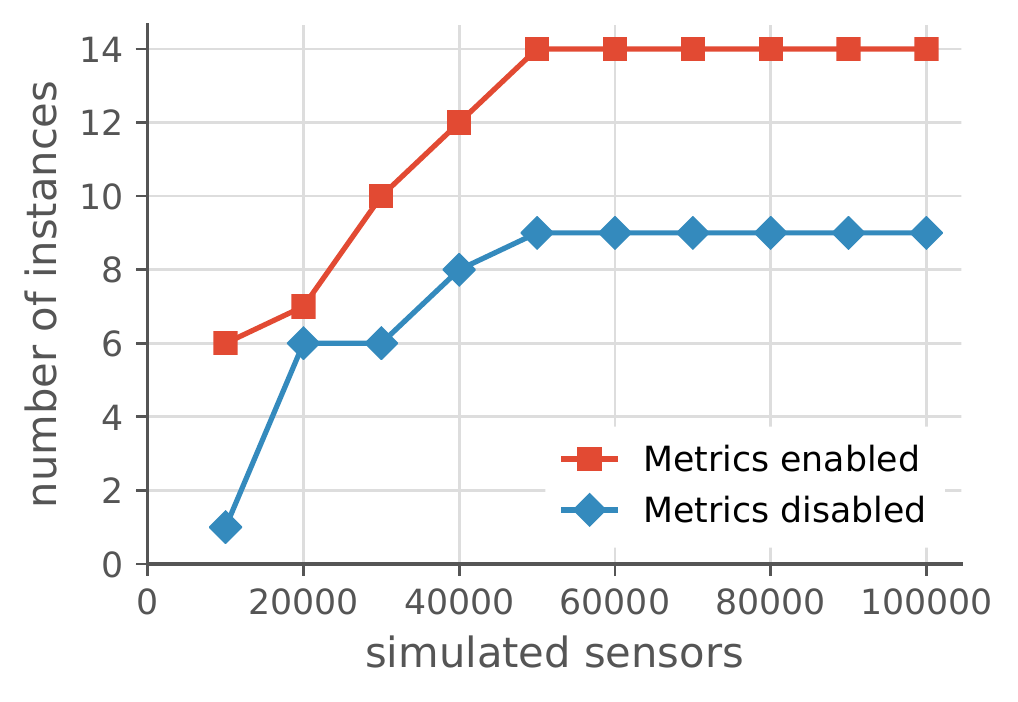}
		\caption{UC1}
		\label{fig:eval:frameworks:beam-config-samza:uc1}
	\end{subfigure}
	\hfill%
	\begin{subfigure}[b]{0.495\linewidth}%
		\centering
		\includegraphics[width=\textwidth]{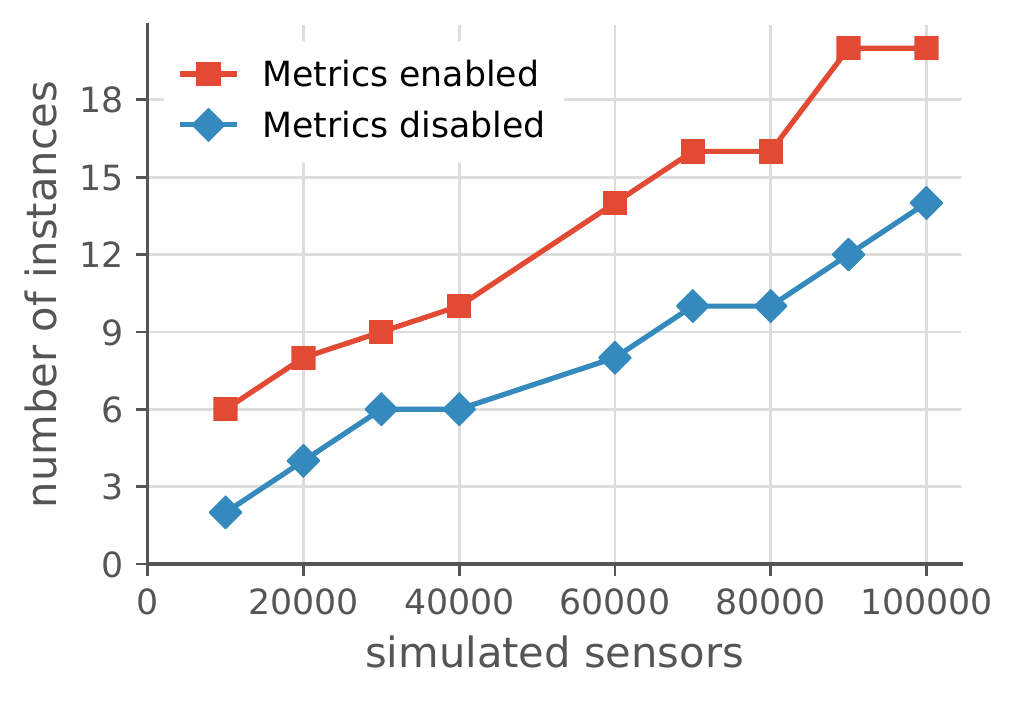}
		\caption{UC2}
		\label{fig:eval:frameworks:beam-config-samza:uc2}
	\end{subfigure}
	
	\vspace{1em}
	
	\begin{subfigure}[b]{0.495\linewidth}%
		\centering
		\includegraphics[width=\textwidth]{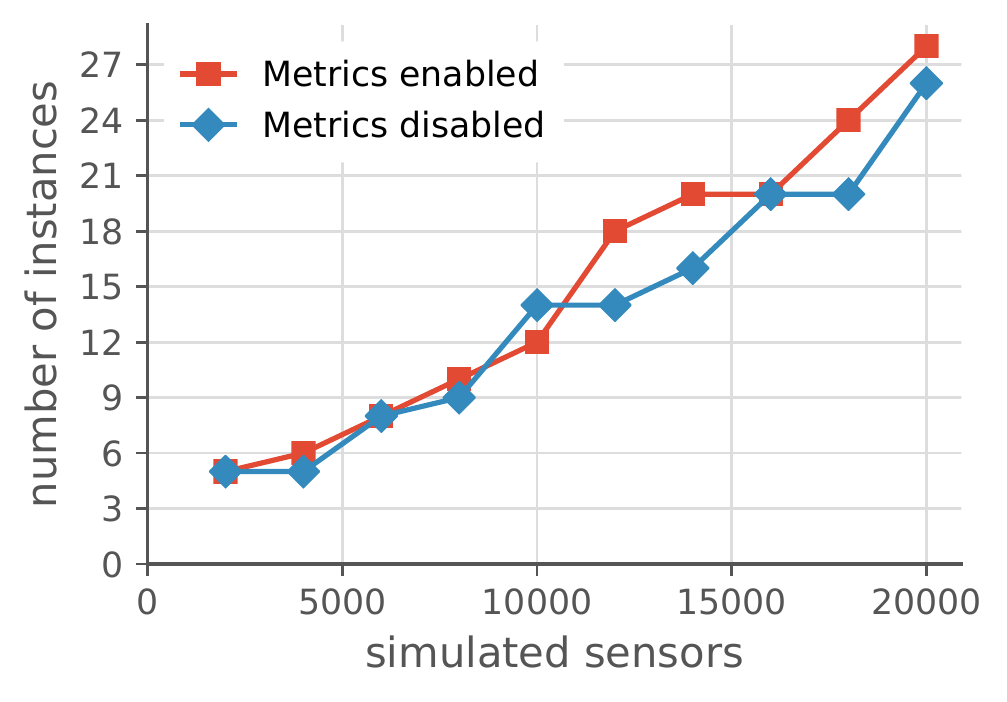}
		\caption{UC3}
		\label{fig:eval:frameworks:beam-config-samza:uc3}
	\end{subfigure}
	\hfill%
	\begin{subfigure}[b]{0.495\linewidth}%
		\centering
		\includegraphics[width=\textwidth]{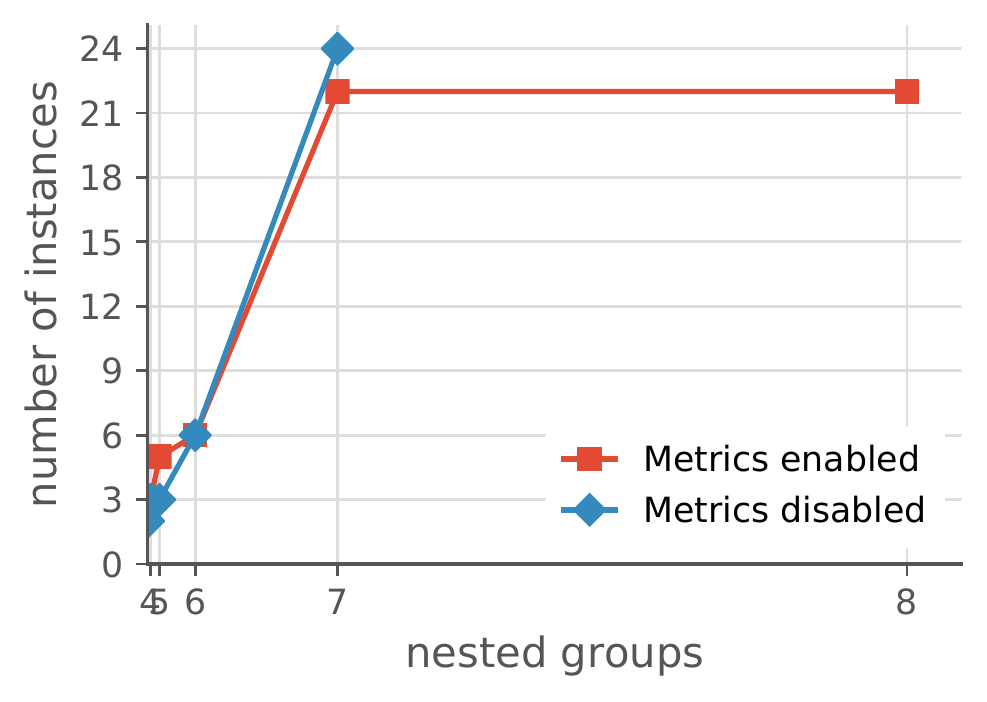}
		\caption{UC4}
		\label{fig:eval:frameworks:beam-config-samza:uc4}
	\end{subfigure}
	\caption{Scalability benchmark results for different configurations of Apache Beam with the Apache Samza runner.}
	\label{fig:eval:frameworks:beam-config-samza}
\end{figure}

\Cref{fig:eval:frameworks:beam-config-samza} shows our results for benchmarking scalability with Beam metrics enabled and disabled. We can observe that independent of the benchmark, disabling metrics results in a similar linear increase in resource demand, yet at a lower level.
However, also with metrics disabled, the resource demand of Apache Beam with the Samza runner is considerable higher compared to most other frameworks (with metrics not disabled).

Worth mentioning is also the bachelor's thesis of \citet{Bensien2021}, who benchmarked Beam with the Samza runner in version 2.22 using Theodolite's UC1 benchmark. This Beam version did not include the performance improvements by \citet{Zhang2020} (released in Beam version 2.27).
\citeauthor{Bensien2021} observed a resource demand more than twice as high with metrics enabled, compared to disabling them.

\begin{tcolorbox}
	\textbf{\rqbeam:}
	The previously discovered negative impact on performance of using Apache Beam as abstraction layer persists.
	Possible configuration options for improving performance (Flink's \textit{FasterCopy}, disabling metrics, and disabling logging) have no significant impact on our results. %
\end{tcolorbox}

\subsection{Scaling the Window Aggregation Duration}\label{sec:experimental-results:windows}

In \cref{sec:experimental-results:base}, we configure benchmark UC3 with a window duration of 3 days to compute an average daily course.
This is a trade-off to still benchmark generated data volume of reasonable size.
However, it is likely that in practice, larger time windows are required to obtain more reasonable results. Therefore, in this section, we evaluate, how different stream processing frameworks scale with increasing benchmark UC3's window duration.
We increase the window duration from 3~days to 30~days, while keeping the number of simulated sensors and, thus, the incoming message rate constant.
For the benchmark deployment depicted in \cref{fig:experimental-setup:benchmark-architecture}, this means the message volume and the state processed by the \textit{aggregate} operation increases.
Again, we use the private cloud environment (see \cref{tab:clouds}).
This evaluation is an example of benchmarking scalability with respect to the work performed for each incoming message in contrast to scaling the load at the framework and, thus, addresses \rqwindows.

\begin{figure}
	\centering
	\begin{subfigure}[b]{0.495\linewidth}%
		\centering
		\includegraphics[width=\textwidth]{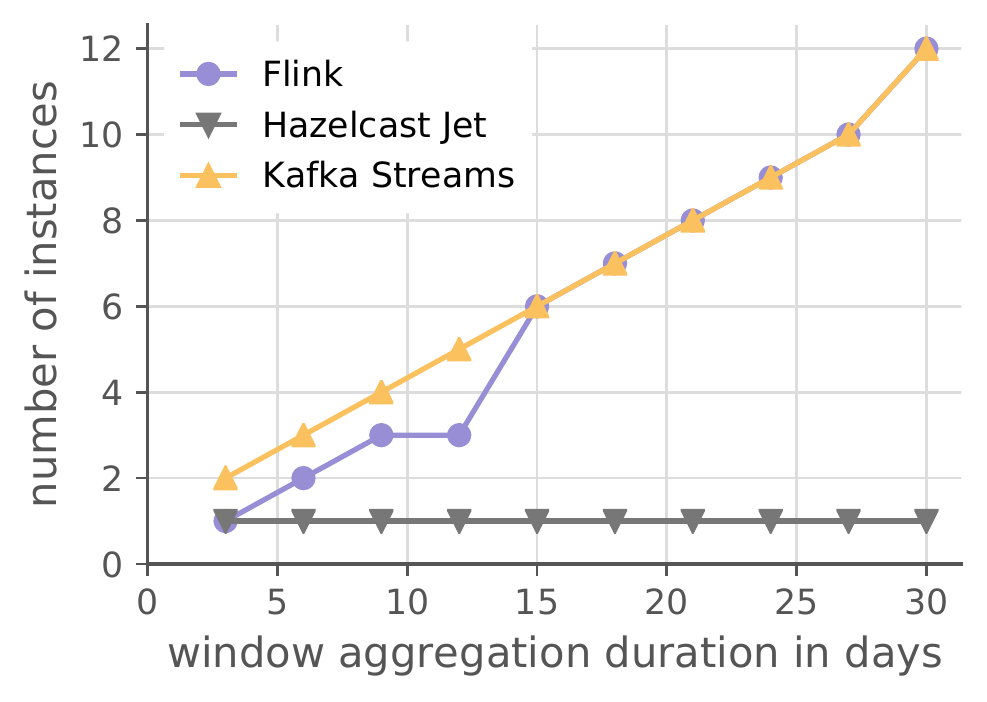}
		\caption{10\,000 simulated sensors}
		\label{fig:eval:frameworks:windows:not-beam}
	\end{subfigure}
	\hfill%
	\begin{subfigure}[b]{0.495\linewidth}%
		\centering
		\includegraphics[width=\textwidth]{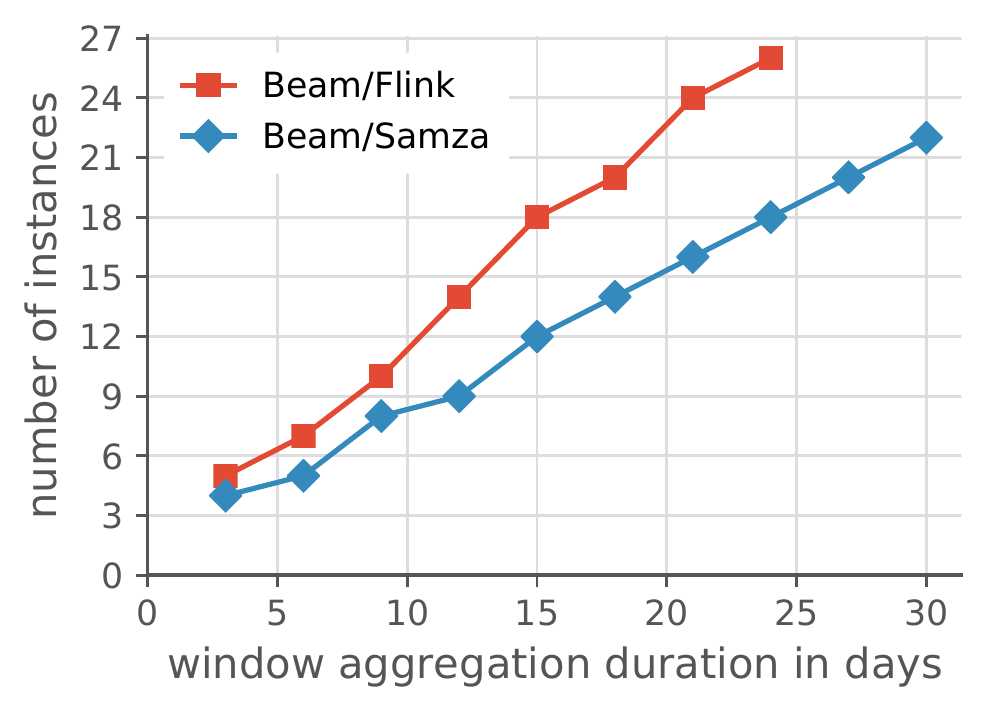}
		\caption{2\,000 simulated sensors}
		\label{fig:eval:frameworks:windows:beam}
	\end{subfigure}
	\caption{Scalability benchmark results of all stream processing frameworks when increasing the duration of aggregation windows and, thus the number of windows maintained simultaneously.}
	\label{fig:eval:frameworks:windows}
\end{figure}

\Cref{fig:eval:frameworks:windows} shows the results of these experiments. According to our previous results in \cref{sec:experimental-results:base}, we simulate 10\,000 sensors for our experiments with Flink, Hazelcast Jet, and Kafka Streams (see \cref{fig:eval:frameworks:windows:not-beam}) and 2\,000 sensors for the Beam SUTs (see \cref{fig:eval:frameworks:windows:beam}).
We can observe that again all frameworks scale approximately linearly.
Remarkable is again the performance of Hazelcast Jet, which only requires a single instance, independently of the window size. We repeat these experiments with a higher load of 100\,000 sensors. As shown in \cref{fig:eval:frameworks:hazelcastjet-uc3:parallel-windows}, Hazelcast Jet also scales approximately linearly in this case.
In contrast to scaling with the number of sensors (see \cref{fig:eval:frameworks:base:uc3}), Kafka Streams and Flink scale now with about the same rate of resource demand increase.
Similar to the results shown in \cref{fig:eval:frameworks:base-low:uc3}, Samza's resource demand increases less steeply compared to Beam's Flink runner.

\begin{tcolorbox}
	\textbf{\rqwindows:}
	Similar to scaling the load intensity the microservice is subject to, all frameworks show linear scalability when scaling the computational work performed inside the microservice. Again, however, the increase in resource demand differs noticeably between frameworks with Hazelcast Jet being superior for the studied case, followed by Flink and Kafka Streams, and significantly higher resource demand of the Beam-based deployments.
\end{tcolorbox}

\subsection{Scaling on a Single Node}\label{sec:experimental-results:vertical}

\begin{figure*}[tb]
	\captionsetup[subfigure]{
		skip=4pt
	}
	\centering
	\begin{subfigure}[b]{0.246\linewidth}%
		\centering
		\includegraphics[width=\textwidth]{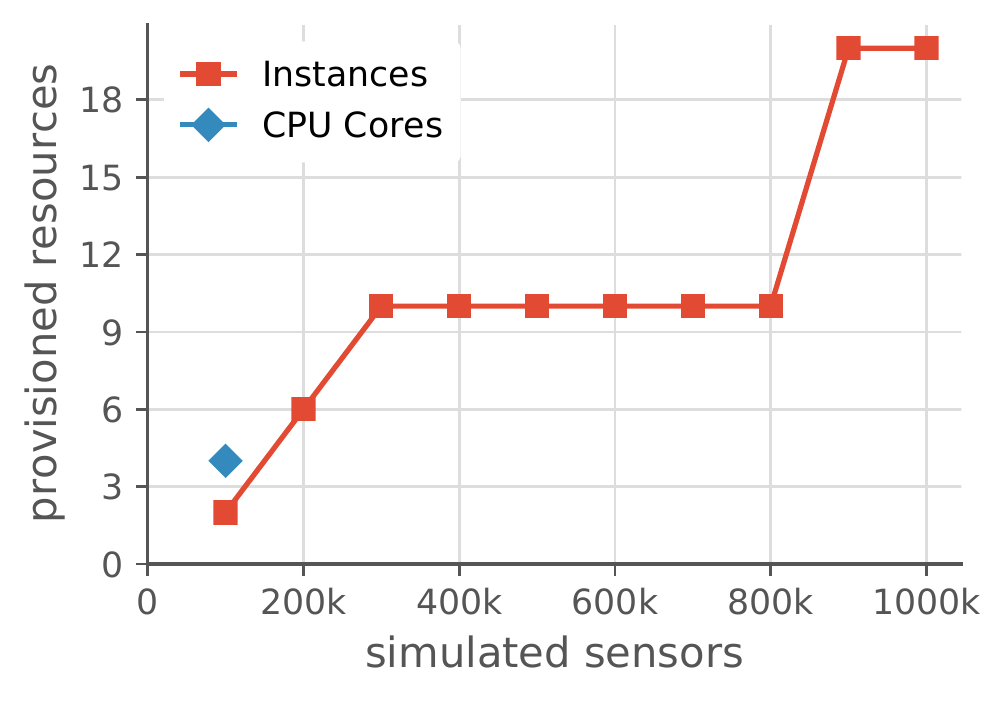}
		\caption{Beam/Flink UC1}
		\label{fig:eval:frameworks:vertical:beamflink-uc1}
	\end{subfigure}
	\hfill%
	\begin{subfigure}[b]{0.246\linewidth}%
		\centering
		\includegraphics[width=\textwidth]{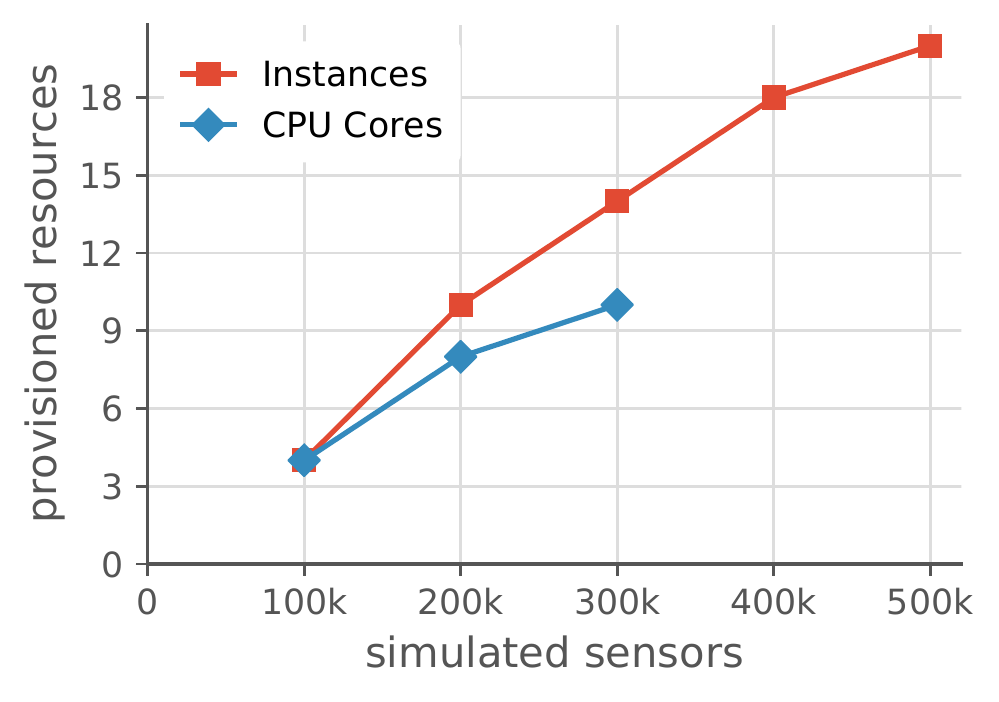}
		\caption{Beam/Flink UC2}
		\label{fig:eval:frameworks:vertical:beamflink-uc2}
	\end{subfigure}
	\hfill%
	\begin{subfigure}[b]{0.246\linewidth}%
		\centering
		\includegraphics[width=\textwidth]{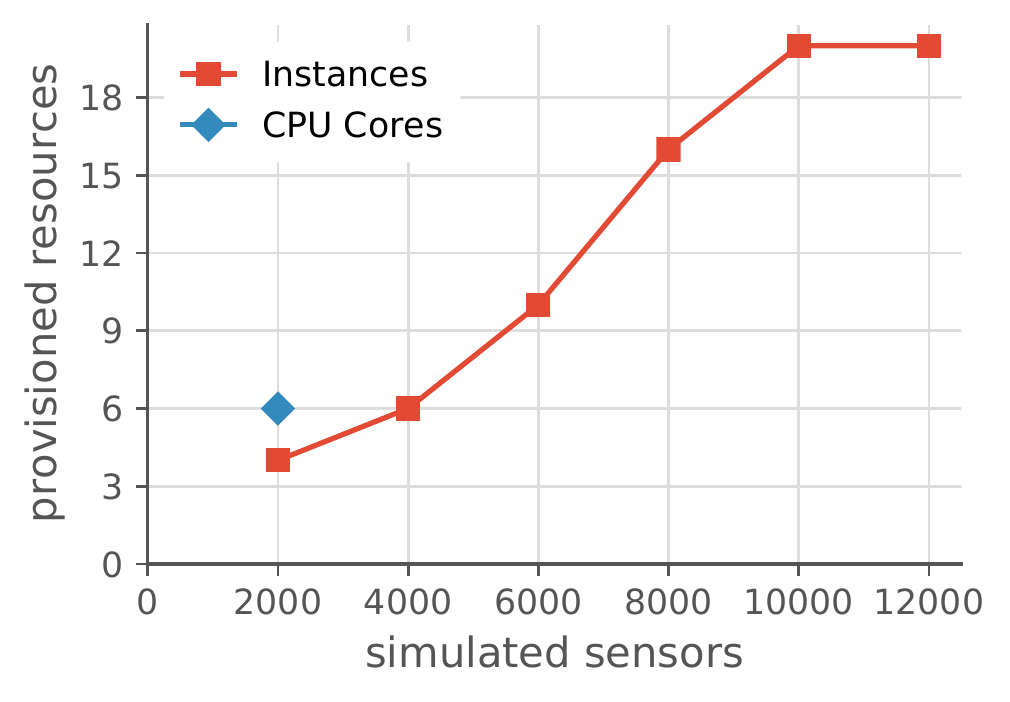}
		\caption{Beam/Flink UC3}
		\label{fig:eval:frameworks:vertical:beamflink-uc3}
	\end{subfigure}
	\hfill%
	\begin{subfigure}[b]{0.246\linewidth}%
		\centering
		\includegraphics[width=\textwidth]{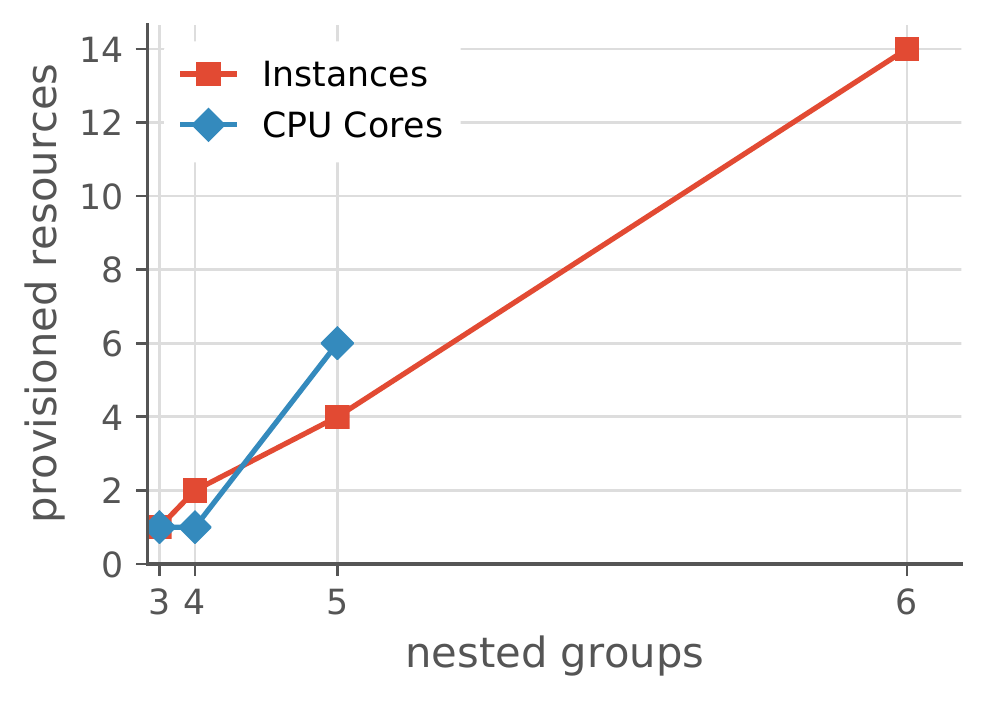}
		\caption{Beam/Flink UC4}
		\label{fig:eval:frameworks:vertical:beamflink-uc4}
	\end{subfigure}
	\vspace{0.5em}
	\begin{subfigure}[b]{0.246\linewidth}%
		\centering
		\includegraphics[width=\textwidth]{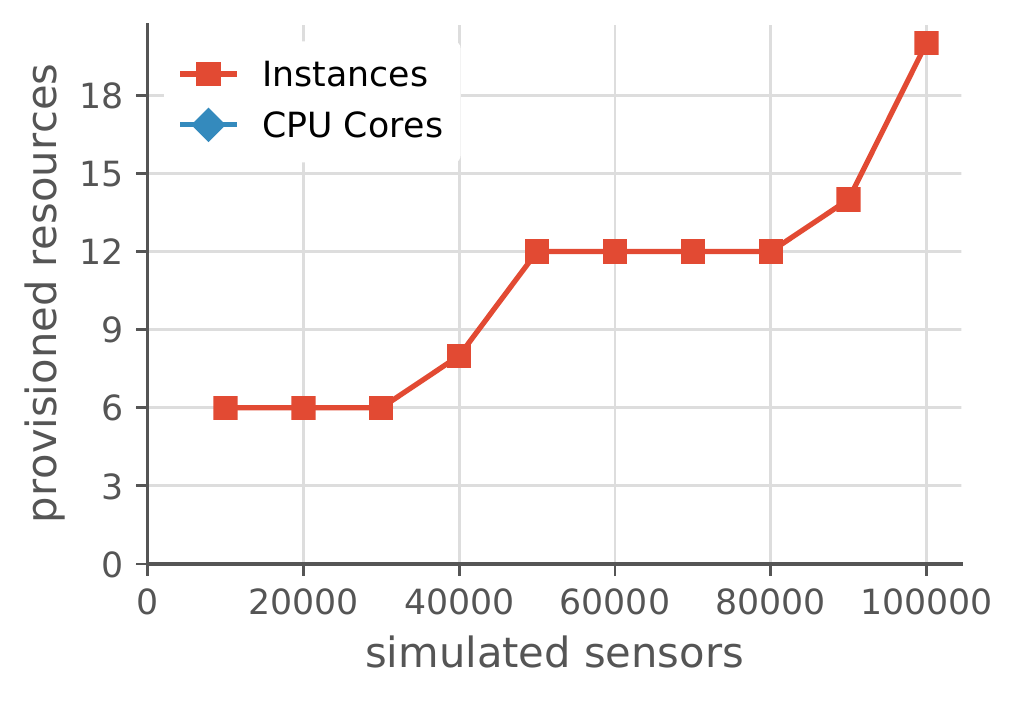}
		\caption{Beam/Samza UC1}
		\label{fig:eval:frameworks:vertical:beamsamza-uc1}
	\end{subfigure}
	\hfill%
	\begin{subfigure}[b]{0.246\linewidth}%
		\centering
		\includegraphics[width=\textwidth]{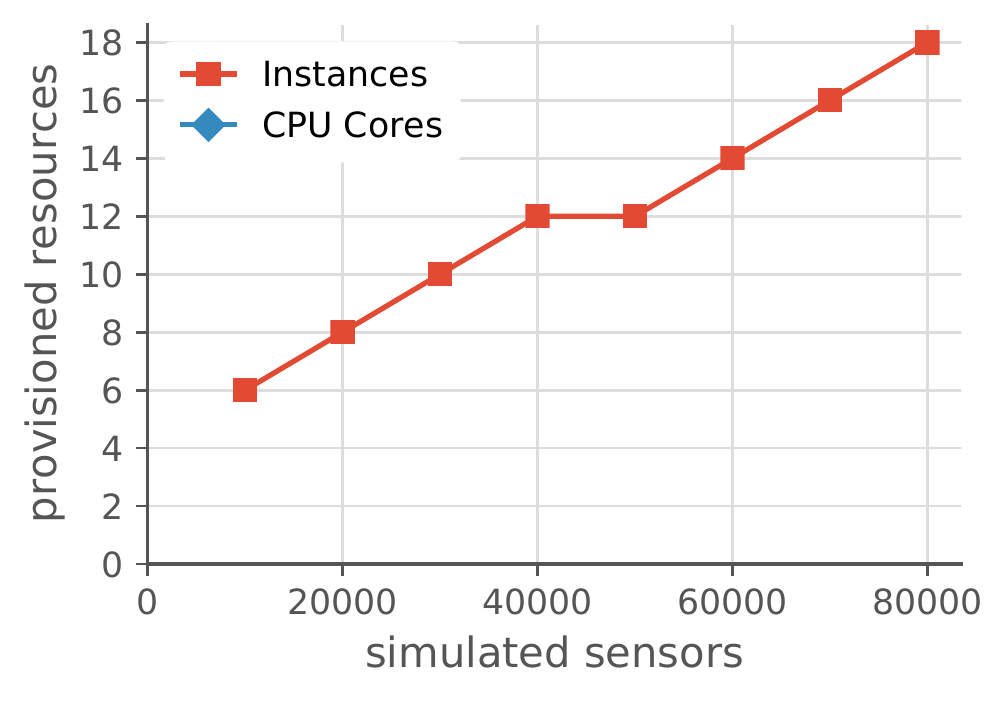}
		\caption{Beam/Samza UC2}
		\label{fig:eval:frameworks:vertical:beamsamza-uc2}
	\end{subfigure}
	\hfill%
	\begin{subfigure}[b]{0.246\linewidth}%
		\centering
		\includegraphics[width=\textwidth]{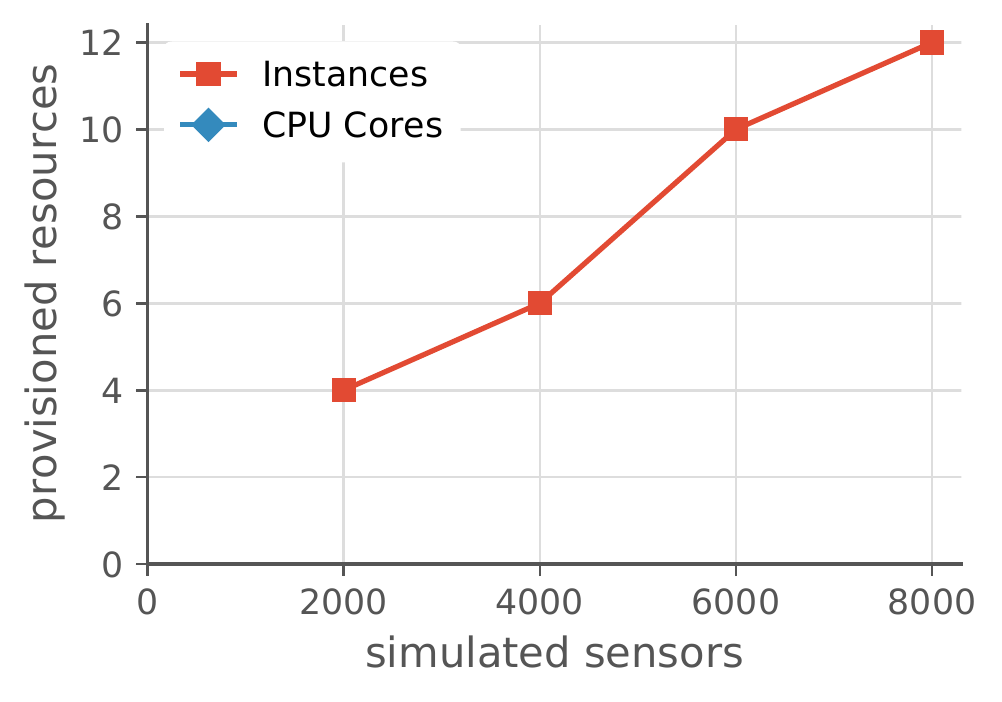}
		\caption{Beam/Samza UC3}
		\label{fig:eval:frameworks:vertical:beamsamza-uc3}
	\end{subfigure}
	\hfill%
	\begin{subfigure}[b]{0.246\linewidth}%
		\centering
		\includegraphics[width=\textwidth]{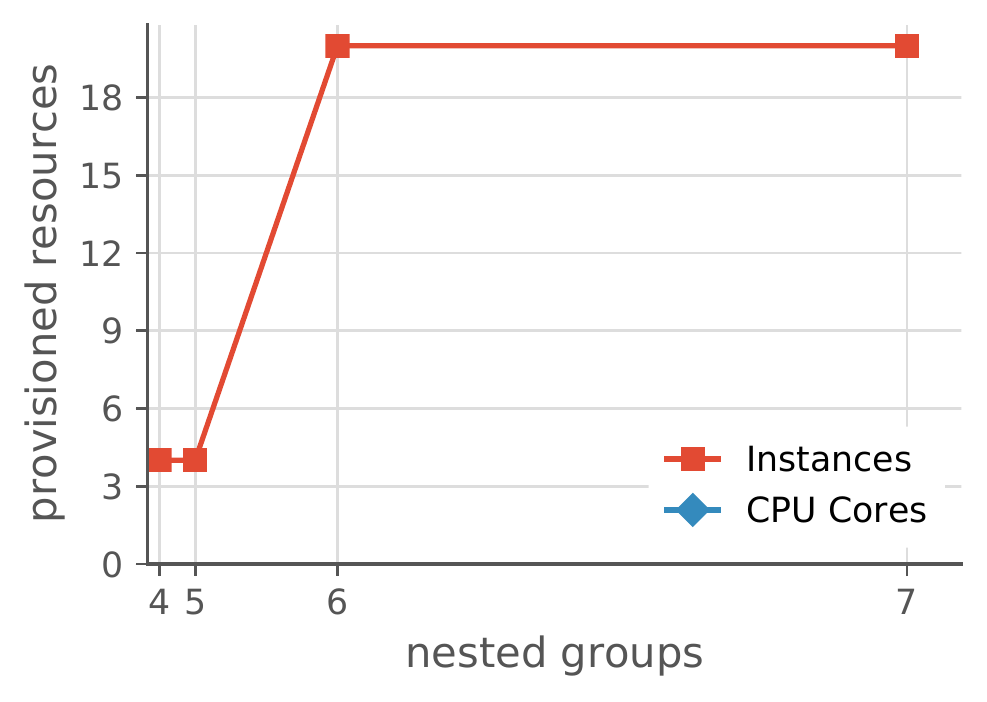}
		\caption{Beam/Samza UC4}
		\label{fig:eval:frameworks:vertical:beamsamza-uc4}
	\end{subfigure}
	\vspace{0.5em}
	\begin{subfigure}[b]{0.246\linewidth}%
		\centering
		\includegraphics[width=\textwidth]{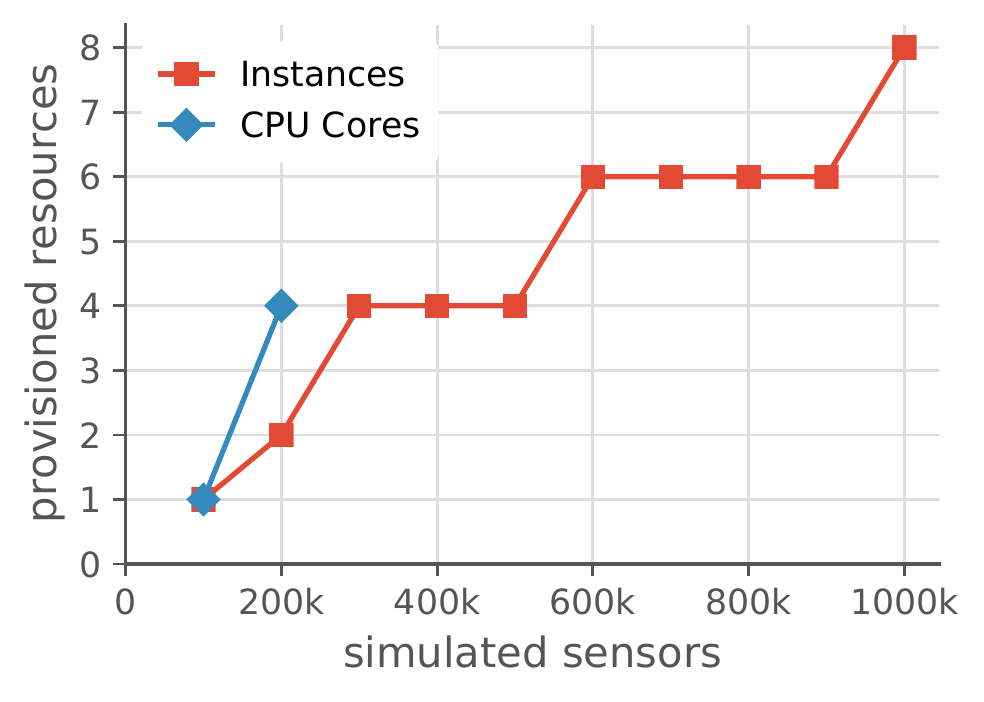}
		\caption{Flink UC1}
		\label{fig:eval:frameworks:vertical:flink-uc1}
	\end{subfigure}
	\hfill%
	\begin{subfigure}[b]{0.246\linewidth}%
		\centering
		\includegraphics[width=\textwidth]{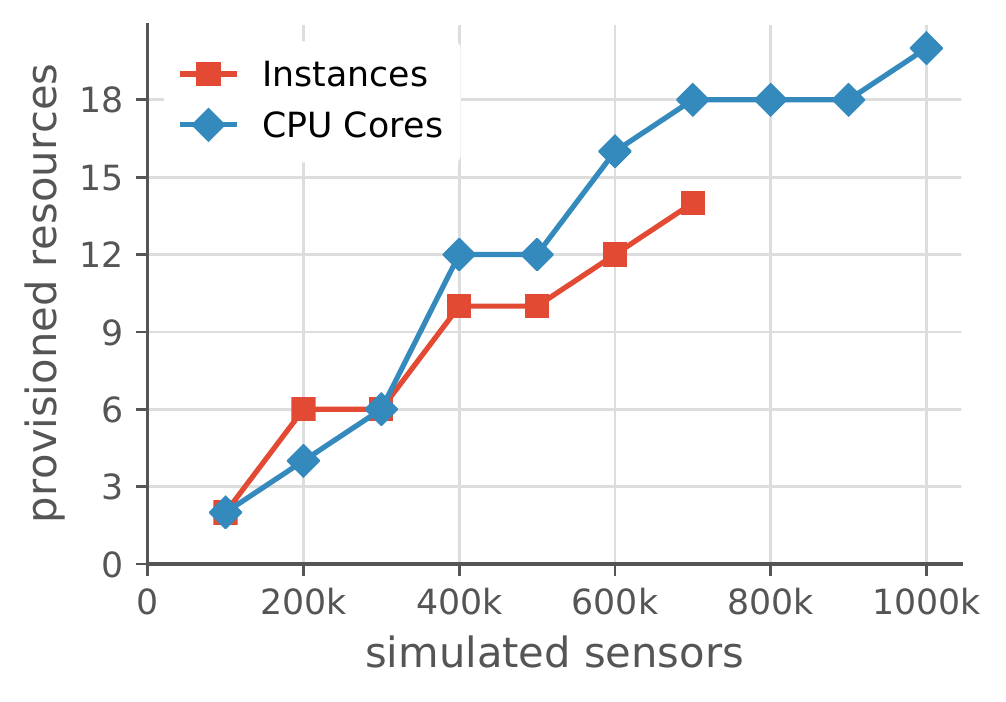}
		\caption{Flink UC2}
		\label{fig:eval:frameworks:vertical:flink-uc2}
	\end{subfigure}
	\hfill%
	\begin{subfigure}[b]{0.246\linewidth}%
		\centering
		\includegraphics[width=\textwidth]{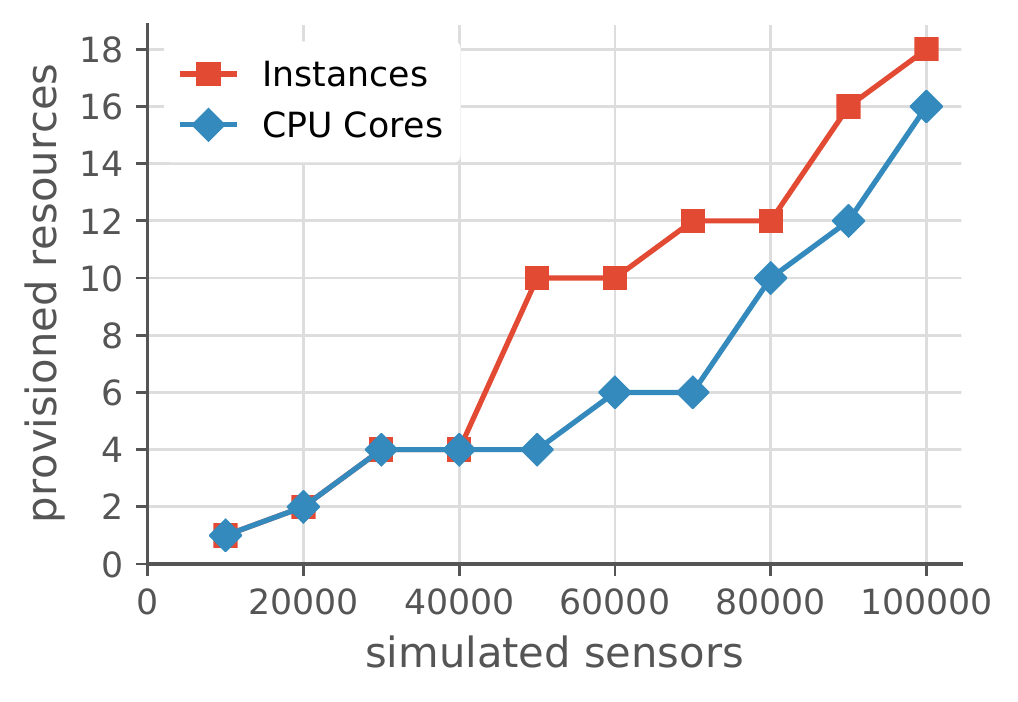}
		\caption{Flink UC3}
		\label{fig:eval:frameworks:vertical:flink-uc3}
	\end{subfigure}
	\hfill%
	\begin{subfigure}[b]{0.246\linewidth}%
		\centering
		\includegraphics[width=\textwidth]{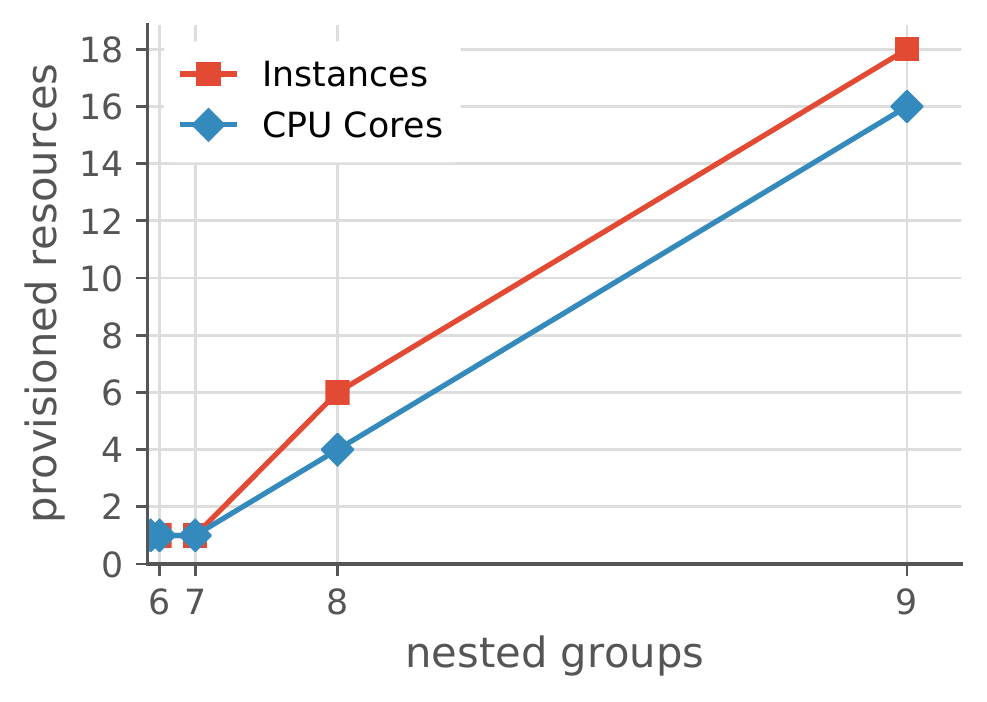}
		\caption{Flink UC4}
		\label{fig:eval:frameworks:vertical:flink-uc4}
	\end{subfigure}
	\vspace{0.5em}
	\begin{subfigure}[b]{0.246\linewidth}%
		\centering
		\includegraphics[width=\textwidth]{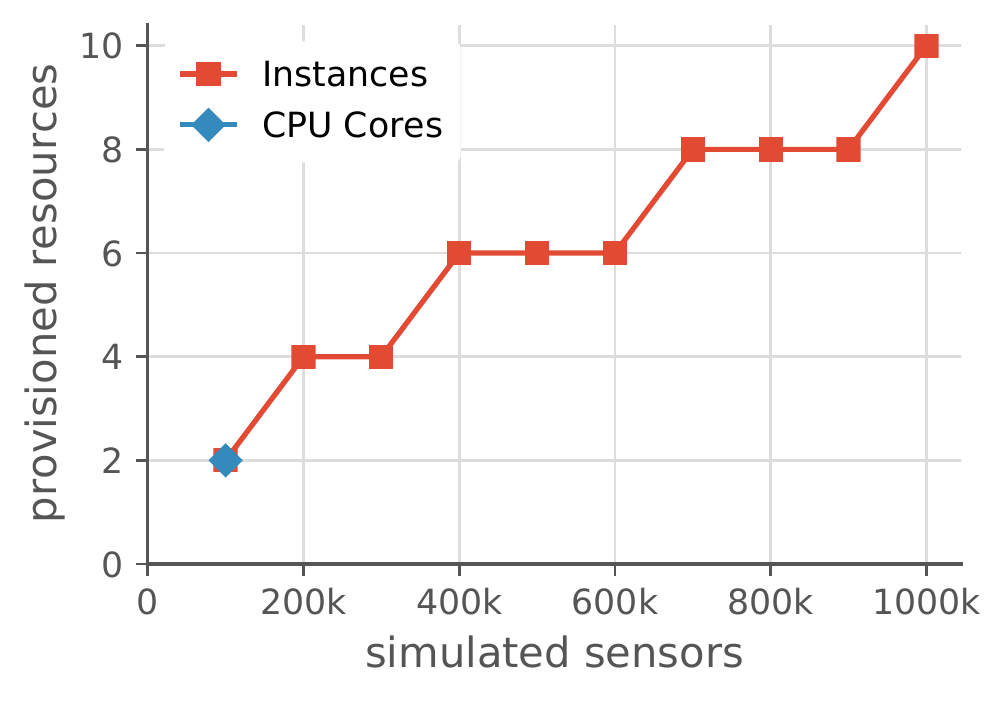}
		\caption{Hazelcast Jet UC1}
		\label{fig:eval:frameworks:vertical:hazelcastjet-uc1}
	\end{subfigure}
	\hfill%
	\begin{subfigure}[b]{0.246\linewidth}%
		\centering
		\includegraphics[width=\textwidth]{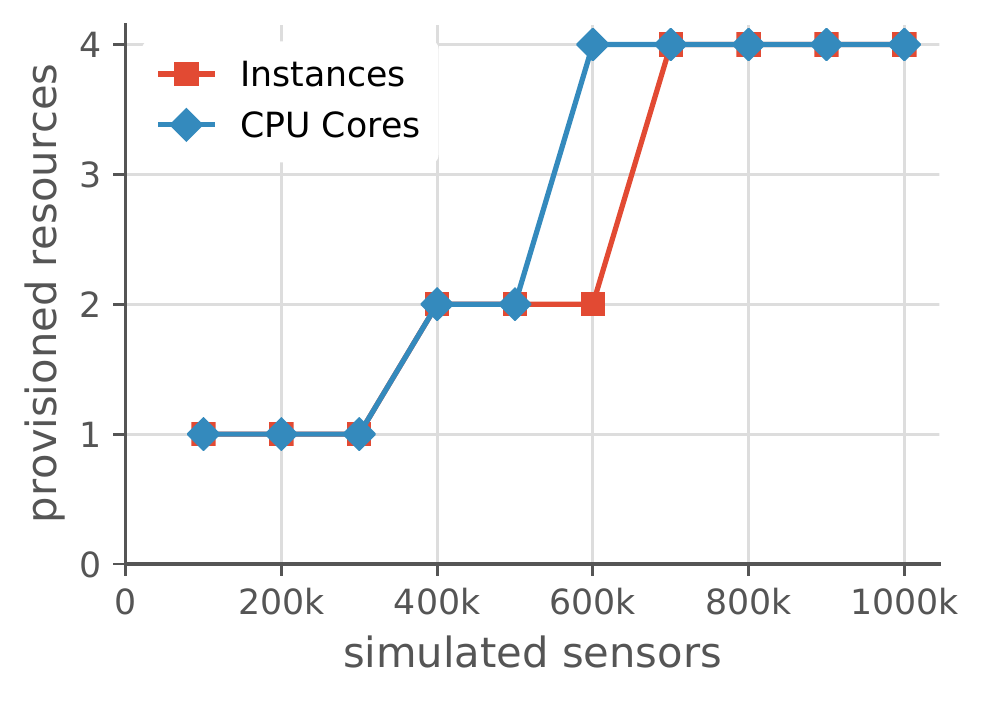}
		\caption{Hazelcast Jet UC2}
		\label{fig:eval:frameworks:vertical:hazelcastjet-uc2}
	\end{subfigure}
	\hfill%
	\begin{subfigure}[b]{0.246\linewidth}%
		\centering
		\includegraphics[width=\textwidth]{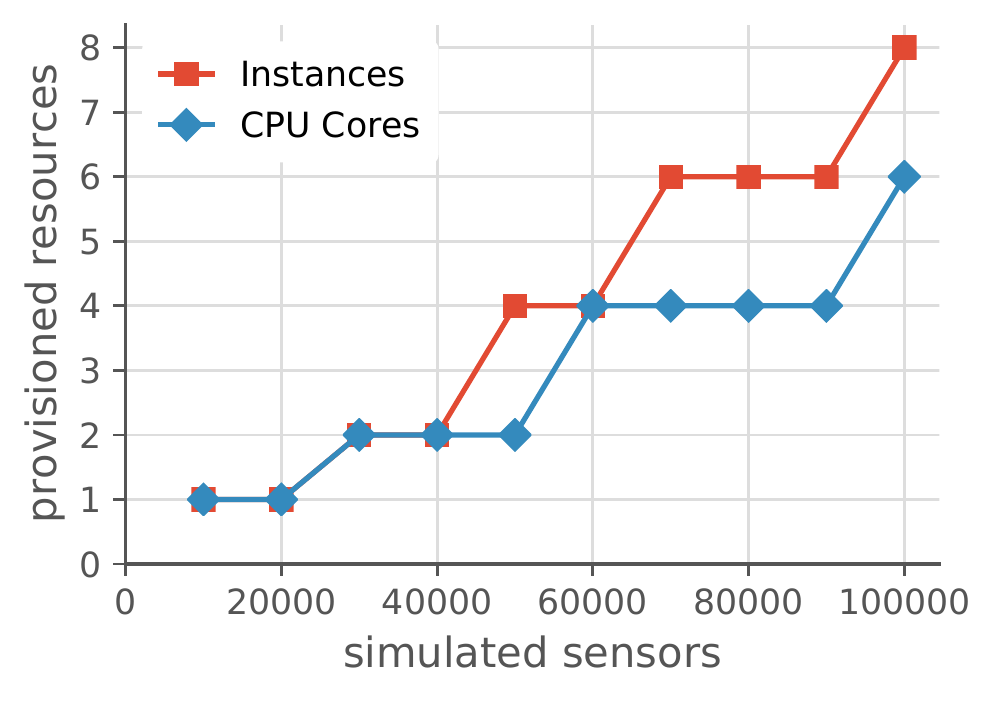}
		\caption{Hazelcast Jet UC3}
		\label{fig:eval:frameworks:vertical:hazelcastjet-uc3}
	\end{subfigure}
	\hfill%
	\begin{subfigure}[b]{0.246\linewidth}%
		\centering
		\includegraphics[width=\textwidth]{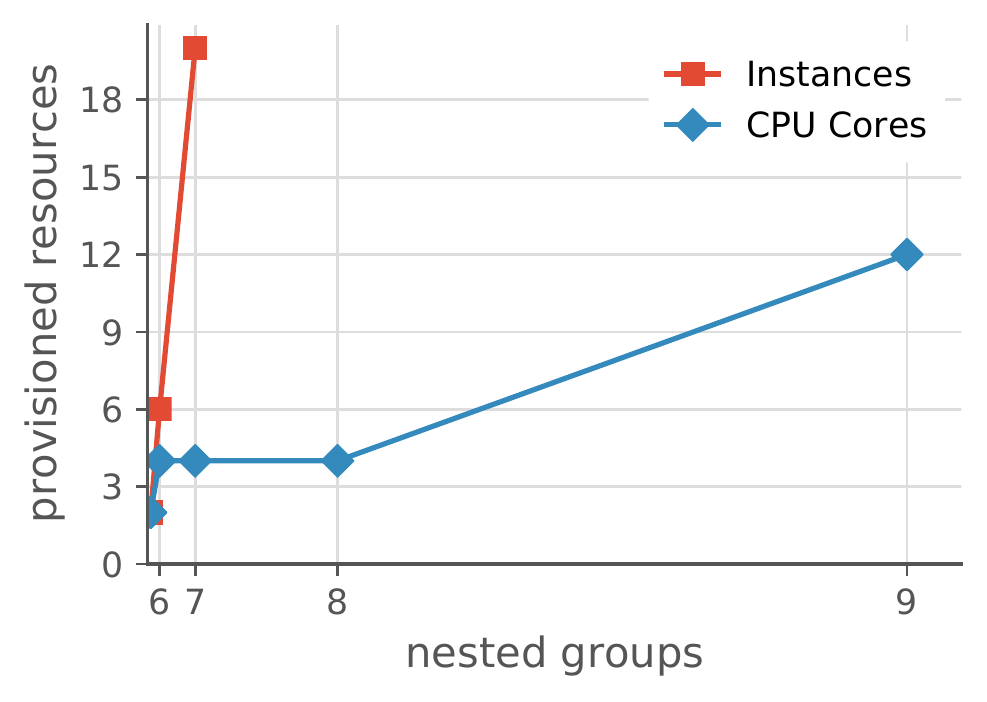}
		\caption{Hazelcast Jet UC4}
		\label{fig:eval:frameworks:vertical:hazelcastjet-uc4}
	\end{subfigure}
	\vspace{0.5em}
	\begin{subfigure}[b]{0.246\linewidth}%
		\centering
		\includegraphics[width=\textwidth]{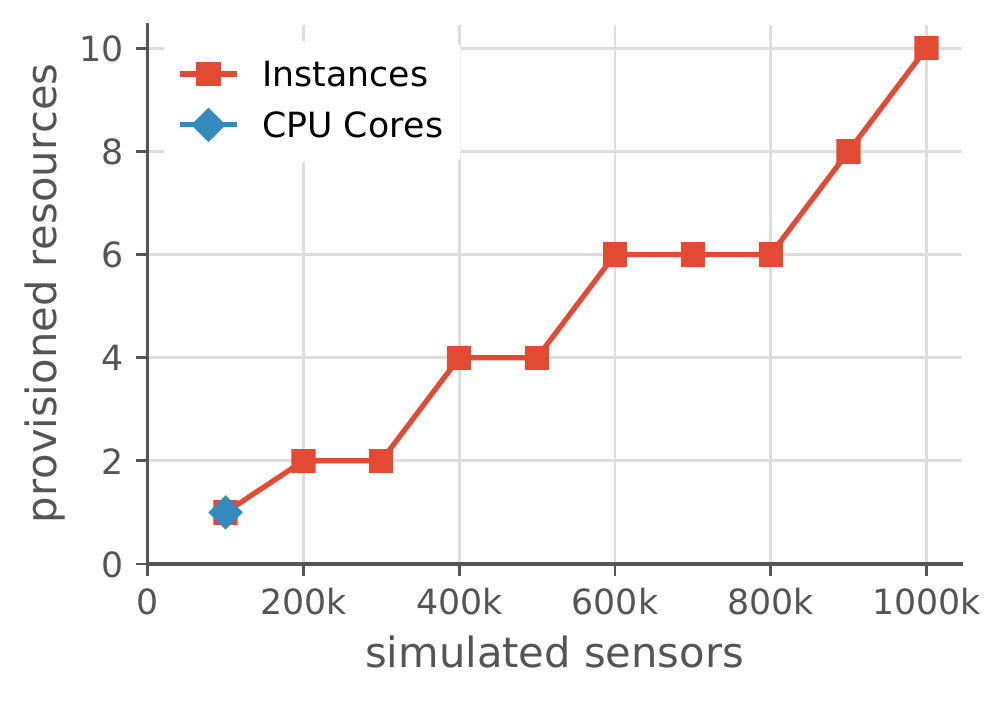}
		\caption{Kafka Streams UC1}
		\label{fig:eval:frameworks:vertical:kstreams-uc1}
	\end{subfigure}
	\hfill%
	\begin{subfigure}[b]{0.246\linewidth}%
		\centering
		\includegraphics[width=\textwidth]{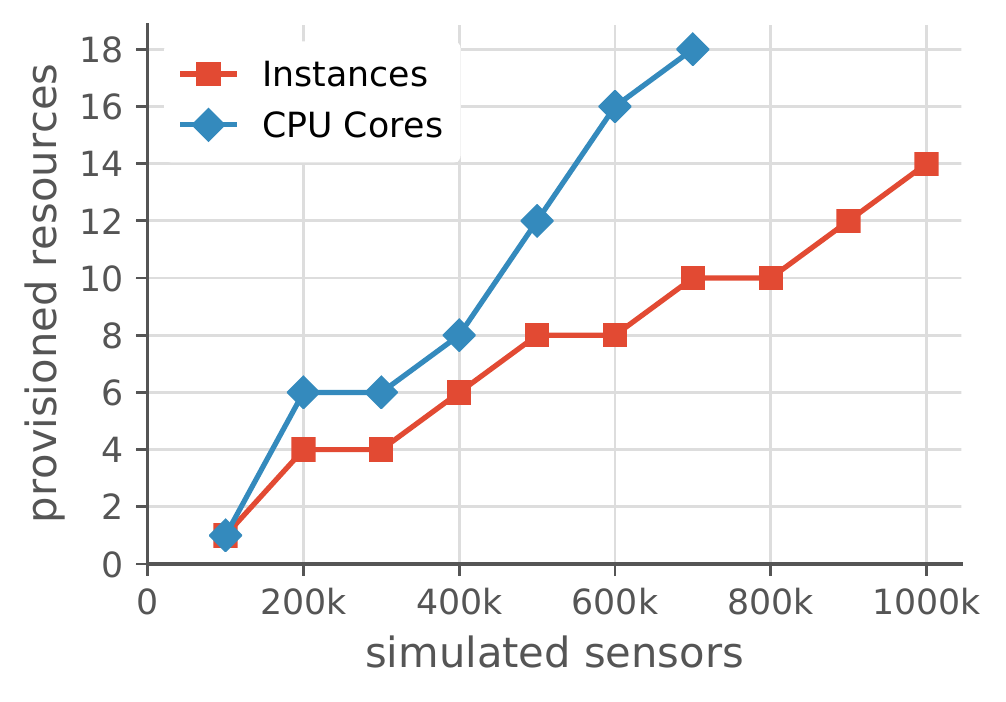}
		\caption{Kafka Streams UC2}
		\label{fig:eval:frameworks:vertical:kstreams-uc2}
	\end{subfigure}
	\hfill%
	\begin{subfigure}[b]{0.246\linewidth}%
		\centering
		\includegraphics[width=\textwidth]{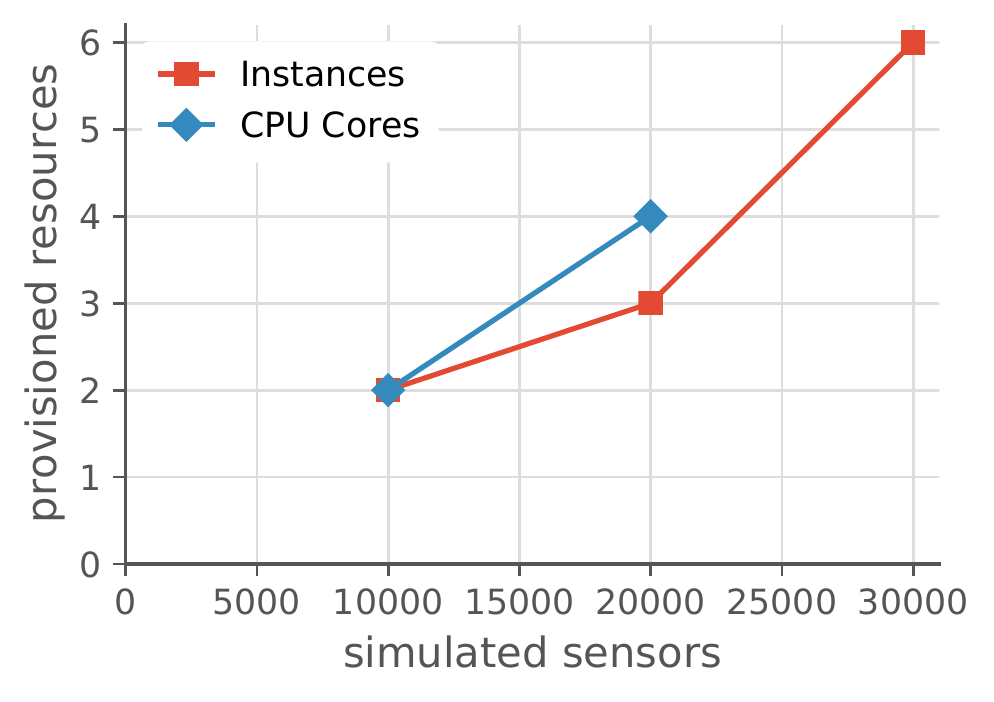}
		\caption{Kafka Streams UC3}
		\label{fig:eval:frameworks:vertical:kstreams-uc3}
	\end{subfigure}
	\hfill%
	\begin{subfigure}[b]{0.246\linewidth}%
		\centering
		\includegraphics[width=\textwidth]{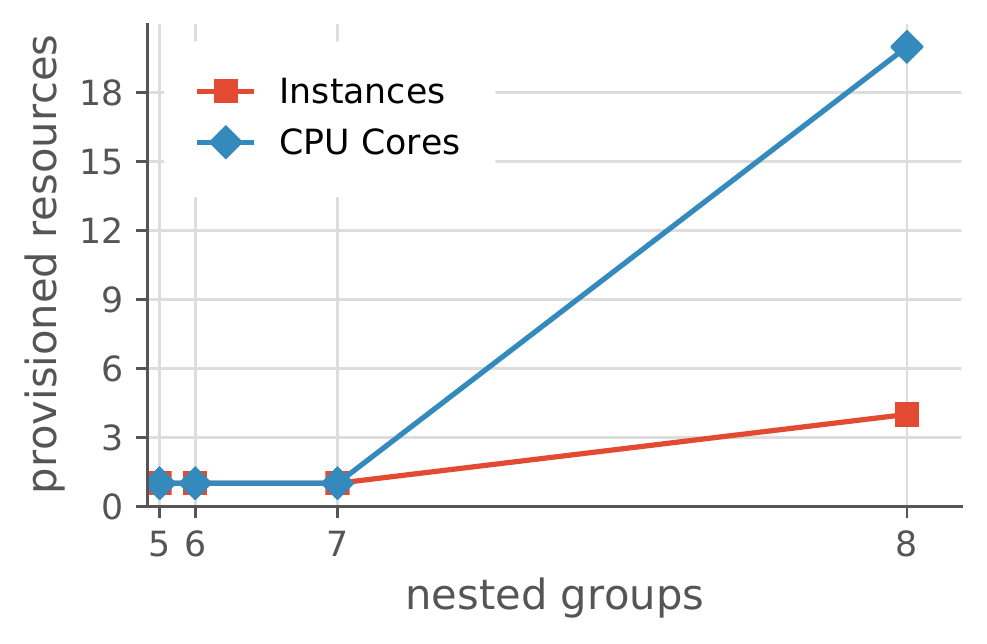}
		\caption{Kafka Streams UC4}
		\label{fig:eval:frameworks:vertical:kstreams-uc4}
	\end{subfigure}
	\caption{Scalability of stream processing frameworks on a single node.}
	\label{fig:eval:frameworks:vertical}
\end{figure*}

In this section, we address \rqvertical\ and evaluate whether vertical scaling can be a complementary measure to achieve scalability of stream processing frameworks.
We, therefore, scale our SUT vertically by varying both the number of instances, which are all deployed to the same cluster node, as well as the CPU and memory provided for a single SUT pod.

For the experiments of this section, we slightly modify our experimental setup (see \cref{tab:clouds}). We only deploy 3~Kafka brokers, which run on three different Kubernetes nodes. Of the other two nodes, we dedicate one to run the load generators and one to run the SUT instances.
Although we only run 3~Kafka brokers and all load generator instances on the same node, we can confirm that the configured load is still successfully generated.

To not fully utilize the SUT node, which also runs some infrastructure and monitoring components, we deploy up to 20 instances with one CPU core each or up to 20~CPU cores for a single instance. For scaling with the number of instances, we keep the same configuration as in the previous experiments. For scaling with the resources per pod, we only refer to the number of CPU cores, but scale memory proportionally. (However, in all our experiments we never observed fully utilized pod memory.)
In order to utilize multiple CPU cores, most stream processing frameworks have to be configured accordingly. For Flink, we scale the number of task slots of the TaskManager equally to the number of CPU cores. In Kafka Streams, we equally scale the number of threads to the number of CPU cores. For Samza, the documentation is inconsistent regarding scaling a standalone application on a single node. We decided to equally scale the \textit{container thread pool} size to the number of CPU cores. Hazelcast Jet does not require additional configuration as an instance configures its cooperative thread pool automatically according to the number of CPU cores provided.

For most experiments, we generate the same load intensities as in the first experiment. However, we use the smaller load intensities from \cref{sec:experimental-results:beam-config} for the Beam experiments and the 30~days window for Hazelcast Jet with benchmark UC3 as introduced in \cref{sec:experimental-results:windows}. %

\Cref{fig:eval:frameworks:vertical} shows the results of our experiments with a single node.
Almost all frameworks show approximately linear scalability when scaling the number of instances.
We observe that Beam with the Samza runner does not scale with increasing the CPU resources per pod. Hence, we assume that scaling the container thread pool is not the right option to increase capacity on a single node. Whether other configuration options exist remains unclear.
Further, we can observe that no framework is able to process load intensities higher than 200\,000 messages per second with a single instance for benchmark UC1.
The reason for this is that we simulate database writes by printing all incoming records to the standard output stream. Running a single instance of the frameworks causes all threads to write to the same stream, which is synchronized and becomes the bottleneck of our evaluation. 

Kafka Streams seems to be more efficient when scaling with the number of instances, compared to scaling with the number of cores.
For Flink and Hazelcast Jet, scaling with the number of cores is more efficient for the more complex dataflows in UC3 and UC4, while with benchmark UC2 both types of scaling yield similar results.
Remarkable are the results for scaling with the number of cores with Beam and the Flink runner. In contrast to native Flink, Beam with the Flink runner seems not to be scalable with respect to the number of cores.
Moreover, the Kafka Streams implementation of benchmark UC3 scales with neither increasing the number of instances nor with increasing the number of cores. As it scales linearly when running on multiple nodes (see \cref{fig:eval:frameworks:base-low:uc3}), our results suggest that underlying hardware resources become exhausted. However, from manually observing system-level metrics, we cannot observe anything conspicuous.

\begin{figure*}[tb]
	\captionsetup[subfigure]{
		skip=4pt
	}
	\centering
	\begin{subfigure}[b]{0.246\linewidth}%
		\centering
		\includegraphics[width=\textwidth]{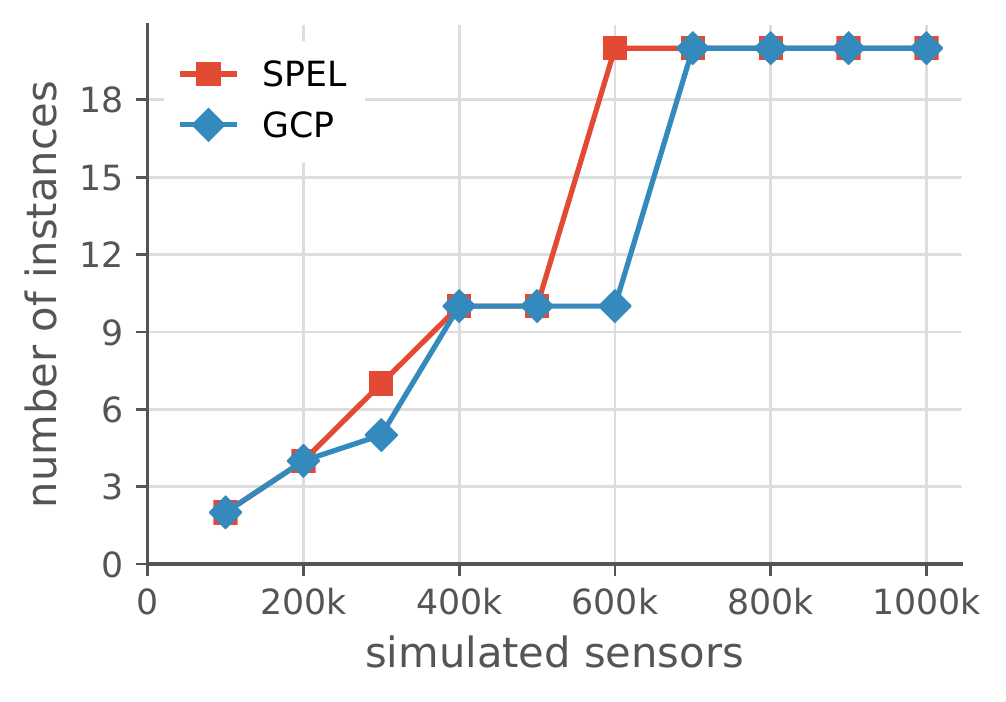}
		\caption{Beam/Flink UC1}
		\label{fig:eval:frameworks:public-private:beamflink-uc1}
	\end{subfigure}
	\hfill%
	\begin{subfigure}[b]{0.246\linewidth}%
		\centering
		\includegraphics[width=\textwidth]{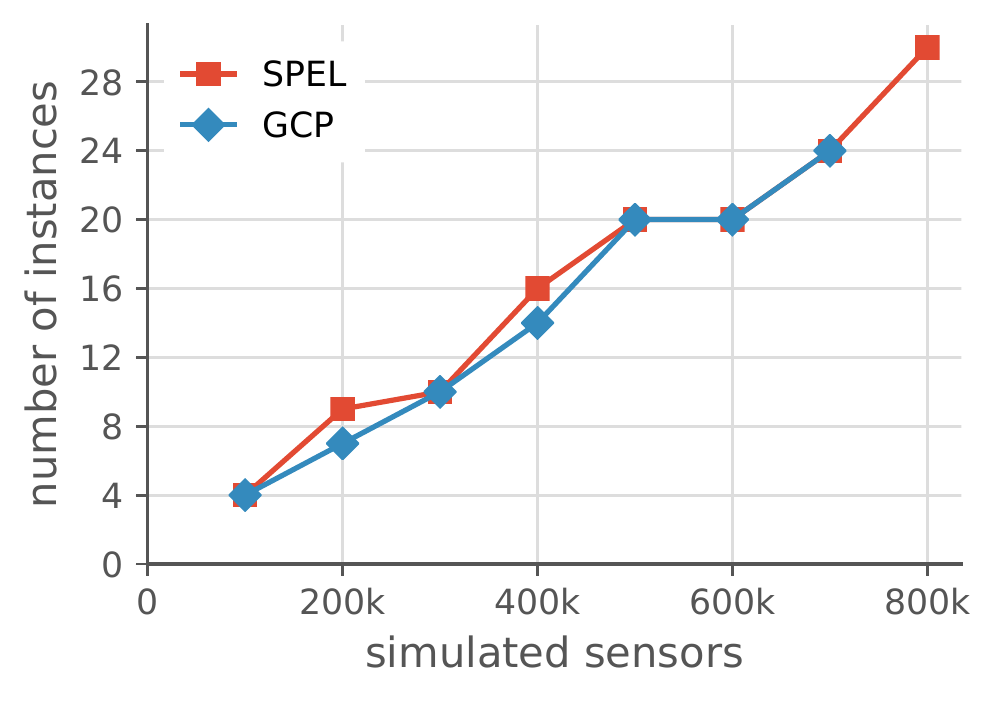}
		\caption{Beam/Flink UC2}
		\label{fig:eval:frameworks:public-private:beamflink-uc2}
	\end{subfigure}
	\hfill%
	\begin{subfigure}[b]{0.246\linewidth}%
		\centering
		\includegraphics[width=\textwidth]{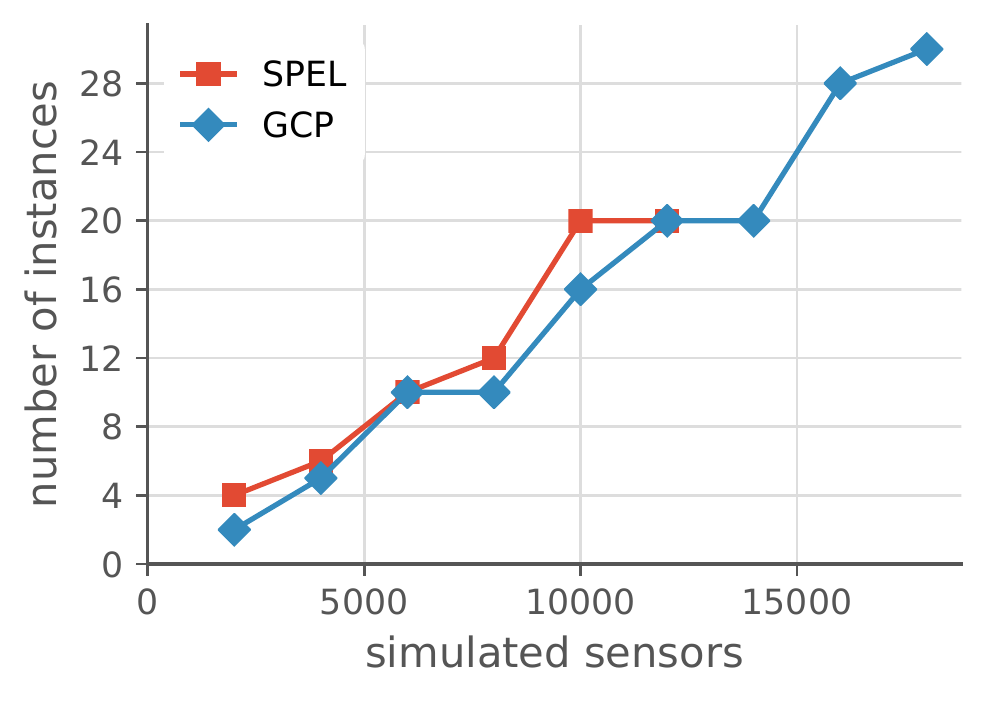}
		\caption{Beam/Flink UC3}
		\label{fig:eval:frameworks:public-private:beamflink-uc3}
	\end{subfigure}
	\hfill%
	\begin{subfigure}[b]{0.246\linewidth}%
		\centering
		\includegraphics[width=\textwidth]{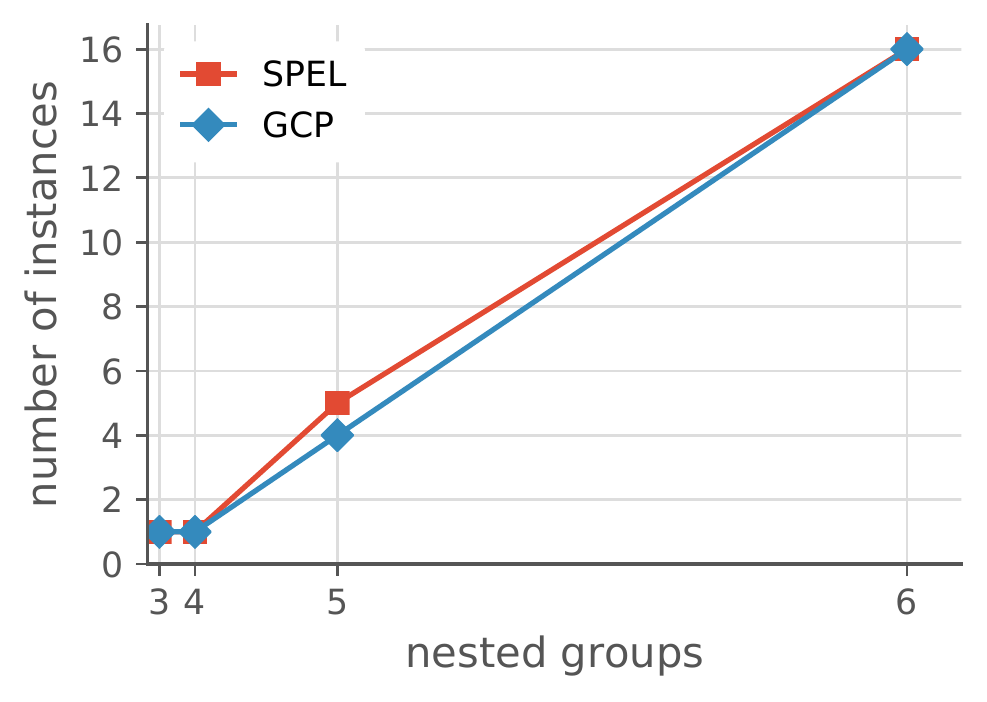}
		\caption{Beam/Flink UC4}
		\label{fig:eval:frameworks:public-private:beamflink-uc4}
	\end{subfigure}
	\vspace{0.5em}
	\begin{subfigure}[b]{0.246\linewidth}%
		\centering
		\includegraphics[width=\textwidth]{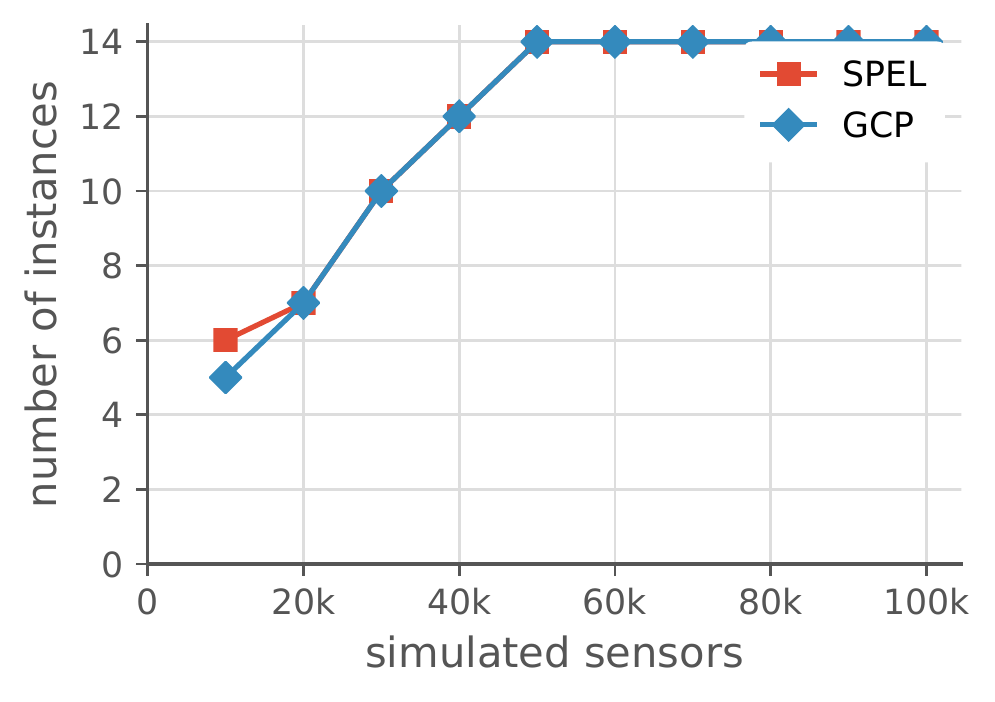}
		\caption{Beam/Samza UC1}
		\label{fig:eval:frameworks:public-private:beamsamza-uc1}
	\end{subfigure}
	\hfill%
	\begin{subfigure}[b]{0.246\linewidth}%
		\centering
		\includegraphics[width=\textwidth]{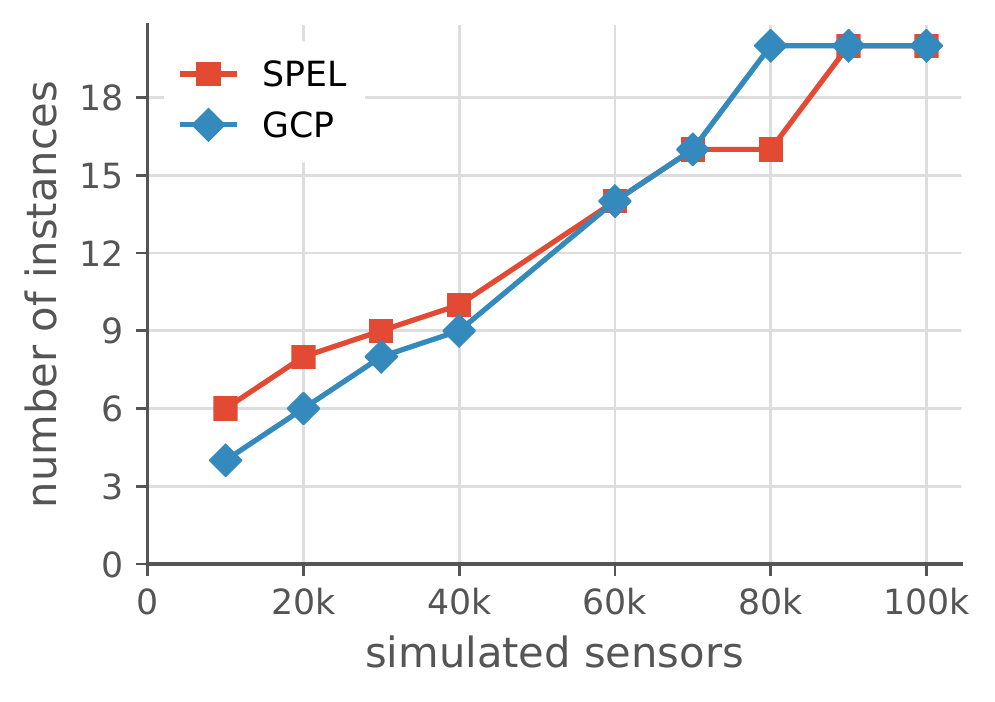}
		\caption{Beam/Samza UC2}
		\label{fig:eval:frameworks:public-private:beamsamza-uc2}
	\end{subfigure}
	\hfill%
	\begin{subfigure}[b]{0.246\linewidth}%
		\centering
		\includegraphics[width=\textwidth]{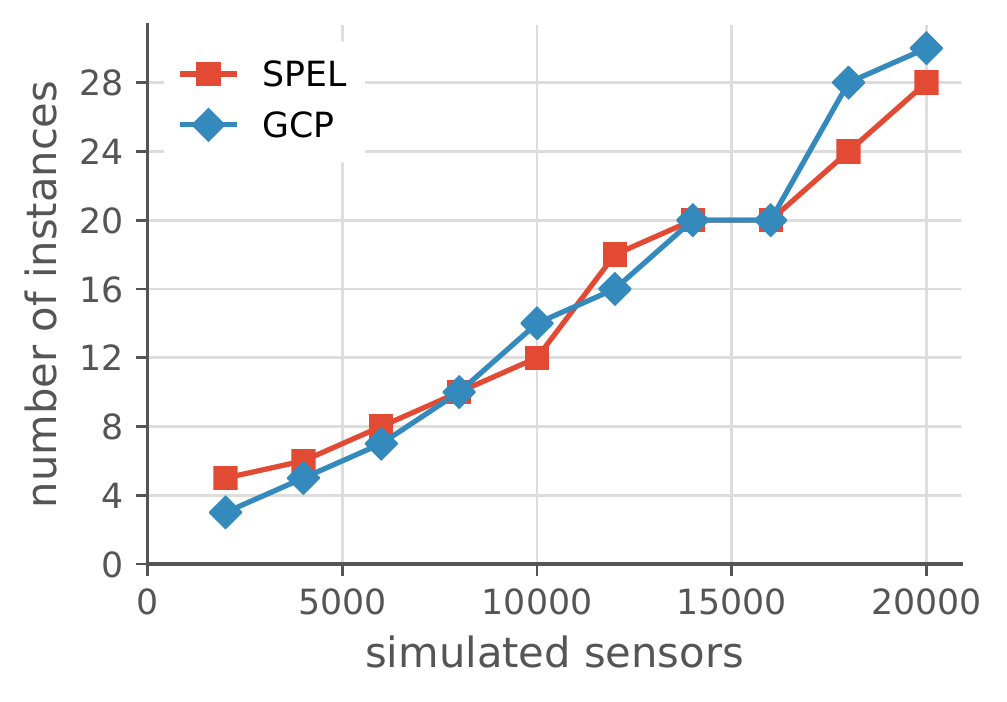}
		\caption{Beam/Samza UC3}
		\label{fig:eval:frameworks:public-private:beamsamza-uc3}
	\end{subfigure}
	\hfill%
	\begin{subfigure}[b]{0.246\linewidth}%
		\centering
		\includegraphics[width=\textwidth]{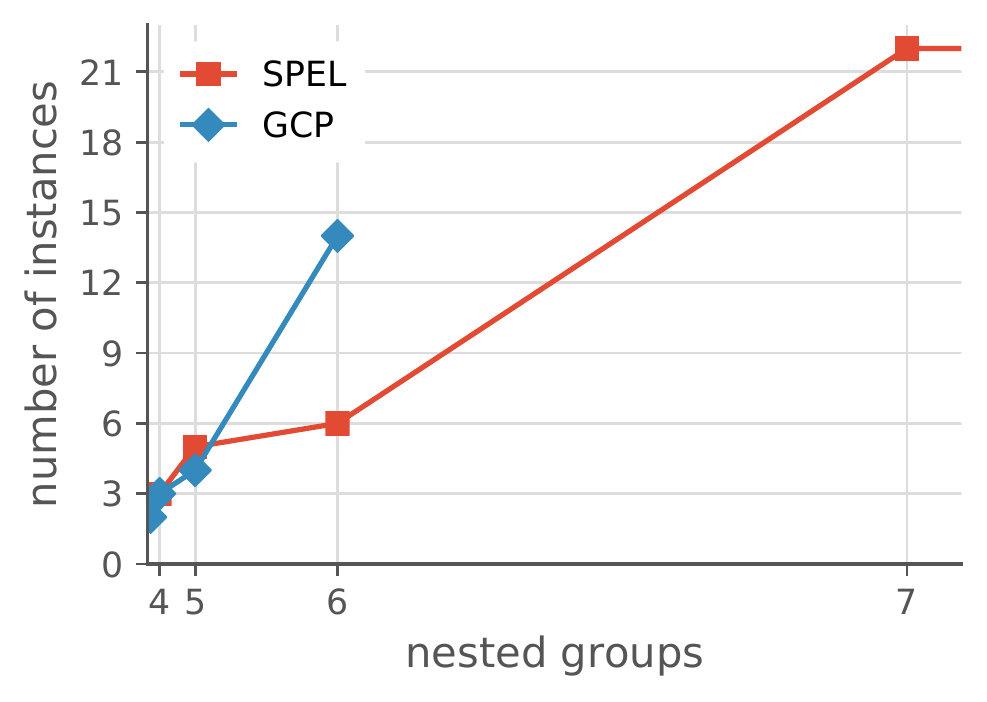}
		\caption{Beam/Samza UC4}
		\label{fig:eval:frameworks:public-private:beamsamza-uc4}
	\end{subfigure}
	\vspace{0.5em}
	\begin{subfigure}[b]{0.246\linewidth}%
		\centering
		\includegraphics[width=\textwidth]{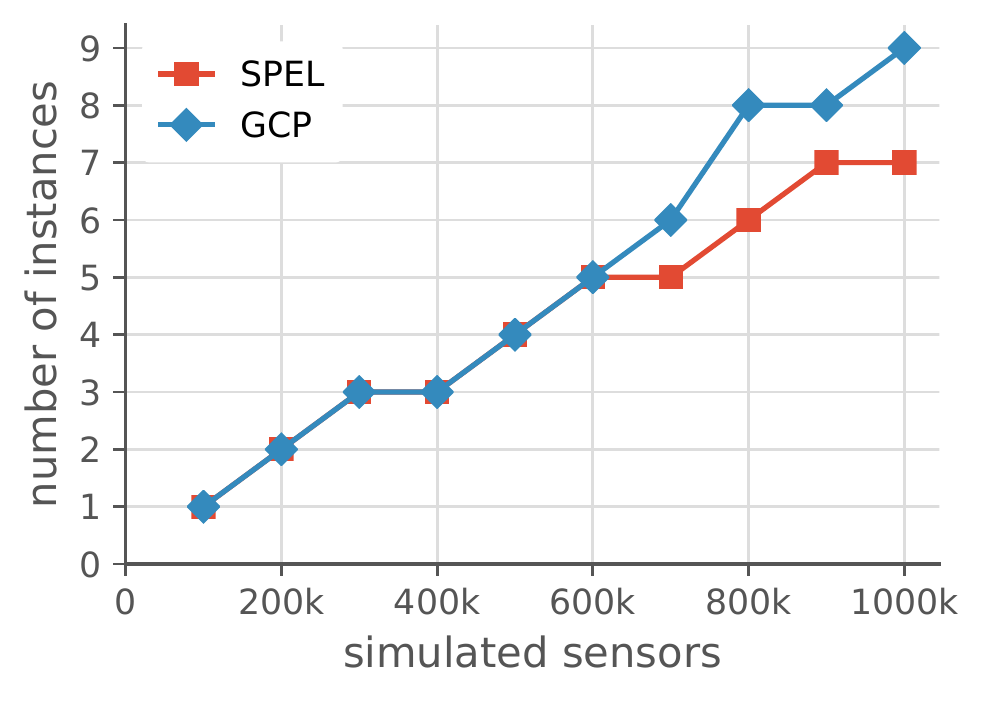}
		\caption{Flink UC1}
		\label{fig:eval:frameworks:public-private:flink-uc1}
	\end{subfigure}
	\hfill%
	\begin{subfigure}[b]{0.246\linewidth}%
		\centering
		\includegraphics[width=\textwidth]{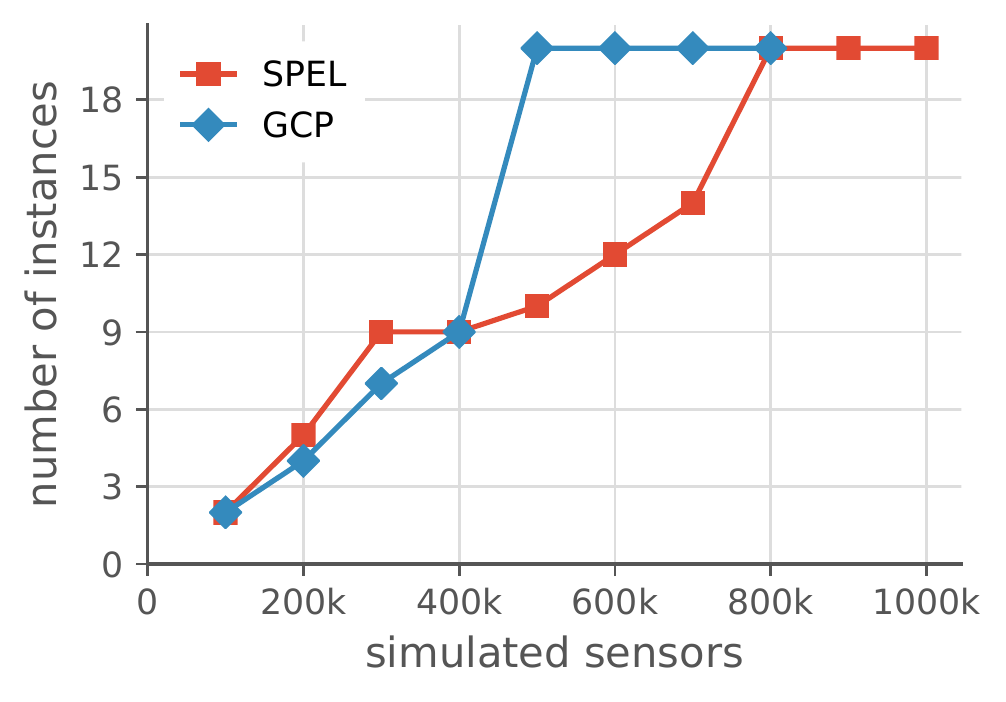}
		\caption{Flink UC2}
		\label{fig:eval:frameworks:public-private:flink-uc2}
	\end{subfigure}
	\hfill%
	\begin{subfigure}[b]{0.246\linewidth}%
		\centering
		\includegraphics[width=\textwidth]{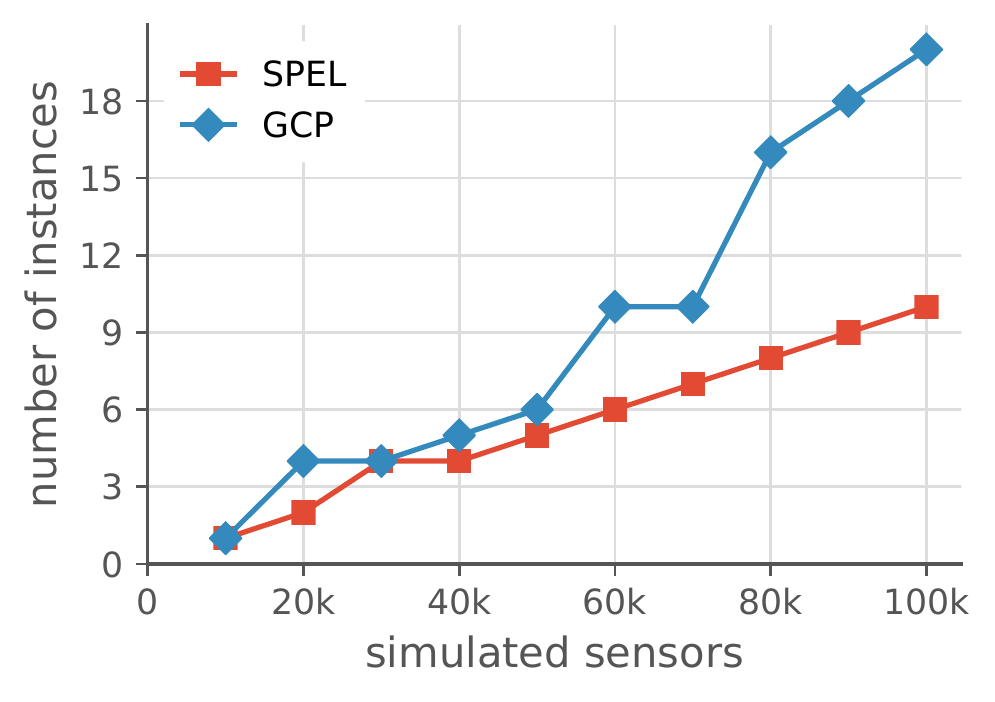}
		\caption{Flink UC3}
		\label{fig:eval:frameworks:public-private:flink-uc3}
	\end{subfigure}
	\hfill%
	\begin{subfigure}[b]{0.246\linewidth}%
		\centering
		\includegraphics[width=\textwidth]{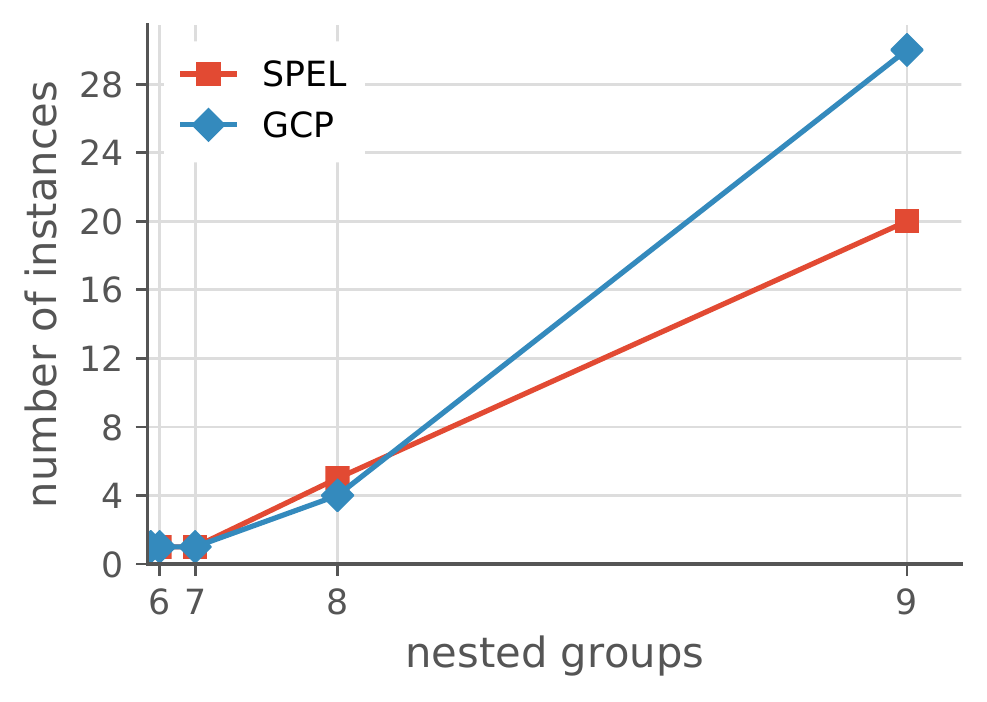}
		\caption{Flink UC4}
		\label{fig:eval:frameworks:public-private:flink-uc4}
	\end{subfigure}
	\vspace{0.5em}
	\begin{subfigure}[b]{0.246\linewidth}%
		\centering
		\includegraphics[width=\textwidth]{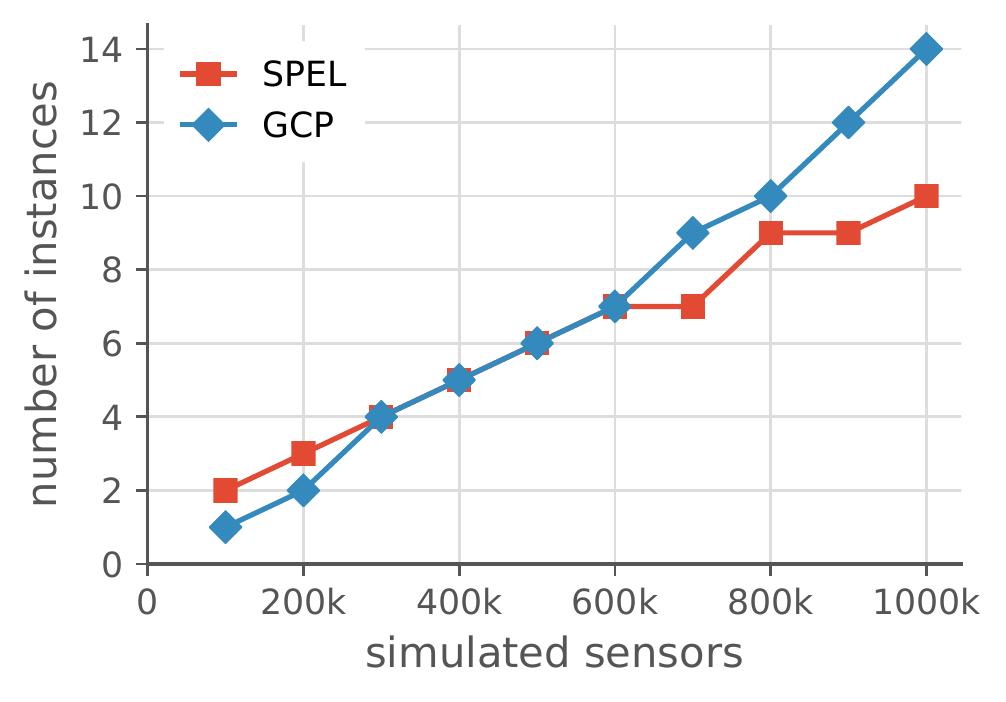}
		\caption{Hazelcast Jet UC1}
		\label{fig:eval:frameworks:public-private:hazelcastjet-uc1}
	\end{subfigure}
	\hfill%
	\begin{subfigure}[b]{0.246\linewidth}%
		\centering
		\includegraphics[width=\textwidth]{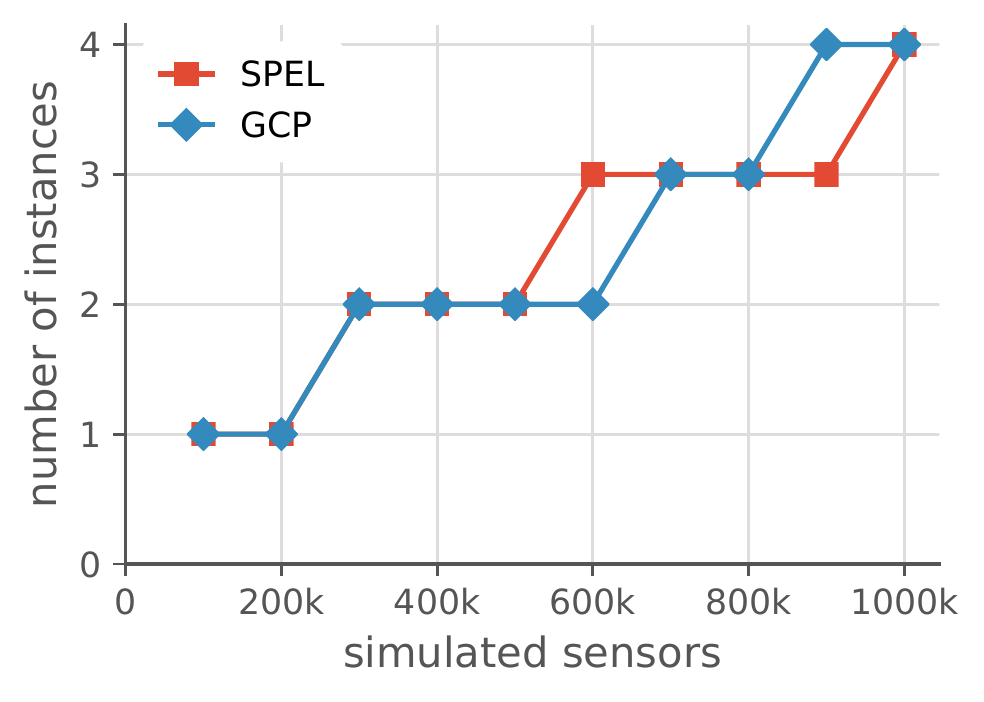}
		\caption{Hazelcast Jet UC2}
		\label{fig:eval:frameworks:public-private:hazelcastjet-uc2}
	\end{subfigure}
	\hfill%
	\begin{subfigure}[b]{0.246\linewidth}%
		\centering
		\includegraphics[width=\textwidth]{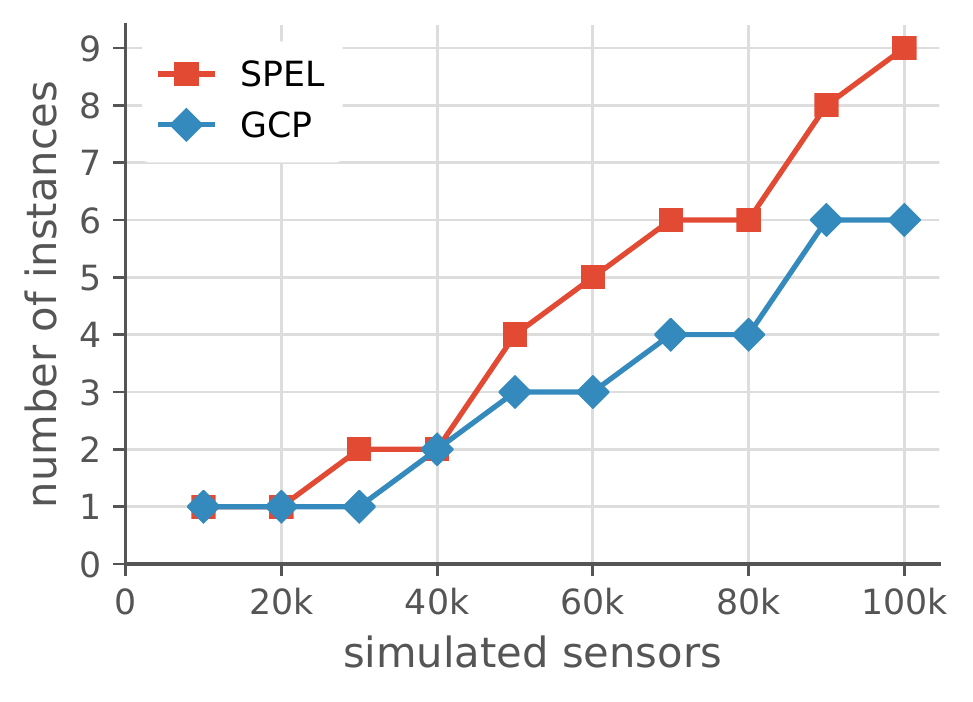}
		\caption{Hazelcast Jet UC3}
		\label{fig:eval:frameworks:public-private:hazelcastjet-uc3}
	\end{subfigure}
	\hfill%
	\begin{subfigure}[b]{0.246\linewidth}%
		\centering
		\includegraphics[width=\textwidth]{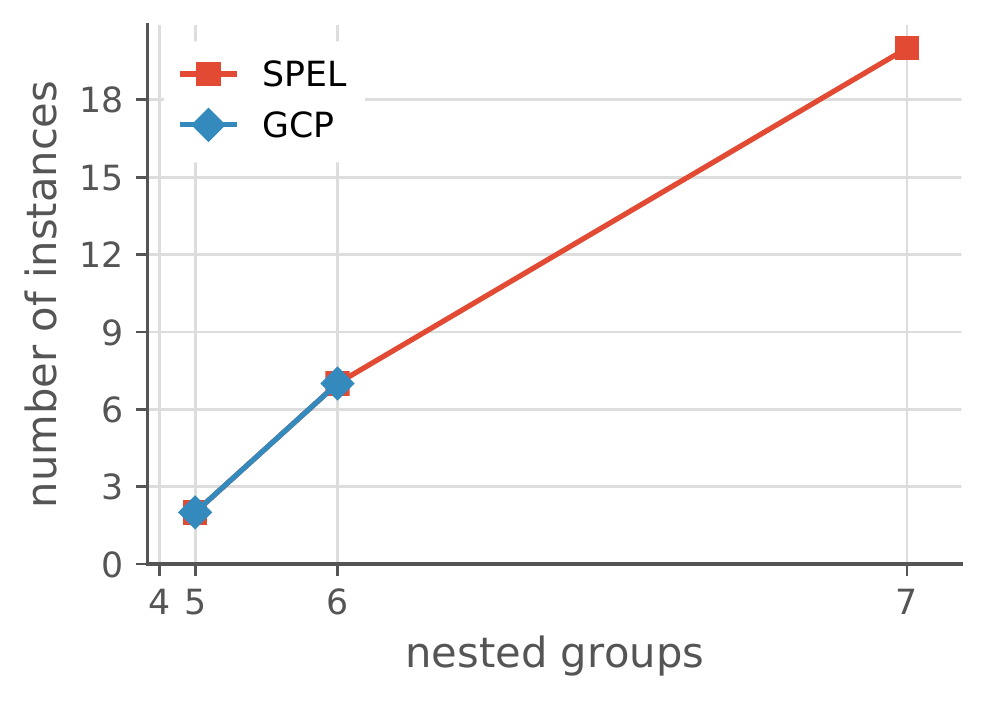}
		\caption{Hazelcast Jet UC4}
		\label{fig:eval:frameworks:public-private:hazelcastjet-uc4}
	\end{subfigure}
	\vspace{0.5em}
	\begin{subfigure}[b]{0.246\linewidth}%
		\centering
		\includegraphics[width=\textwidth]{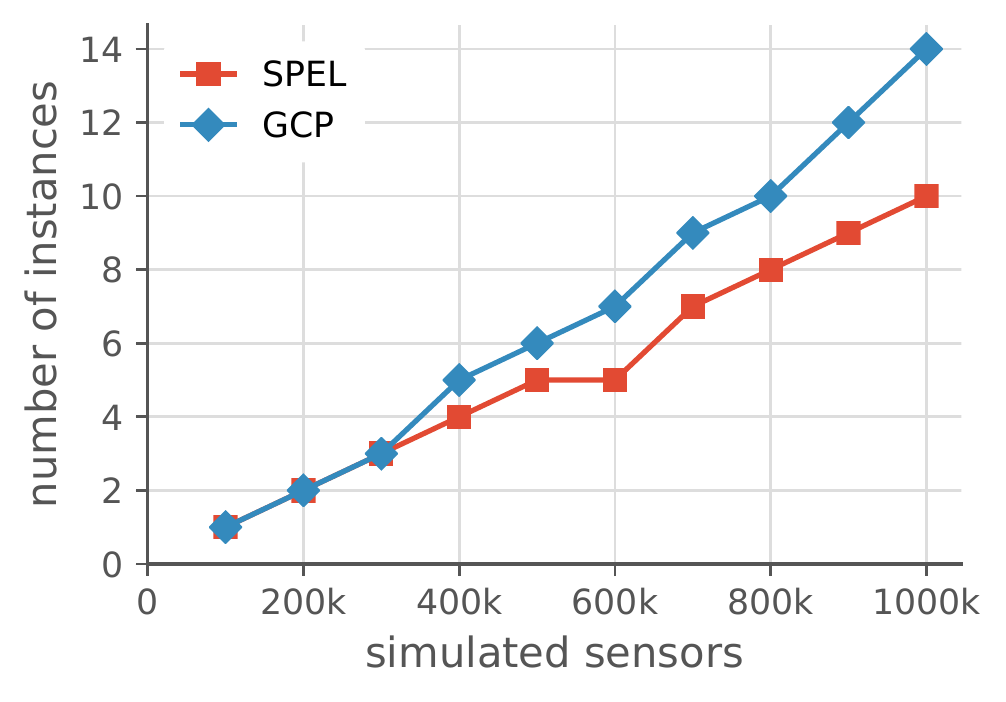}
		\caption{Kafka Streams UC1}
		\label{fig:eval:frameworks:public-private:kstreams-uc1}
	\end{subfigure}
	\hfill%
	\begin{subfigure}[b]{0.246\linewidth}%
		\centering
		\includegraphics[width=\textwidth]{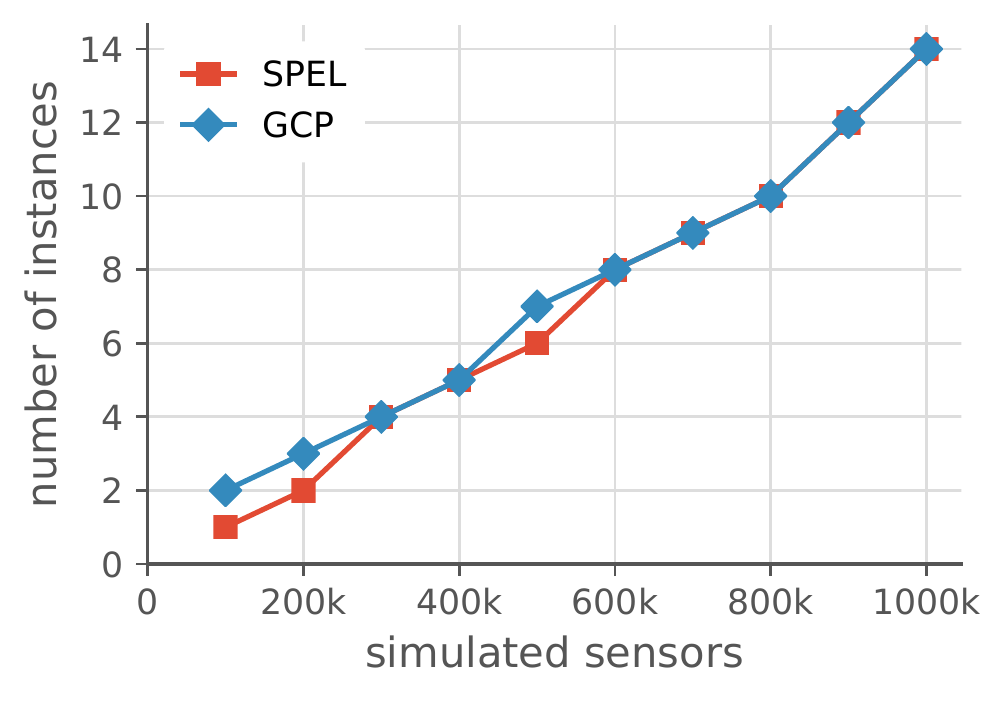}
		\caption{Kafka Streams UC2}
		\label{fig:eval:frameworks:public-private:kstreams-uc2}
	\end{subfigure}
	\hfill%
	\begin{subfigure}[b]{0.246\linewidth}%
		\centering
		\includegraphics[width=\textwidth]{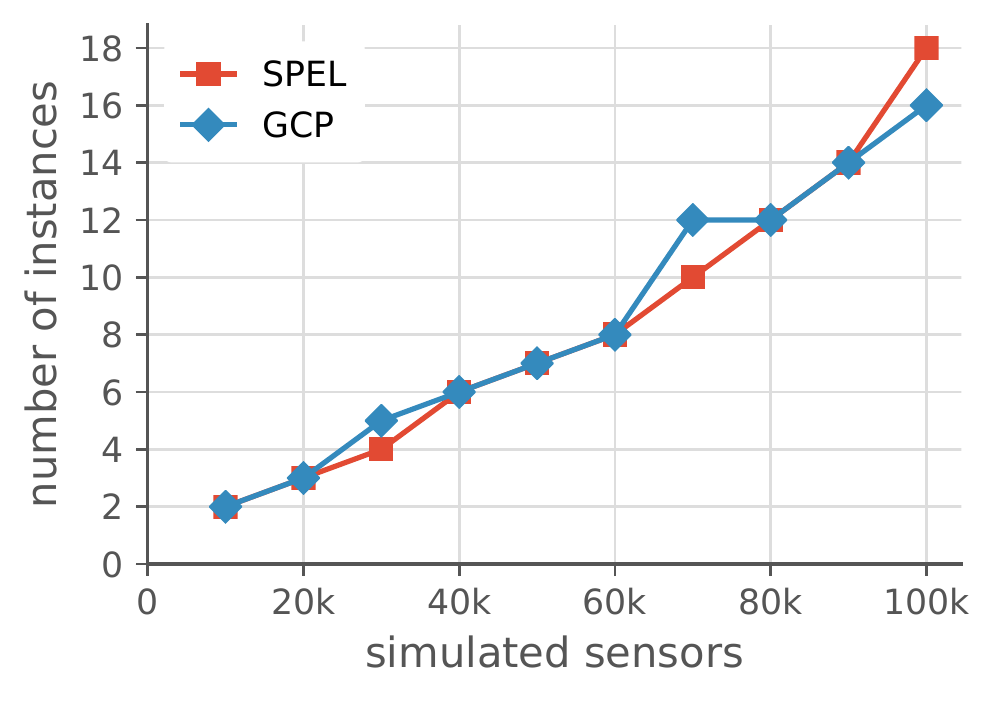}
		\caption{Kafka Streams UC3}
		\label{fig:eval:frameworks:public-private:kstreams-uc3}
	\end{subfigure}
	\hfill%
	\begin{subfigure}[b]{0.246\linewidth}%
		\centering
		\includegraphics[width=\textwidth]{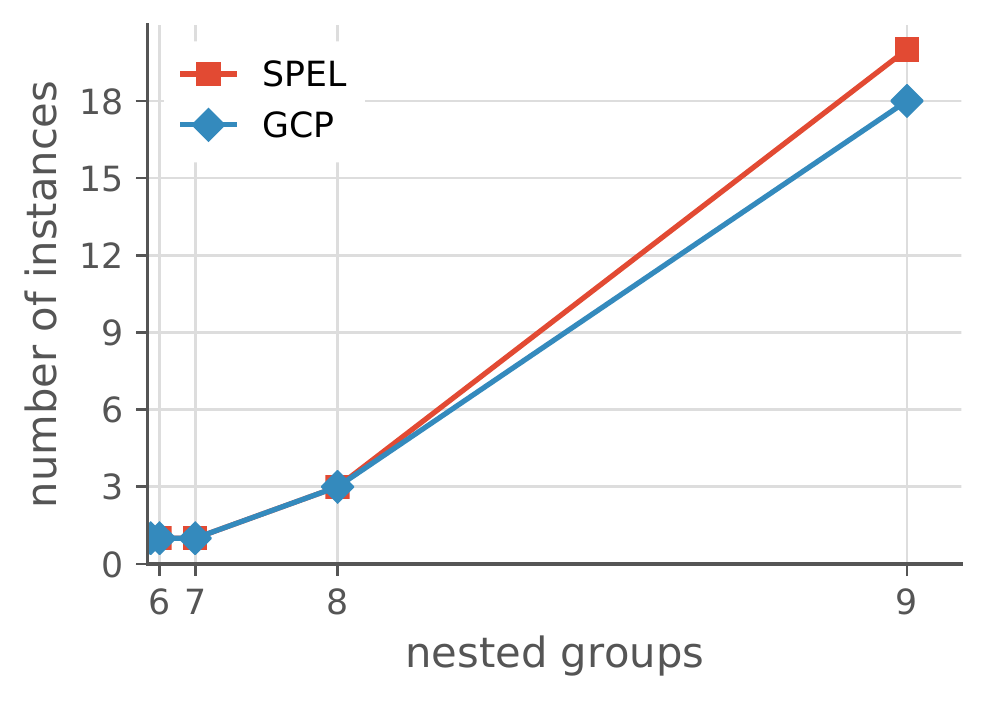}
		\caption{Kafka Streams UC4}
		\label{fig:eval:frameworks:public-private:kstreams-uc4}
	\end{subfigure}
	\caption{Scalability of stream processing frameworks in our private cloud (SPEL) compared to the Google cloud (GCP).}
	\label{fig:eval:frameworks:public-private}
\end{figure*}

\begin{tcolorbox}
	\textbf{\rqvertical:}
	Despite some exceptions for specific task samples, stream processing frameworks allow scaling them on a single node, making vertical scaling a complementary measure to achieve scalability of stream processing frameworks.
	For Flink and Hazelcast Jet, it might be more resource efficient to scale with the CPU cores per microservice instance instead of scaling the number of instances.
	However, caution should be exercised when accessing shared resources.
\end{tcolorbox}

\subsection{Comparing Scalability in Public and Private Clouds}\label{sec:experimental-results:public-private}

\Cref{sec:experimental-results:base} presented our baseline results of benchmarking scalability in the private cloud Kubernetes cluster. We repeat these experiments in an identically configured Kubernetes cluster in the Google cloud to address \rqpublic\ (see \cref{tab:clouds}).
As in the previous section, %
we generate the same load intensities as in the first experiment, yet using the smaller load intensities from \cref{sec:experimental-results:beam-config} for the Beam experiments and the 30~days window for Hazelcast Jet with benchmark UC3.

\Cref{fig:eval:frameworks:public-private} compares the scalability results in the private and in the public cloud for each benchmark and stream processing framework. In general, we can observe that the results for both clouds are similar and all frameworks show linear scalability, independent of the cloud platform.
We can see a tendency that for higher loads (Flink, Hazelcast Jet, and Kafka Streams), the resource demand in the Google cloud increases at a slightly steeper rate. We expect that this is due to the fact that we used Google's general-purpose \textit{E2} virtual machines with potentially less powerful resources.
Our benchmark UC3 execution of Hazelcast stands out as its resource demand increases at a lower rate for the Google cloud. Further experiments would be required to investigate whether this combination of framework and task sample is special in some way.

\begin{tcolorbox}
	\textbf{\rqpublic:}
	Our previous results apply independently of whether a public or private cloud environment is used.
	In our experiments, resource demand increases steeper in the public cloud, yet we expect this to be due to the specific machine types selected.
\end{tcolorbox}

\subsection{Scaling the Cluster Size}\label{sec:experimental-results:large-cluster}

In the previous experiments, we deployed up to 30~SUT instances since for larger numbers we observed interference of the SUT, load generation, messaging system, and infrastructure components causing unstable results. In this section, we address \rqshift\ and evaluate whether stream processing frameworks can be scaled further when also increasing the underlying computing resources.
To obtain a more stable infrastructure, we modify our experimental setup in this section (see \cref{tab:clouds}).
We compose our Kubernetes cluster of two node groups in Google Cloud:
The first consists of four 16-core machines, which run the load generators, four Kafka brokers, and additional benchmarking infrastructure.
The second node group only runs the SUTs. We evaluate two sizes of this node group, namely four and eight 16-core machines.
As we observed linear scalability for all frameworks and independent of the benchmark, we focus on benchmark UC3 in these experiments. We simulate up to 1\,000\,000 sensors for Flink, Hazelcast Jet, and Kafka Streams and up to 100\,000 sensors for the Beam SUTs.
For the cluster with four SUT nodes, we deploy up to 55~instances, while for the cluster with eight SUT nodes, we deploy up to 110~instances. For all frameworks, we increase the number of instances in steps of five.

\begin{figure}
	\centering
	\begin{subfigure}[b]{0.495\linewidth}%
		\centering
		\includegraphics[width=\textwidth]{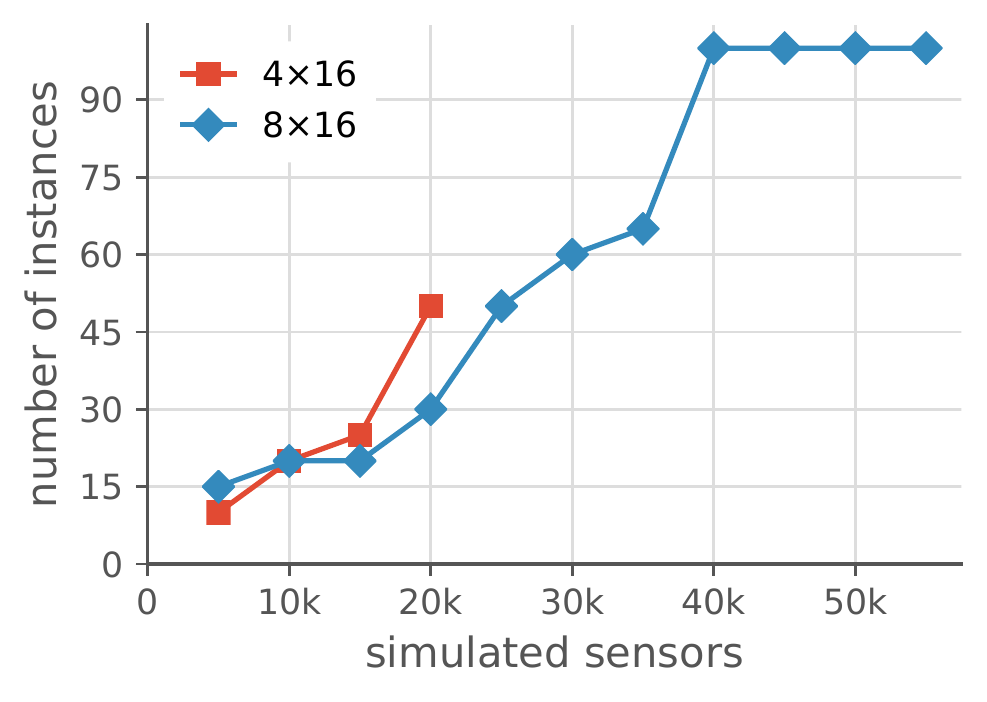}
		\caption{Beam/Flink}
		\label{fig:experimental-results:large-cluster:beam_flink_uc3}
	\end{subfigure}
	\hfill%
	\begin{subfigure}[b]{0.495\linewidth}%
		\centering
		\includegraphics[width=\textwidth]{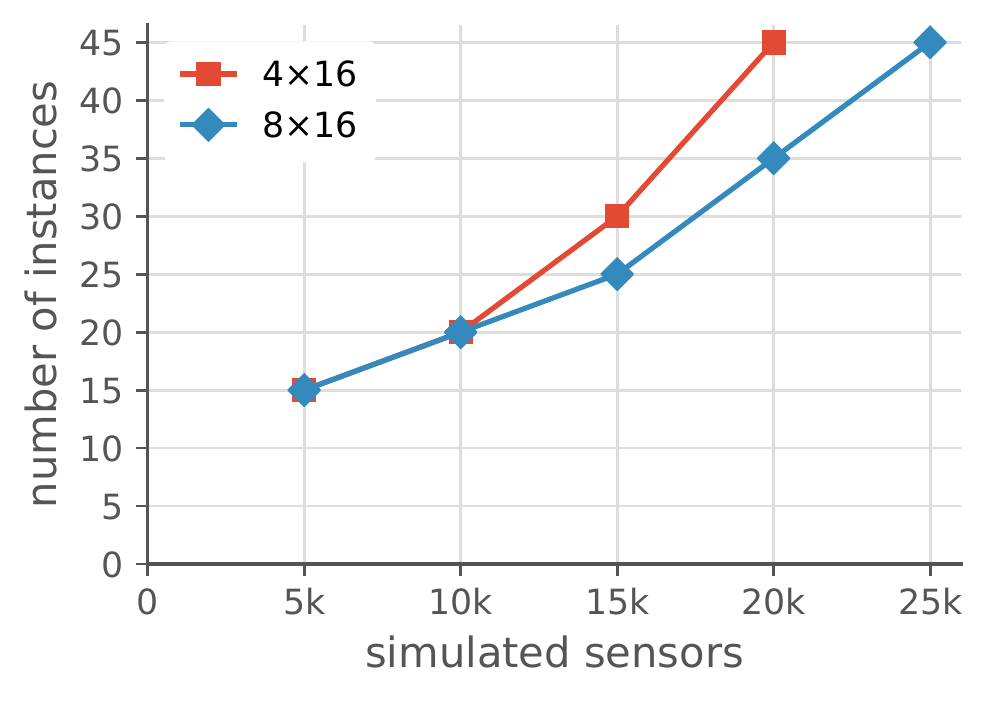}
		\caption{Beam/Samza}
		\label{fig:experimental-results:large-cluster:beam_samza_uc3}
	\end{subfigure}
	
	\vspace{1em}
	
	\begin{subfigure}[b]{0.495\linewidth}%
		\centering
		\includegraphics[width=\textwidth]{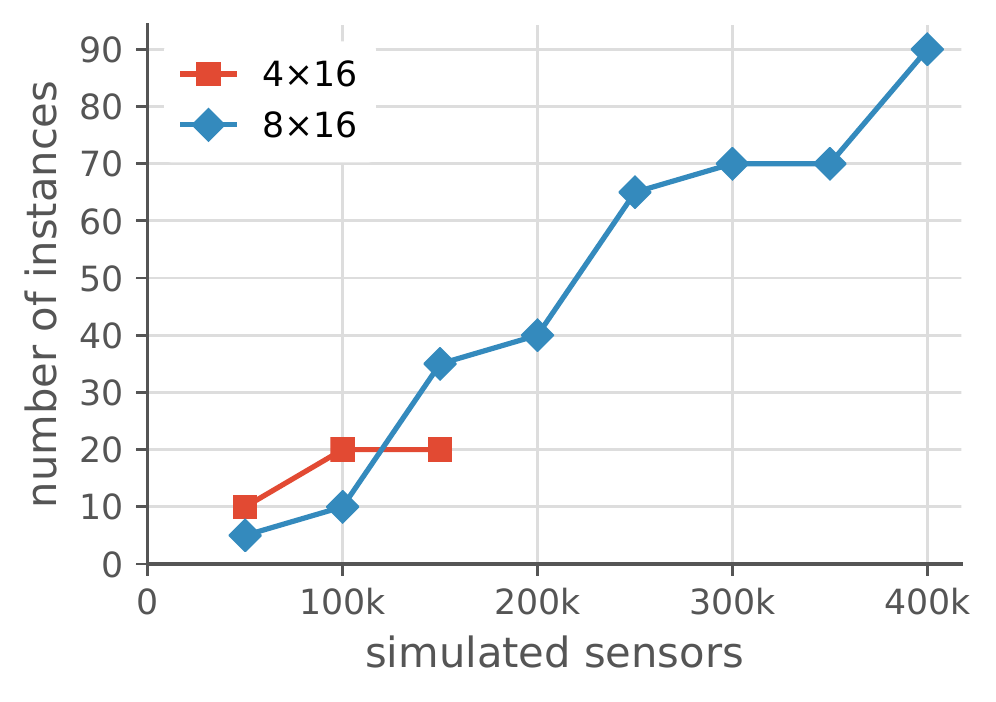}
		\caption{Flink}
		\label{fig:experimental-results:large-cluster:flink_uc3}
	\end{subfigure}
	\hfill%
	\begin{subfigure}[b]{0.495\linewidth}%
		\centering
		\includegraphics[width=\textwidth]{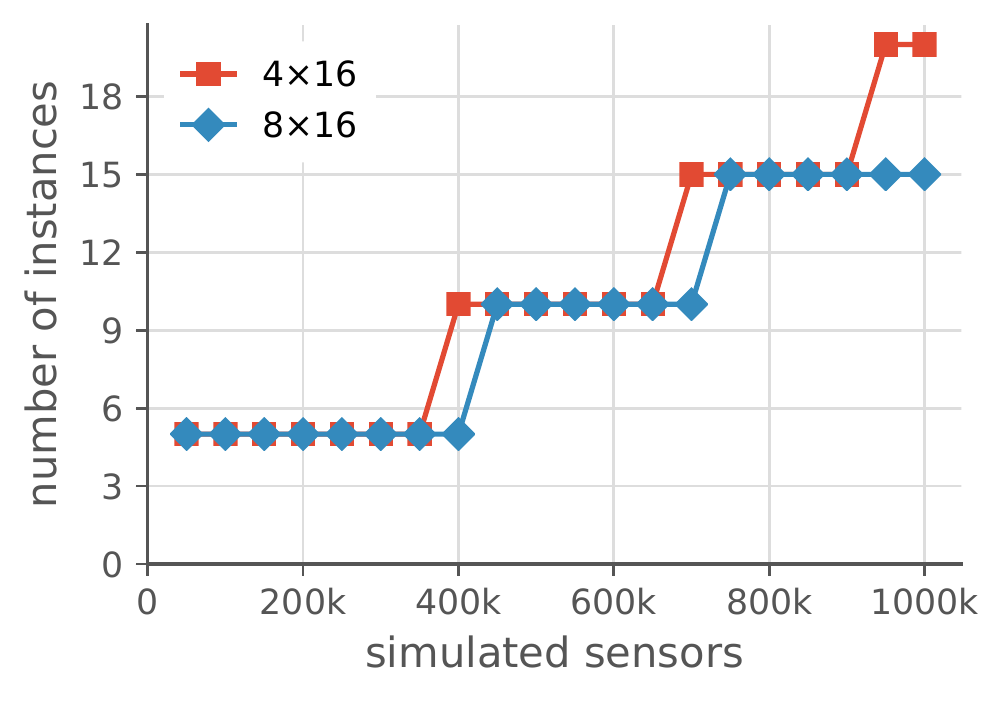}
		\caption{Hazelcast Jet}
		\label{fig:experimental-results:large-cluster:hazelcastjet_uc3}
	\end{subfigure}
	
	\vspace{1em}
	
	\begin{subfigure}[b]{0.495\linewidth}%
		\centering
		\includegraphics[width=\textwidth]{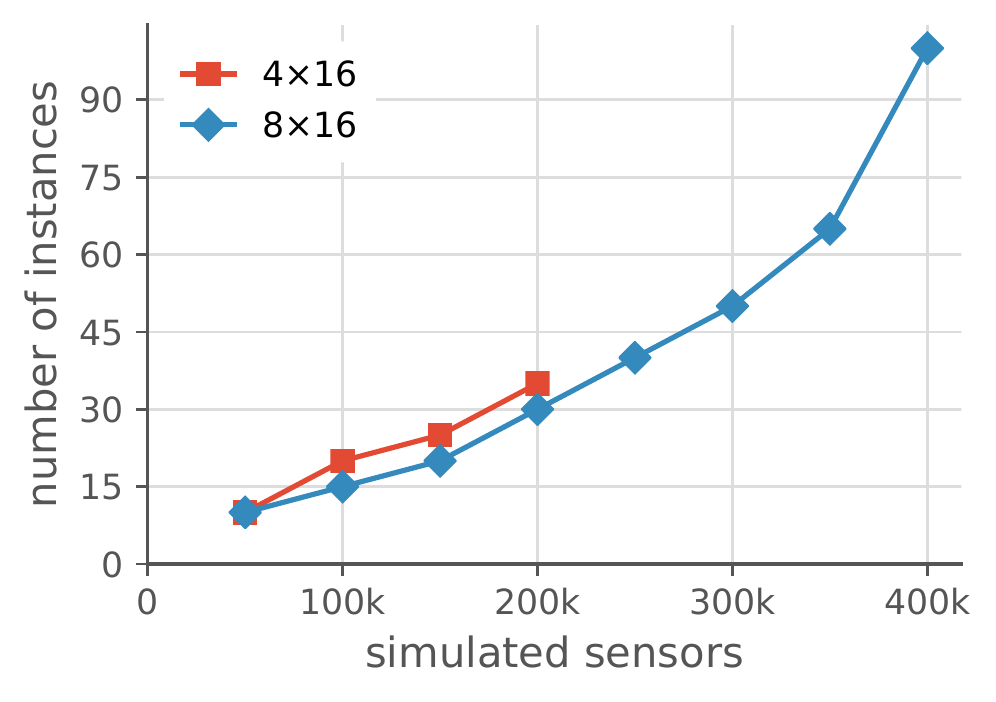}
		\caption{Kafka Streams}
		\label{fig:experimental-results:large-cluster:kstreams_uc3}
	\end{subfigure}
	\caption{Scalability benchmark results for benchmark UC3 in Google Cloud for different cluster sizes (machines$\times$cores).}
	\label{fig:experimental-results:large-cluster}
\end{figure}

\Cref{fig:experimental-results:large-cluster} shows our scalability benchmark results for larger clusters. For Beam/Flink, Flink, and Kafka Streams, we can observe that the maximum processable load increases when using a cluster of eight nodes instead of four nodes. For Hazelcast Jet, already the four-node cluster is able to process the highest generated load. For Beam with the Samza runner, only 20\,000 messages per second can be processed, independent of the cluster size.
We observe frequent crashes with large Beam/Samza deployments, but due to the many issues observed in our Samza experiments, do not further investigate this.
An important observation is that for all frameworks, fewer SUT instances are required when using the larger Kubernetes cluster.

\begin{tcolorbox}
	\textbf{\rqshift:}
	Observed scalability limits seem to be caused by utilized hardware and not by the streaming frameworks themselves. Hence, limits can be raised by using larger clusters.
\end{tcolorbox}

\section{Threats to Validity}\label{sec:threats-to-validity}

Despite careful research design, there exist threats and limitations to the validity of our evaluation, which we report below.

\paragraph{Threats to Internal Validity}
Cloud platforms in general allow only making little assumptions regarding the underlying hardware or software infrastructure~\cite{Bermbach2017}. Cloud-native containerized deployments as in this study further abstract this.
We consciously chose a representative, Kubernetes-based execution environment, which, however, limits control of possible influences on our result. 
To obtain statistically grounded benchmarking results, we build these evaluations upon the results of our benchmarking method's evaluation~\cite{EMSE2022}.
Nevertheless, we only found that the selected configuration options provide good estimates.
Hence, resource demands obtained in these evaluations should also only be considered  estimates. Repeating individual experiments more often and for longer time periods as well as using our \textit{full search} strategy is likely to produce very similar, but not necessarily identical results.
Moreover, we evaluate a larger set of SUTs, load types, resource types, SLOs, and benchmarks in this study compared to our method's evaluation.
We address this limitation by carefully observing the benchmark execution but do not conduct as extensive experiments as in our method's evaluation.
We refer to our previous publication for further discussion on balancing statistical grounding and time-efficient execution for scalability benchmarking~\cite{EMSE2022}.

\paragraph{Threats to External Validity}

We conduct all experiments in this evaluation with our Theodolite benchmark task samples.
As discussed in previous work~\cite{BDR2021}, these benchmarks represent relevant use cases.
However, we cannot directly generalize our findings to arbitrary other applications, where other frameworks, configuration options, or deployment options might perform differently than our experiments.
We conduct all experiments in this evaluation in our private cloud environment and in Google Cloud.
For the private cloud, we use comparatively large bare metal nodes, while for Google Cloud we use large virtual machines.
Although our evaluations in \cref{sec:experimental-results:public-private} show only small deviations between both environments, we cannot necessarily conclude that we would obtain the same results in other cloud environments.
We deliberately assess all frameworks primarily using their default configurations. This should mitigate any bias stemming from different levels of our experience with the different frameworks. However, it does not allow us to make any conclusions about the potential performance and scalability improvements when fine-tuned for specific scenarios.
When choosing between different stream processing frameworks, practitioners might also consider other attributes beyond scalability, such as end-to-end latency, the ability to recover from failures, or ease of use and development.

\section{Related Work}\label{sec:related-work}

Over the last few years, several benchmarks for stream processing frameworks have been proposed and stream processing benchmarking studies have been conducted. The differentiation between benchmarks and experimental studies applying them is sometimes blurry. Many publications that present benchmarks perform also an experimental study with them. On the other hand, many experimental studies utilize existing benchmarks, but modify them.
Nevertheless, we structure this section into two parts: First, we give an overview of stream processing benchmarks to justify our benchmark selection for this study. Second, we discuss related stream processing benchmarking studies.
For a systematic mapping of the literature on performance in stream processing, we refer to the recent study by \citet{SEAA2023}.

\subsection{Related Work on Stream Processing Benchmarks}

Besides the Theodolite stream processing benchmarks used in this study, several other benchmarks for stream processing frameworks have been proposed.
\cref{tab:related-benchmarks} summarizes characteristics of the discussed benchmarks.

\begin{table*}
	\begin{threeparttable}[b]
		\caption{Overview of the characteristics and implementations of stream processing benchmarks~\cite{PhDHenning23}.}
		\label{tab:related-benchmarks}
		\footnotesize
		\newcommand{\cmark}{\ding{51}}%
		\newcommand{\xmark}{\ding{55}}%
		\newcommand{\qmark}{\makebox[0pt][l]{\textbf{\textit{?}}}\phantom{\cmark}}%
		
		\newcommand{\txnote}[1]{\makebox[0pt][l]{\tnote{#1}}}
		
		\newcommand\undefcolumntype[1]{\expandafter\let\csname NC@find@#1\endcsname\relax}
		\newcommand\forcenewcolumntype[1]{\undefcolumntype{#1}\newcolumntype{#1}}
		
		\newcommand*\rot{\rotatebox{90}}
		\newcolumntype{L}{>{\raggedright\arraybackslash}X}
		\newcolumntype{R}{>{\raggedleft\arraybackslash}X}
		\newcolumntype{C}{>{\centering\arraybackslash}X}
		\newcolumntype{o}{p{0pt}}
		\renewcommand{\arraystretch}{1.2}
		\newcommand{\fnoptional}{a}
		\newcommand{\fnbeam}{b}
		\newcommand{\fnriottasksamples}{d}
		\newcommand{\fnbeamnexmark}{c}
		\begin{tabularx}{\textwidth}{ll o C o C o CCC o CCCCCCC o C o CCC}
			\toprule
			&&& && && \multicolumn{3}{c}{Messaging} && \multicolumn{7}{c}{Stream processing framework} && && \multicolumn{3}{c}{Cloud-native} \\
			\cmidrule{8-10} \cmidrule{12-18} \cmidrule{22-24}
			Benchmark & Published && \rot{Task samples} && \rot{Open source} && \rot{Kafka} & \rot{Others} & \rot{None} && \rot{Flink} & \rot{Spark} & \rot{Storm} & \rot{Samza} & \rot{Kafka Streams} & \rot{Hazelcast Jet} & \rot{Others} && \rot{Database} && \rot{Containers} & \rot{Kubernetes} & \rot{Others} \\
			\midrule
			Theodolite \cite{BDR2021} & \citeyear{BDR2021}
			& %
			& 4
			& %
			& \cmark %
			& %
			& \cmark %
			& %
			& %
			& %
			& \cmark %
			& %
			& %
			& \cmark\txnote{\fnbeam} %
			& \cmark %
			& \cmark %
			& \cmark\txnote{\fnbeam} %
			& %
			& \phantom{\cmark}\txnote{\fnoptional} %
			& %
			& \cmark %
			& \cmark %
			& %
			\\
			Beam Nexmark \cite{BeamNexmark2022} & \citeyear{BeamNexmark2022}\txnote{\fnbeamnexmark}
			& %
			& 13
			& %
			& \cmark %
			& %
			& \cmark %
			& \cmark %
			& %
			& %
			& \cmark\txnote{\fnbeam} %
			& \cmark\txnote{\fnbeam} %
			& %
			& \qmark\txnote{\fnbeam} %
			& %
			& \qmark\txnote{\fnbeam} %
			& \cmark\txnote{\fnbeam} %
			& %
			& %
			& %
			& %
			& %
			& %
			\\
			ESPBench \cite{Hesse2021} & \citeyear{Hesse2021}
			& %
			& 5
			& %
			& \cmark %
			& %
			& \cmark %
			& %
			& %
			& %
			& \cmark\txnote{\fnbeam} %
			& \cmark\txnote{\fnbeam} %
			& %
			& \qmark\txnote{\fnbeam} %
			& %
			& \cmark\txnote{\fnbeam} %
			& \qmark\txnote{\fnbeam} %
			& %
			& \cmark %
			& %
			& %
			& %
			& %
			\\
			OSPBench \cite{vanDongen2020} & \citeyear{vanDongen2020}
			& %
			& 5
			& %
			& \cmark %
			& %
			& \cmark %
			& %
			& %
			& %
			& \cmark %
			& \cmark %
			& %
			& %
			& \cmark %
			& %
			& %
			& %
			& %
			& %
			& \cmark %
			& %
			& \cmark %
			\\
			DSPBench \cite{Bordin2020} & \citeyear{Bordin2020}
			& %
			& 5
			& %
			& \cmark %
			& %
			& \cmark %
			& %
			& %
			& %
			& %
			& \cmark %
			& \cmark %
			& %
			& %
			& %
			& %
			& %
			& \cmark %
			& %
			& %
			& %
			& %
			\\
			\citet{Shahverdi2019} & \citeyear{Shahverdi2019}
			& %
			& 1
			& %
			& \cmark %
			& %
			& \cmark %
			& %
			& %
			& %
			& \cmark %
			& \cmark %
			& \cmark %
			& %
			& \cmark %
			& \cmark %
			& %
			& %
			& \cmark %
			& %
			& %
			& %
			& %
			\\
			\citet{Karimov2018} & \citeyear{Karimov2018}
			& %
			& 2
			& %
			& %
			& %
			& %
			& %
			& \cmark %
			& %
			& \cmark %
			& \cmark %
			& \cmark %
			& %
			& %
			& %
			& %
			& %
			& %
			& %
			& %
			& %
			& %
			\\
			RIoTBench \cite{Shukla2017} & \citeyear{Shukla2017}
			& %
			& 4\txnote{\fnriottasksamples} %
			& %
			& \cmark %
			& %
			& %
			& \cmark %
			& %
			& %
			& %
			& %
			& \cmark %
			& %
			& %
			& %
			& %
			& %
			& \cmark %
			& %
			& %
			& %
			& %
			\\
			YSB \cite{Chintapalli2016} & \citeyear{Chintapalli2016}
			& %
			& 1
			& %
			& \cmark %
			& %
			& \cmark %
			& %
			& %
			& %
			& \cmark %
			& \cmark %
			& \cmark %
			& %
			& %
			& %
			& %
			& %
			& \cmark %
			& %
			& %
			& %
			& %
			\\
			SparkBench \cite{Li2015} & \citeyear{Li2015}
			& %
			& 10
			& %
			& \cmark %
			& %
			& %
			& %
			& \cmark %
			& %
			& %
			& \cmark %
			& %
			& %
			& %
			& %
			& %
			& %
			& %
			& %
			& %
			& %
			& %
			\\
			StreamBench \cite{Lu2014} & \citeyear{Lu2014}
			& %
			& 7
			& %
			& %
			& %
			& \cmark %
			& %
			& %
			& %
			& %
			& \cmark %
			& \cmark %
			& %
			& %
			& %
			& %
			& %
			& %
			& %
			& %
			& %
			& %
			\\
			Linear Road \cite{Arasu2004} & \citeyear{Arasu2004}
			& %
			& 5
			& %
			& %
			& %
			& %
			& %
			& \cmark %
			& %
			& %
			& %
			& %
			& %
			& %
			& %
			& \cmark %
			& %
			& %
			& %
			& %
			& %
			& %
			\\
			\bottomrule
		\end{tabularx}
		\begin{tablenotes}\footnotesize
			\item[\fnoptional] optional
			\item[\fnbeam] using Apache Beam
			\item[\fnbeamnexmark] the Beam Nexmark benchmarks are based on the Nexmark paper \cite{Tucker2010} published in \citeyear{Tucker2010}
			\item[\fnriottasksamples] RIoTBench's 4 application benchmarks are composed of 27 microbenchmarks
		\end{tablenotes}
	\end{threeparttable}
\end{table*}

StreamBench~\cite{Lu2014} is one of the earliest benchmarks for modern stream processing frameworks. While originally only implemented for Spark and Storm, it has later been used to benchmark Apache Apex, Beam, Flink, and Samza as well \cite{Hesse2019, Qian2016}.
As its name suggests, SparkBench~\cite{Li2015} is a benchmark tailored to Apache Spark.
The Yahoo Streaming Benchmark (YSB) \cite{Chintapalli2016} is frequently used and adapted in research \cite{Lopez2016, Yang2017, Karakaya2017, Nasiri2019, Zeuch2019, Chu2020, vanDongen2020}.
Worth highlighting is the work of \citet{Shahverdi2019}, who extend YSB with implementations for the frameworks Kafka Streams and Hazelcast Jet. As discussed in \cref{sec:frameworks}, these frameworks are particularly suited to be deployed as microservices.
RIoTBench \cite{Shukla2017} provides four application benchmarks for Storm composed of 27~small task samples. \citet{Nasiri2019} adopt RIoTBench for Flink and Spark.
\citet{Karimov2018} present a benchmark with two task samples, derived from a real industrial context, yet without providing open-source implementations.

More recently, DSPBench \cite{Bordin2020}, OSPBench~\cite{vanDongen2020}, and ESPBench \cite{Hesse2021} have been proposed.
DSPBench contains 15~benchmarks, which resample typical stream processing applications, derived from reviewing the literature.
OSPBench provides benchmarks for analyzing traffic sensor data. Besides evaluations of latency, throughput, and resource usage, \citeauthor{vanDongen2020} used OSPBench to also evaluate scalability~\cite{vanDongen2021b} and fault recovery~\cite{vanDongen2021a}.
In contrast to most other benchmarks, OSPBench provides implementations for the rather new framework Kafka Streams, which is also evaluated in this study.
The Enterprise Stream Processing Benchmark (ESPBench) builds upon the Senska benchmark \cite{Hesse2018}.
It is special in the sense that it integrates a relational database management system.
In contrast to most other benchmarks, ESPBench's task samples are implemented with Apache Beam. While \citet{Hesse2021} only perform evaluations with Spark, Flink, and Hazelcast Jet, we expect that also other Beam runners can be used to run the benchmark.

The Nexmark benchmark \cite{Tucker2010} has originally been proposed as the \textit{Niagara Extension to the XMark benchmark} addressed to first-generation stream processing systems (see the survey of \citet{Fragkoulis2023} for a discussion of first and second-generation stream processing systems).
The Apache Beam community adapted and extended Nexmark with implementations for Beam to benchmark the performance of different runners~\cite{BeamNexmark2022}.
Documentation and benchmark results are provided for the direct runner as well as for the Flink, the Spark, and the Google Cloud Dataflow runners.
However, running the benchmark with other runners should be possible as well.
Recently, there seems to be an effort to implement the Nexmark task samples with other frameworks in an open-source project.\footnote{\url{https://github.com/nexmark/nexmark}}
However, currently this project only provides implementations for Apache Flink.
Moreover, \citet{Gencer2021} implemented the Nexmark benchmark for their performance evaluation of Hazelcast Jet.

Worth mentioning is also the Linear Road benchmark presented by \citet{Arasu2004}. Although published years before all modern stream processing frameworks considered in this work have been released, it is still used in research \cite{Zhang2017,Zeuch2019,Sax2020} and compared to newer benchmarks \cite{Bordin2020,Hesse2021}.
\citet{Pagliari2020} and \citet{Garcia2022a, Garcia2022b} present approaches to generate benchmarks.

From \cref{tab:related-benchmarks}, we can see that a lot of open-source benchmarks have been proposed. Apart from the Theodolite benchmarks, none of these benchmarks is particularly addressed to scalability.
Often originating in data management research, many benchmarks are defined as ``queries'' over data streams~\cite{Tucker2010,Karimov2018,Hesse2021}.
Most benchmarks include a messaging system as a middleware component between workload generation and stream processing framework. In the vast majority of cases, this is Apache Kafka.
\citet{Karimov2018} exclude such a system to not let it become the benchmark's bottleneck. Our Theodolite benchmarks purposely include Kafka to represent more realistic microservice deployments~\cite{BDR2021}.
Flink, Spark, and Storm are by far the most supported frameworks. Only a few benchmarks exist for Samza, Kafka Streams, and Hazelcast Jet, which are frameworks particularly suited to be deployed as microservices. Our Theodolite benchmarks are the only ones providing implementations for all of them.
While some benchmarks include an interaction with a database in their setup, others do not.
With the Theodolite benchmarks, a database can optionally be used as we did in a previous study~\cite{IC2E2022FaaSStreaming}.
Besides our Theodolite benchmarks, there is only one other benchmark (OSPBench) that is provided as container images to be used in a cloud-native setting. No other benchmark provides Kubernetes manifests.

\subsection{Related Work on Stream Processing Benchmarking}

\begin{table*}
	\begin{threeparttable}[b]
		\caption{Overview of employed benchmarks, frameworks, and experimental setup of stream processing benchmarking studies~\cite{PhDHenning23}.}
		\label{tab:related-experiments}
		\footnotesize
		\newcommand{\cmark}{\ding{51}}%
		\newcommand{\xmark}{\ding{55}}%
		\newcommand{\qmark}{\makebox[0pt][l]{\textbf{\textit{?}}}\phantom{\cmark}}%
		
		\newcommand{\txnote}[1]{\makebox[0pt][l]{\tnote{#1}}}
		
		\newcommand\undefcolumntype[1]{\expandafter\let\csname NC@find@#1\endcsname\relax}
		\newcommand\forcenewcolumntype[1]{\undefcolumntype{#1}\newcolumntype{#1}}
		
		\newcommand*\rot{\rotatebox{90}}
		\newcolumntype{L}{>{\raggedright\arraybackslash}X}
		\newcolumntype{R}{>{\raggedleft\arraybackslash}X}
		\newcolumntype{C}{>{\centering\arraybackslash}X}
		\newcolumntype{o}{p{0pt}}
		\renewcommand{\arraystretch}{1.2}
		\newcommand{\fnvandenpoel}{a}
		\newcommand{\fnbeam}{b}
		\begin{tabularx}{\textwidth}{ll o CCCCCCCCCCCCC o CCCCCCC o CCCCCC}
			\toprule
			&&& \multicolumn{13}{c}{Benchmark} && \multicolumn{7}{c}{Framework} && \multicolumn{6}{c}{Execution} \\
			\cmidrule{4-16} \cmidrule{18-24} \cmidrule{26-31}
			Publication & Year &&
			\rot{Theodolite \cite{BDR2021}} &
			\rot{Beam Nexmark \cite{BeamNexmark2022}} &
			\rot{ESPBench \cite{Hesse2021}} &
			\rot{OSPBench \cite{vanDongen2020}} &
			\rot{DSPBench \cite{Bordin2020}} &
			\rot{\citet{Shahverdi2019}} &
			\rot{\citet{Karimov2018}} &
			\rot{RIoTBench \cite{Shukla2017}} &
			\rot{YSB \cite{Chintapalli2016}} &
			\rot{SparkBench \cite{Li2015}} &
			\rot{StreamBench \cite{Lu2014}} &
			\rot{Linear Road \cite{Arasu2004}} &
			\rot{Others}
			&&
			\rot{Flink} &
			\rot{Spark} &
			\rot{Storm} &
			\rot{Samza} &
			\rot{Kafka Streams} &
			\rot{Hazelcast Jet} &
			\rot{Others}
			&&
			\rot{Cloud environment} &
			\rot{Distributed} &
			\rot{Different resource amounts} &
			\rot{\dots in isolated experiments} &
			\rot{Different load intensities} &
			\rot{\dots in isolated experiments}
			\\
			\midrule
			This work &
				& %
				& \cmark %
				& %
				& %
				& %
				& %
				& %
				& %
				& %
				& %
				& %
				& %
				& %
				& %
				& %
				& \cmark %
				& %
				& %
				& \cmark\txnote{\fnbeam} %
				& \cmark %
				& \cmark %
				& %
				& %
				& \cmark %
				& \cmark %
				& \cmark %
				& \cmark %
				& \cmark %
				& \cmark %
			\\
			\citet{IC2E2022FaaSStreaming} & \citeyear{IC2E2022FaaSStreaming}
				& %
				& \cmark %
				& %
				& %
				& %
				& %
				& %
				& %
				& %
				& %
				& %
				& %
				& %
				& %
				& %
				& \cmark\txnote{\fnbeam} %
				& %
				& %
				& \cmark\txnote{\fnbeam} %
				& %
				& %
				& \cmark\txnote{\fnbeam} %
				& %
				& \cmark %
				& \cmark %
				& \cmark %
				& \cmark %
				& \cmark %
				& \cmark %
			\\
			\citet{Hesse2021} & \citeyear{Hesse2021}
				& %
				& %
				& %
				& \cmark %
				& %
				& %
				& %
				& %
				& %
				& %
				& %
				& %
				& %
				& %
				& %
				& \cmark\txnote{\fnbeam} %
				& \cmark\txnote{\fnbeam} %
				& %
				& %
				& %
				& \cmark\txnote{\fnbeam} %
				& %
				& %
				& %
				& \cmark %
				& \cmark %
				& \cmark %
				& %
				& %
			\\
			van Dongen\tnote{\fnvandenpoel} \cite{vanDongen2021b} & \citeyear{vanDongen2021b}
				& %
				& %
				& %
				& %
				& \cmark %
				& %
				& %
				& %
				& %
				& %
				& %
				& %
				& %
				& %
				& %
				& \cmark %
				& \cmark %
				& %
				& %
				& \cmark %
				& %
				& %
				& %
				& \cmark %
				& \cmark %
				& \cmark %
				& %
				& \cmark %
				& \cmark %
			\\
			van Dongen\tnote{\fnvandenpoel} \cite{vanDongen2021a} & \citeyear{vanDongen2021a}
				& %
				& %
				& %
				& %
				& \cmark %
				& %
				& %
				& %
				& %
				& %
				& %
				& %
				& %
				& %
				& %
				& \cmark %
				& \cmark %
				& %
				& %
				& \cmark %
				& %
				& %
				& %
				& \cmark %
				& \cmark %
				& %
				& %
				& \cmark %
				& %
			\\
			\citet{Bordin2020} & \citeyear{Bordin2020}
				& %
				& %
				& %
				& %
				& %
				& \cmark %
				& %
				& %
				& %
				& %
				& %
				& %
				& %
				& %
				& %
				& %
				& \cmark %
				& \cmark %
				& %
				& %
				& %
				& %
				& %
				& \cmark %
				& \cmark %
				& %
				& %
				& \cmark %
				& \cmark %
			\\
			\citet{Chu2020} & \citeyear{Chu2020}
				& %
				& %
				& %
				& %
				& %
				& %
				& %
				& %
				& %
				& \cmark %
				& %
				& %
				& %
				& \cmark %
				& %
				& \cmark %
				& %
				& \cmark %
				& %
				& %
				& %
				& \cmark %
				& %
				& %
				& \cmark %
				& \cmark %
				& %
				& %
				& %
			\\
			\citet{Vikash2020} & \citeyear{Vikash2020}
				& %
				& %
				& %
				& %
				& %
				& %
				& %
				& %
				& %
				& %
				& %
				& %
				& %
				& \cmark %
				& %
				& \cmark %
				& \cmark %
				& \cmark %
				& %
				& %
				& %
				& \cmark %
				& %
				& %
				& \cmark %
				& %
				& %
				& \cmark %
				& \cmark %
			\\
			van Dongen\tnote{\fnvandenpoel} \cite{vanDongen2020} & \citeyear{vanDongen2020}
				& %
				& %
				& %
				& %
				& \cmark %
				& %
				& %
				& %
				& %
				& %
				& %
				& %
				& %
				& %
				& %
				& \cmark %
				& \cmark %
				& %
				& %
				& \cmark %
				& %
				& %
				& %
				& \cmark %
				& \cmark %
				& \cmark %
				& %
				& %
				& %
			\\
			\citet{Nasiri2019} & \citeyear{Nasiri2019}
				& %
				& %
				& %
				& %
				& %
				& %
				& %
				& %
				& \cmark %
				& \cmark %
				& %
				& %
				& %
				& %
				& %
				& \cmark %
				& \cmark %
				& \cmark %
				& %
				& %
				& %
				& %
				& %
				& %
				& \cmark %
				& \cmark %
				& \cmark %
				& \cmark %
				& \cmark %
			\\
			\citet{Shahverdi2019} & \citeyear{Shahverdi2019}
				& %
				& %
				& %
				& %
				& %
				& %
				& \cmark %
				& %
				& %
				& %
				& %
				& %
				& %
				& %
				& %
				& \cmark %
				& \cmark %
				& \cmark %
				& %
				& \cmark %
				& \cmark %
				& %
				& %
				& \cmark %
				& \cmark %
				& \cmark %
				& \cmark %
				& %
				& %
			\\
			\citet{Zeuch2019} & \citeyear{Zeuch2019}
				& %
				& %
				& %
				& %
				& %
				& %
				& %
				& %
				& %
				& \cmark %
				& %
				& %
				& \cmark %
				& \cmark %
				& %
				& \cmark %
				& \cmark %
				& \cmark %
				& %
				& %
				& %
				& \cmark %
				& %
				& %
				& \cmark %
				& %
				& %
				& \cmark %
				& \cmark %
			\\
			\citet{Karimov2018} & \citeyear{Karimov2018}
				& %
				& %
				& %
				& %
				& %
				& %
				& %
				& \cmark %
				& %
				& %
				& %
				& %
				& %
				& %
				& %
				& \cmark %
				& \cmark %
				& \cmark %
				& %
				& %
				& %
				& %
				& %
				& %
				& \cmark %
				& \cmark %
				& %
				& \cmark %
				& \cmark %
			\\
			\citet{Truong2018} & \citeyear{Truong2018}
				& %
				& %
				& %
				& %
				& %
				& %
				& %
				& %
				& %
				& %
				& %
				& %
				& %
				& \cmark %
				& %
				& %
				& %
				& %
				& %
				& %
				& %
				& \cmark %
				& %
				& \cmark %
				& \cmark %
				& %
				& %
				& \cmark %
				& \cmark %
			\\
			\citet{Karakaya2017} & \citeyear{Karakaya2017}
				& %
				& %
				& %
				& %
				& %
				& %
				& %
				& %
				& %
				& \cmark %
				& %
				& %
				& %
				& %
				& %
				& \cmark %
				& \cmark %
				& \cmark %
				& %
				& %
				& %
				& %
				& %
				& %
				& \cmark %
				& %
				& %
				& \cmark %
				& \cmark %
			\\
			\citet{Shukla2017} & \citeyear{Shukla2017}
				& %
				& %
				& %
				& %
				& %
				& %
				& %
				& %
				& \cmark %
				& %
				& %
				& %
				& %
				& %
				& %
				& %
				& %
				& \cmark %
				& %
				& %
				& %
				& %
				& %
				& \cmark %
				& \cmark %
				& \cmark %
				& %
				& %
				& %
			\\
			\citet{Yang2017} & \citeyear{Yang2017}
				& %
				& %
				& %
				& %
				& %
				& %
				& %
				& %
				& %
				& \cmark %
				& %
				& %
				& %
				& \cmark %
				& %
				& \cmark %
				& \cmark %
				& \cmark %
				& %
				& %
				& %
				& %
				& %
				& \cmark %
				& \cmark %
				& %
				& %
				& %
				& %
			\\
			\citet{Chintapalli2016} & \citeyear{Chintapalli2016}
				& %
				& %
				& %
				& %
				& %
				& %
				& %
				& %
				& %
				& \cmark %
				& %
				& %
				& %
				& %
				& %
				& \cmark %
				& \cmark %
				& \cmark %
				& %
				& %
				& %
				& %
				& %
				& %
				& \cmark %
				& \cmark %
				& \cmark %
				& %
				& %
			\\
			\citet{Lopez2016} & \citeyear{Lopez2016}
				& %
				& %
				& %
				& %
				& %
				& %
				& %
				& %
				& %
				& %
				& %
				& %
				& %
				& \cmark %
				& %
				& \cmark %
				& \cmark %
				& \cmark %
				& %
				& %
				& %
				& %
				& %
				& %
				& \cmark %
				& %
				& %
				& \cmark %
				& \cmark %
			\\
			\citet{Qian2016} & \citeyear{Qian2016}
				& %
				& %
				& %
				& %
				& %
				& %
				& %
				& %
				& %
				& %
				& %
				& \cmark %
				& %
				& %
				& %
				& %
				& \cmark %
				& \cmark %
				& \cmark %
				& %
				& %
				& %
				& %
				& %
				& \cmark %
				& \cmark %
				& \cmark %
				& %
				& %
			\\
			\citet{Lu2014} & \citeyear{Lu2014}
				& %
				& %
				& %
				& %
				& %
				& %
				& %
				& %
				& %
				& %
				& %
				& \cmark %
				& %
				& %
				& %
				& %
				& \cmark %
				& \cmark %
				& %
				& %
				& %
				& %
				& %
				& %
				& \cmark %
				& \cmark %
				& \cmark %
				& %
				& %
			\\
			\bottomrule
		\end{tabularx}
		\begin{tablenotes}\footnotesize
			\item[\fnvandenpoel] and van den Poel
			\item[\fnbeam] using Apache Beam
		\end{tablenotes}
	\end{threeparttable}
\end{table*}

\cref{tab:related-experiments} provides an overview of experimental performance evaluation and benchmarking studies. It indicates the applied benchmark, the evaluated stream processing, and information regarding the experiment setup and method. The latter includes whether the respective study was performed in a cloud environment, in a distributed fashion with multiple instances of the framework deployed. Moreover, it shows whether the benchmarks have been executed with different resource amounts and different load intensities and whether different resource amounts and load intensities are evaluated in isolated experiments. In previous work, we argued that scalability should be evaluated with isolated experiments for different combinations of load and resources~\cite{LTB2021,EMSE2022}.

We can observe that there is no established stream processing benchmark. Only YSB is used in several studies. However, YSB can be considered a micro-benchmark~\cite{Bermbach2017} and, hence, is less suited to benchmark entire microservices.
Except for the preliminary evaluation of our Theodolite benchmarks~\cite{BDR2021}, there is no benchmarking study addressed to stream processing frameworks employed within microservice architectures.

Flink, Spark, and Strom are by far the most frequently benchmarked frameworks. Kafka Stream, Hazelcast Jet, and Samza, which are particularly suited to be deployed as microservices, are only benchmarked in a few studies and there is no study benchmarking all of them.

9 out of 20 studies report on experiments in public or private clouds.
Except for this and our previous study~\cite{IC2E2022FaaSStreaming}, there are no evaluations in Kubernetes.
Likewise, there are no further studies evaluating scalability with a systematic approach as we do in this study. \citet{Vikash2020}, \citet{Nasiri2019}, \citet{Karakaya2017}, and \citet{vanDongen2021b} explicitly evaluate scalability, however, without testing different load intensities against different resource amounts in isolated experiments. \citet{Nasiri2019} conduct independent evaluations of scaling load and computing resources and, thus, address another aspect than our study.
Our previous study~\cite{IC2E2022FaaSStreaming} applies our Theodolite method as well, but benchmarks scalability with respect to costs and is addressed to comparing stream processing deployments against Function-as-a-Service offerings.
While this work particularly addresses scalability, there are other recent studies that benchmark qualities such as throughput~\cite{Bordin2020,Chu2020,vanDongen2020}, latency~\cite{Bordin2020,vanDongen2020,Hesse2021}, and fault-tolerance~\cite{vanDongen2021a}.

\section{Conclusions}\label{sec:conclusions}

In this study, we benchmark the scalability of distributed stream processing frameworks particularly suited to be used within microservices.
We experimentally evaluate the frameworks  Apache Flink, Apache Kafka Streams, Hazelcast Jet, and Apache Beam with the Apache Flink runner and the Apache Samza runner in a private cloud environment and in the Google cloud.
We find that all frameworks show linear scalability for most use cases, however with partially significantly different resource demands.
There is no clear superior framework. Instead, depending on the use case Flink, Hazelcast Jet, or Kafka Streams show the lowest increase in resource demand.
Using Apache Beam as abstraction layer still comes with a significant negative impact on performance, leading to a significantly steeper increase in resource demand, independent of the use case.
We observe our results regardless of scaling load on a microservice, scaling the computational work performed inside the microservice, the selected cloud environment, and whether the microservice is scaled over multiple nodes or on a single node.
The latter means that vertical scaling distributed stream processing frameworks can---to some extent---also complement horizontal scaling.
In summary we found that while scalable microservices can be designed with all evaluated frameworks, the choice of a framework and its deployment has a considerable impact on the cost of operating it.
This emphasizes also a key benefit of designing stream processing applications using microservice-based architectures: the ability for each service to select the most suitable stream framework based on its specific use case, requirements for scalability, and other relevant considerations.
Our findings leave room for future research on exploring the multi-dimensional space of cloud deployment options and finding an optimal one.
Moreover, our work paves the way for deeper analysis of the underlying reasons for our observed results by benchmarking further task samples, deployments, and configuration options.

\section*{Acknowledgments}
This works is supported by the Google Cloud Research Credits program under award no.\ 211920173.
We would like to thank JKU and Dynatrace for co-funding one of the authors (S.~Henning) during the revision of this article.

\bibliography{references}

\end{document}